\documentclass[pre,showpacs,amsmath,amssymb,floatfix,onecolumn]{revtex4-1}
\usepackage{graphicx}
\usepackage{epsfig}
\usepackage{dcolumn}
\usepackage{bm}
\usepackage{color}
\usepackage{epstopdf}
\usepackage{subfigure}
\usepackage{amsmath,amssymb,amsthm,mathrsfs,amsfonts,dsfont} 
\usepackage{algorithmic}
\usepackage{algorithm}

\theoremstyle{plain}

\theoremstyle{definition}

\theoremstyle{remark}

\def\atanh{{\rm atanh}}

\def\bs{\bf s}
\def\bx{{\bf x}}
\def\bJ{{\bf J}}
\def\bS{{\bf S}}
\def\by{{\bf y}}
\def\bh{{\bf h}}
\def\bz{{\bf z}}
\def\bJ{{\bf J}}
\def\by{{\bf y}}
\def\bh{{\bf h}}

\def\(({\left(}
\def\)){\right)}                       
\def\[[{\left[}
\def\]]{\right]}
\def\e{{\rm e}}

\newcommand{\E}{\mathbb{E}}

\newcommand{\<}{\langle}
\renewcommand{\>}{\rangle}

\newcommand{\beq}{\begin{equation}}
\newcommand{\eeq}{\end{equation}}
\newcommand{\be}{\begin{equation}}
\newcommand{\ee}{\end{equation}}
\newcommand{\bea}{\begin{eqnarray}}
\newcommand{\eea}{\end{eqnarray}}
\newcommand{\sign}{\text{sign}}

\newcommand{\di}{\partial i}

\def\dimj{{\partial i \setminus j}}

\begin{document}


\title{Statistical physics of inference: Thresholds and algorithms}

\author{Lenka Zdeborov\'a$^{1,*}$, and Florent Krzakala$^{2,*}$}

\affiliation{
  $^1$  Institut de Physique Th\'eorique, CNRS, CEA, Universit\'e Paris-Saclay, F-91191, Gif-sur-Yvette, France \\
  $^2$ Laboratoire de Physique Statistique, CNRS, PSL Universit\'es \\
  Ecole Normale Sup\'erieure. Sorbonne Universit\'es \\
  Universit\'e Pierre \& Marie Curie, 75005, Paris, France.\\
  $^*$ lenka.zdeborova@cea.fr and florent.krzakala@ens.fr\\
}


\begin{abstract}
  Many questions of fundamental interest in today's science can be
  formulated as inference problems: Some partial, or noisy,
  observations are performed over a set of variables and the goal is
  to recover, or infer, the values of the variables based on the
  indirect information contained in the measurements. For such
  problems, the central scientific questions are: Under what
  conditions is the information contained in the measurements
  sufficient for a satisfactory inference to be possible? What are the
  most efficient algorithms for this task?  A growing body of work has
  shown that often we can understand and locate these fundamental
  barriers by thinking of them as phase transitions in the sense of
  statistical physics. Moreover, it turned out that we can use the
  gained physical insight to develop new promising algorithms.  The
  connection between inference and statistical physics is currently
  witnessing an impressive renaissance and we review here the current
  state-of-the-art, with a pedagogical focus on the Ising model which,
  formulated as an inference problem, we call the planted spin glass.
  In terms of applications we review two classes of problems: (i)
  inference of clusters on graphs and networks, with community
  detection as a special case and (ii) estimating a signal from its
  noisy linear measurements, with compressed sensing as a case of
  sparse estimation. Our goal is to provide a pedagogical review for
  researchers in physics and other fields interested in this
  fascinating topic.
\end{abstract}

\maketitle




\tableofcontents

\section{Introduction}
Our goal in this review is to describe recent developments in a
rapidly evolving field at the interface between statistical inference
and statistical physics. Statistical physics was developed to derive
macroscopic properties of material from microscopic physical
laws. Inference aims to discover structure in data. On the first look,
these goals seem quite different.  Yet, it was the very same man,
Pierre Simon, Marquis de Laplace (1749--1827), who did one of the
first sophisticated derivations of the gas laws within the caloric
theory, and who also created the field of statistical inference
\cite{INSR:INSR179_3}. This suggests there may be a link after all. In
fact, the methods and tools of statistical physics are designed to
describe large assemblies of small elements - such as atoms or
molecules. These atoms can quite readily be replaced by other
elementary constituents - bits, nodes, agents, neurons, data points.
This simple fact, and the richness and the power of the methods
invented by physicists over the last century, is at the roots of the
connection, and the source of many fascinating work in the last
decades.

There is a number of books discussing this interdisciplinary
connection.  Without claim of exhaustivity, see for instance
\cite{grassberger1994statistical,NishimoriBook01,Engel:2001:SML:558792,MezardMontanari07,LesHouches2013}.
Our focus here is to present some more recent contributions that were
obtained using methodology that originated in the field of spin
glasses \cite{MezardParisi87b}.  In particular the replica and cavity
methods, with the related message passing algorithms, are at the roots
of our toolbox. We focus on the analogy between phase transitions in
thermodynamics, and the threshold phenomena that arise in the ability to infer something useful from data. The most
interesting outcome of these considerations are deep consequences for the
possibility of development of new algorithms, and their relations to
computational feasibility. This focus is summarized in the title by
the words: ``threshold and algorithms''.

As we shall see, the natural setup in which the connection between
inference and the statistical physics of disordered systems is
particularly transparent is the so-called {\it teacher-student}
scenario. In a nutshell, the teacher-student scenario can be described
as follows: The teacher uses some ground truth information and a
probabilistic model to generate data that he then hands to the student

who is supposed to recover the ground truth as well as he can only
from the knowledge of the data and the model. This formulation of
inference problems is particularly popular in neural networks
\cite{Engel:2001:SML:558792,seung1992statistical,watkin1993statistical},
where the teacher uses a rule to generate output, and the student is
then trying to learn the rule to reproduce the same output. The
advantage of this setting is that it is easily amenable to the
Bayesian approach and strategies that recover the ground truth in an
optimal way can be analyzed.

Let us make several comments: First we do not make this Bayesian
choice because we think the Bayesian approach is more correct than
other lines of thought common in statistics. We do it because it is in
this setting that the connection between statistical inference and
statistical physics becomes the most transparent. Our main point is to
explain and illustrate the insight that can be obtained from this
connection in terms of phase transitions.  In a sense the strategy
followed in this review is closer to the approach in information
theory where we know the properties of the noise that corrupted the
signal than to traditional approach in statistics where we aim to
assume as little as possible.

Second, we focus on the teacher-student scenario and consequently we
{\it do not} analyze problems where one is given data or observations
and has absolutely no idea of the underlying system that generated
them. Nor do we discuss model selection where one needs to judge from
data which model is more suitable. There is a range of textbooks on
statistics where such problems are studied
\cite{hastie2005elements,barber2012bayesian,MacKay03}.  In theoretical
physics it is common to study abstract models that display interesting
behavior and use gained knowledge to generalize about what is
happening in the real world.

Third, while the subject is certainly algorithmic in nature, we shall
concern ourselves primarily, as the title shows, with phase transitions
and threshold phenomena, where the connection with statistical physics
is clear.  This means that while we certainly are interested
in analyzing and discussing algorithms, and ask if it is possible
for them to reach the phase transition boundaries, we do {\it not}
review, for any given problem, the vast amount of literature used to
actually solve it in practice. Our goal,
instead, is (in a typical statistical physics attitude) to reach
some ``universal'' understanding of the problem.

The typical reader we have in mind has some knowledge of statistical
physics, knows e.g. about the Ising model, phase transition and free
energy. We have in mind a reader who is curious to know how methods
and concepts from statistical physics such as phase transitions
translate into the context of machine learning and inference. Perhaps
the most important message we want to pass in this review is that many
calculations that physics uses to understand the world can be
translated into algorithms that machine learning uses to understand
data.  We also have in mind readers from statistics and computer
science who might want to learn more about the statistical physics
approach. While we do explain some basic physics concepts, such readers
could further read a short introduction to statistical physics, as for
instance the appendix B (``some physics'') in the book of David McKay
\cite{MacKay03}, or the Chap II in \cite{MezardMontanari07}.

\subsection{Organization of the manuscript}
The present review is organized as follows.

In section~\ref{sec:inference} we introduce the basic concepts of
statistical inference and the terminology. Most of this section is
standard knowledge in statistics and the main purpose of this section
is to set the notations and present the concepts to readers from
physics who are not particularly familiar with probability and
statistics. The only part of sec.~\ref{sec:inference} that is perhaps
not standard, but certainly crucial for the present discussion, is the
idealized teacher-student setting of
Sec.~\ref{teacher_student}. Traditional statistics would typically not
assume knowledge of the model that generated the data and large part
of statistics is precisely concerned with determining a good
model. The main focus of this review is, however, on the connections
to statistical physics. In the teacher-student setting, the analysis
of exact phase diagrams and phase transitions is possible and this is
our primary goal.

In Sec.~\ref{phys_inf} we present some basic concepts of statistical
physics that non-physics readers might not be familiar with. The
emphasis is given on formulation of the teacher-student setting as a
planted statistical ensemble -- this {\it planted} ensemble is not
standard in physics but we do believe it is a really fundamental
concept, that has become popular in the last decades, and that
introducing it greatly elucidates the deep connection between
properties of high-dimensional inference problems, optimization
problem, and disordered systems.

Having introduced the basic concepts we then choose a bottom-up
approach and present in sec.~\ref{planted_Ising} a prototypical
inference problem that we call the planted spin glass. We discuss
how the physics of this inference problem can be deduced using the
existing knowledge from the physics of spin glasses. The main
highlight of this section is that Bayes-optimal inference can be
translated into thermodynamics on the Nishimori line in the underlying
planted model. The physical properties of the Nishimori line are well
understood for a wide range of models, this connection hence provides
a unifying picture for a large class of inference
problems. Consequently, concepts discussed in
section~\ref{planted_Ising} bring a lot of insight into many inference
problems other than the planted spin glass.  We conclude section
\ref{planted_Ising} with a discussion on how the concept of planting
turns out to be innovative in study of glasses in
sec.~\ref{sec:glasses}. All the results of this section are presented
in a way that should be understandable without going into any
calculations. The purpose of this is twofold: Readers who are not
familiar with the replica and the cavity calculations can still
understand the reasoning and the main results. Readers who are
familiar with the calculations can appreciate the reasoning,
connections and consequences without having to separate them from the
technicalities.

Section~\ref{sec:cavity}, however, is a more analytical chapter, that
should be read with a pencil so that one can redo the computations. It
summarizes the basic analytical technique based on belief propagation
and the cavity method that we can use to analyse the properties and
phase diagrams of high-dimensional inference problems. We do not go
into many details and the related methods are presented in
greater depths in other existing literature and textbooks,
e.g.~\cite{MezardMontanari07}. Readers familiar with these techniques
can skip this section, except perhaps for its last part \ref{phase_various} where we present the
zoology of the main phase transitions that are encountered in
inference problems.

Armed with the statistical physics' understanding of inference from
sections~\ref{planted_Ising}, \ref{sec:cavity}, we discuss related
recent algorithmic ideas in Sec.~\ref{sec:algs}.  The various
algorithmic ideas are discussed again on the example of the planted
spin glass model or its simple variants. This section can be read
separately from the previous ones by readers interested mainly in the
algorithmic consequences stemming from physics. 

The planted spin glass problem is a simple example. To
illustrate how to generalize from it to other, more practically
relevant, inference problems we present in
section~\ref{chap:clustering} the problem of clustering of networks
with its recent development, and in section~\ref{chap:CS} the
generalized linear estimation problem with the compressed sensing as
its special case. The two examples of applications in
sections~\ref{chap:clustering},~\ref{chap:CS} are chosen with respect
to author's own work. Each of these last two sections can be seen as a
concise introduction into these practically relevant problems and
is therefore mainly for readers that are interested in various
applications of physics concepts or who want to specifically start
learning about clustering or compressed sensing. 

Finally, a word of caution about the subjects that we have chosen to cover
(and more importantly {\it not to cover}). Historically, one of the
most prominent example of statistical inference studied in physics are
the error correcting codes. Their studies in physics were pioneered by
Sourlas~\cite{Sourlas89} and the results are reviewed nicely e.g. in
\cite{NishimoriBook01,montanari2007modern,MezardMontanari07}. 
We do not discuss error correction in this review, except for a couple
of interesting and/or historical connections.

Another very prominent problem that can be formulated as statistical
inference and that has been extensively studied by tools of
statistical physics are neural networks. The book
\cite{NishimoriBook01} covers very nicely the early results. A very
detailed presentation, with an emphasis on teacher-student protocols,
can be found in the book \cite{Engel:2001:SML:558792}. The reviews \cite{seung1992statistical,watkin1993statistical} are also
very good references. While we do not focus on this application that
is rather well documented, we will see in section
\ref{chap:CS} that the problem of perceptron is a special case of the
generalized linear estimation setting.

The two applications, network clustering and compressed sensing, that
we chose to present in this review have beed studied only more
recently (in the last decade) and there are interesting results
stemming from the connection between inference and physics that are
not (yet) well covered in other reviews.

\subsection{What is statistical inference?}
\label{sec:inference} 
Inference is the act or process of deriving logical conclusions from
premises known or assumed to be true. Statistical inference is the
process of deducing properties of an underlying distribution by
analysis of data. A way to think about inference problems, that is
particularly relevant in the context of today's data deluge (often
referred to as {\it big data}), are procedures to extract useful
information from large amounts of data. As the boundaries between
fields are breaking, nowadays inference problems appear literally
everywhere, in machine learning, artificial intelligence, signal
processing, quantitative biology, medicine, neuroscience and many
others.

A more formal way to set an inference problem is the following: One
considers a set of variables ${\bf x} = \{x_i \}_{i=1,\dots,N}$ on
which one is able to perform some partial observations or measurements
${\bf y} = \{y_\mu\}_{\mu = 1,\dots, M}$. The goal is to deduce, or
infer, the values of the variables~${\bf x}$ based on the perhaps
indirect and noisy information contained in the observed
data~${\bf y}$. To give a couple of examples, the variables may
correspond for instance to image pixels, an acoustic signal or to an
assignment of data points into clusters, while the corresponding
observations would correspond for instance to a blurred photograph, a
recording of a song performed on a platform while the train was
arriving, or positions of points in space.

%
For inference problems the two main
scientific questions of interest are:
\begin{itemize}
\item The question of sufficient information -- Under what conditions is
  the information contained in the observations sufficient for
  satisfactory recovery of the variables?
\item The question of computational efficiency -- Can the inference be
  done in an algorithmically efficient way? What are the
  optimal algorithms for this task?
\end{itemize}
These two questions are of very different nature. The first one is a
question that belongs to the realm of statistics and information
theory. The second one, instead, is in the realm of computer science
and computational complexity. Before the era of big data, statistics
was often not concerned with computational tractability, because the
number of parameters to be estimated was usually small. Nowadays the tradeoff
between statistical and computational efficiency has changed, making it most reasonable to study the two above questions together.

We will see that these two questions have a very clear interpretation
in terms of phase transitions in physics. More specifically, the first one is
connected to a genuine phase transition while the second one is
connected to the type of the transition (first or second order), and
metastability associated with first order phase transitions is related
to algorithmic hardness.

\subsubsection{Terminology and simple examples}
\label{sec:example}

In this review, we will follow the approach of Bayesian
inference. Developed by Bayes and Laplace, it concentrates on
conditional probabilities.  The probability of having an event $A$
conditioned on the event $B$ is denoted $P(A|B)$ and defined as
\begin{equation}
    P(A|B) = \frac{ P(A,B) }{P(A)} \, , \label{eq:cond}
\end{equation}
where $P(A,B)$ is the joint probability of both events $A$ and $B$,
and $P(A)=\sum_{B} P(A,B)$ is the probability of event $A$ (the sum is
taken over all the possible realizations of event $B$). 
When events $A$, $B$ are taken from a continuous domain the $P(A)$ denotes
probability distribution function, whereas $P(A) {\rm d} A$ denotes
the probability of observing an outcome in the interval $(A,A+{\rm d}
A)$. For discrete variables the probability of one event, and the
probability distribution are interchangeable. Note that in this review
we discuss both models with discrete variables (sec.~\ref{planted_Ising} or \ref{chap:clustering})
and with continuous variables (sec.~\ref{chap:CS}), the concepts that we
build hold for both these cases. 

The central point of the approach is the so-called Bayes formula that
follows directly from the definition of the conditional probability (\ref{eq:cond})
\begin{equation}
P({\bx}|{\by})=\frac{P({\by}|{\bx})}{P({\by})}P({\bx})\, .
\label{posterior}
\end{equation}
There is a precise terminology for all the terms entering this
equation, so we shall start by discussing the Bayesian inference vernacular:
\begin{itemize}
\item $P({\bx}|{\by})$ is called the {\it posterior} probability distribution for $\bx$
  given all the data $\by$. 
\item $P({\by}|{\bx})$ is called the {\it likelihood}. 
\item $P({\by})$, also denoted as a normalisation $Z({\by})$,
  is called the {\it evidence}.
\item $P({\bx})$ is called the {\it prior} probability distribution for $\bx$. It
  reflects our belief of what $\bx$ should be before we
  received any data.
\end{itemize}

\paragraph*{ Example of decaying particles.} 
Our first example is borrowed from Ref. \cite{MacKay03}. Unstable particles are emitted from a
source and decay at a distance $y$ from the source. Unfortunately, we
cannot build a perfect infinite length detector, and decay events can
only be observed if they occur in a window extending from distances
$y_{\rm min} = 1~\text{cm}$ to $y_{\rm max} = 20~\text{cm}$ from the
source.

We run the experiment for some time, and we observe $M$ decays at
locations $\by = \{y_1 , . . . , y_M \}$.  We also know that we should
expect an exponential probability distribution with characteristic
length $x$ for the event.  Writing things in the inference language,
${\bf y}$ are the observations and $x$ is the unknown signal (of
dimension $N=1$).  The question we would like to answer is: Given
the observations, how to estimate $x$? 

In this particular example physics tells us exactly what is
the probability distribution of a particle decaying at distance $y$
given the characteristic length $x$. In terms of Bayesian probability
this means that we know exactly the model that created the observations. This is of
course not true in general, but when it is true we can
set up the so-called Bayes-optimal inference: The likelihood for one observation is
\begin{equation}
  P(y|x)=\begin{cases}
    \frac{1}{x {\cal M}(x) }e^{-y/x} &\text{if } 1<y<20\, ,\\
    0 &\text{otherwise}\, ,
  \end{cases}
  \label{Eq:emissions}
\end{equation}
with a normalisation
${\cal M}(x)=\int_1^{20
}dz\frac{1}{x}e^{-z/x}=e^{-1/x}-e^{-20/x}$.
The observations are independent, and therefore in this case their joint probability
distribution is simply a product of the single observation distributions. From eq. (\ref{Eq:emissions}) we can now extract $P(x|\by)$
using the Bayes' formula and get
\begin{equation} 
 P(x|\by)= \frac{P(x)}{Z({\bf y})\left[x {\cal
       M}(x)\right]^M}e^{-\sum_{i=1}^My_i/x} \, .
\end{equation}
Looking at this equation, we see that we have $P(x)$ on the
right hand side. If we do not have any additional information on $x$, we assume
than the prior distribution $P(x)$ is a constant that just enters in the $x$-independent
normalization, denoting then $\tilde Z({\bf y})=Z({\bf y})/P(x)$, we obtain
\begin{equation}
 P(x|\by)=\frac{1}{\tilde{Z}({\bf y})\left[x {\cal
       M}(x)\right]^M}e^{-\sum_{i=1}^My_i/x}\, .
\end{equation}
This is the final answer from Bayesian statistics. It contains all the
information that we have on $x$ in this approach. For a dataset
consisting of several points, for instance the six points
${y}={1.5, 2, 3, 4, 5, 12}$ cm, one can compute the most probable
value of $x$ to be around $\hat x = 3.7$ cm. Note that this is
considerably different from the mere average of the points which is
$\overline y = 4.6$ cm. The reader can convince herself (for instance
performing simulation and comparing to the true correct value) that
the presented approach gives a way better result that the naive average.

\paragraph*{Example of linear regression.} Most physicists are used to fit a
theoretical model to experimental data. Consider for instance that we
measure the height of an object falling due to Newton's gravity,
we assume no initial speed and no friction. We expect that the
height $h(t)$ of the object at time $t$ will follow
$h(t,g,h_0)=h_0-gt^2/2$, where $g$ is the acceleration of gravity. We can
do an experiment and measure heights of the object $y_i$ for many values of $t_i$, with $i=1,\ldots,M$.
We expect, if we plot $y_i$ against $t_i^2/2$, to get a straight line
that will give us the value of the acceleration $g$.

Due to small but real measurement errors, the measured points
will not lie exactly on a line. Used to such a situation, a physicist
would in this case perform a linear regression to minimize the so-called sum of squares of errors. The ``best" value of $g$, or to use
the statistics language the {\it estimator} $\hat{g}$ of $g$, would
simply be the one that minimizes the cost function
$\sum_{i=1}^M [ h(t_i,g,h_0)-y_i]^2$.

In order to get more familiarity with the concepts of inference we discuss how this familiar procedure can be cast
into the Bayesian language.  In general we may not know much about the nature of the measurement
noise. In such cases it is often reasonable to expect that the noise
is a Gaussian random variable of zero mean and variance
$\Delta$. Under this assumption the likelihood $P(y_i|g,\Delta)$ of a single data point is
a Gaussian with mean $h(t_i,g,h_0)$ and variance $\Delta]$. We assume that
the noise for different measurements is independent which leads to the
total likelihood
\be
      P({\bf y}|g,h_0,\Delta)=  \frac{1}{(2\pi\Delta)^{\frac{M}{2}}} \prod_{i=1}^M  e^{-\frac{[  h(t_i,g,h_0) - y_i ]^2}{2\Delta}}  
\ee
As in the previous example we assume a
uniform prior distribution on the acceleration constant $P(g)$, initial
height $P(h_0)$ and the
variance $P(\Delta)$, the posterior probability distribution is then given by
\be
    P(\Delta,g,h_0|\by) =  \frac{1}{  Z(\by) (2\pi \Delta)^{\frac{M}{2}}} \prod_{i=1}^M  e^{-\frac{[  h(t_i,g,h_0) - y_i ]^2}{2\Delta}} 
\ee
Looking for the most probable value of $g$ indeed amounts to minimizing
the sum of squares $\sum_{i=1}^M [h(t_i,g,h_0)-y_i]^2$, which is the standard
solution. Note, however, that the assumption of Gaussian noise
independent for every measurement might not be justified in
general. Also, depending on the precise setting, there might be better prior distributions for the
parameters $g$, $h_0$, $\Delta$. For more advanced discussion of
classical linear
regression see for instance \cite{wasserman2013all}. In section
\ref{chap:CS} of this review we will discuss
a high-dimensional version of linear regression, when the number of
parameters to be estimated is comparable to the number of observations
$M$. 


\medskip

Let us make some important comments on the above considerations, which
have been at the roots of the Bayesian approach to inference from the
very start. First, we notice that probabilities are used here to quantify degrees
of belief. To avoid possible confusions, it must thus be emphasized
that the true ${\bx}$ in the real world {\it is not} a random variable, and the fact that
Bayesians use a probability distribution $P({\bx}|{\by})$ does not mean
that they think of the world as stochastically changing its nature
between the states described by the different hypotheses. The
notation of probabilities is used here to represent the beliefs about
the mutually exclusive hypotheses (here, values of $x$), of
which only one is actually true. The fact that probabilities can denote degrees
of belief is at the heart of Bayesian inference.
The reader interested in these considerations is referred to the very
influential work of Bruno de Finetti \cite{de1979theory}.

Another subtle point is how to choose the prior distribution
$P({\bx})$. In fact, discussions around this point have been rather
controversial in the beginning of the history of statistical
inference, and even now there are many schools of thought.
It is not the goal of this review to present these discussions and
debates. In the spirit of a reductionism that is usually done in physics, we
introduce in the next section the so-called teacher-student scenario  in which we know what
the model and the prior distribution really are, so that the use of the Bayes formula is
straightforward and fully justified. Majority of this review is then
set within this teacher-student framework. This can be seen as a simplified
model of the generic situation. As we will see the picture of what
is happening is interestingly rich already in this
framework. A first step towards the generic situation of unknown models
that generated the data
and unknown prior probabilities is the case when the teacher hides
some parameters of the model or/and of the prior distribution from the
student. Analyzing this case provides insight into the generic case,
but still within the realm of a solvable simplified teacher-student framework.

In sec. \ref{planted_intro} we then argue that the teacher-student scenario for a generic
inference problem can be seen as a so-called {\it planted 
ensemble} of the corresponding statistical physics model.  This is 
a known but not widely used way of presenting inference in statistical
physics, but the authors feel that it is an extremely useful and
insightful way. It allows to extend the intuition gathered over several
decades in studies of disordered systems such as glasses and spin
glasses into many currently studied inference problems that otherwise
appear highly non-trivial.
One of the goals of the present review is to bring this insight forward
and establish the way of thinking about inference problems as planted
statistical ensembles.

\subsubsection{Teacher-student scenario}
\label{teacher_student}
%
%
The {\it teacher-student
  scenario} is defined as follows:

\begin{itemize}
\item Teacher: 
In a first step the teacher generates a realization of variables/parameters $\bx^*$ from some probability
distribution $P_{\rm tp}(\bx)$, where 'tp' stands for teacher's prior.
In a second step he uses these {\it ground truth}
values $\bx^*$ to
generate the data $\by$ from some statistical model characterized by a
likelihood of $\by$ given $\bx^*$, denoted as $P_{\rm tm}(\by|\bx^*)$, where 'tm' stands for teacher's
model. Finally, the teacher hands the data $\by$ to
the student together with some information about the distributions
$P_{\rm tp}(\bx)$ and $P_{\rm tm}(\by|\bx)$. 
\item Student: The goal of the student is to infer as precisely as
  possible (and tractably, when computational complexity is included
  into considerations) the original values $\bx^*$ of the variables from the
  information provided by the teacher, i.e. from the data $\by$ and
  the available information about the distributions $P_{\rm tp}(\bx)$ and $P_{\rm tm}(\by|\bx)$. 
\end{itemize}

The variables with a $*$ (e.g. $\bx^*$) denote the ground truth that was used by the
teacher, but that is not available to the student.  
In the present manuscript we will consider cases where both
the $\bx$ and $\by$ are high-dimensional vectors, and where the
teacher's prior $P_{\rm
       tp}(\bx)$ and model $P_{\rm tm}(\by|\bx)$ are parameterized by
     some scalar (or low-dimensional vector) values 
     $\theta^*$ and $\Delta^*$, respectively. This is a case
     relevant to many real-world problems where we assume existence of
     many latent variables (collected into the vector $\bx$), and we
     assume that the model and prior distribution is parametrized by a small number of
     unknown parameters.   

We will in particular be interested in inference
problems where the teacher's prior distribution is separable, i.e. can be written as a product over
components
\be
       P_{\rm tp}(\bx)  = \prod_{i=1}^N  P_{\rm tp}(x_i)\, .
\ee 
Further, in this review, we are interested in cases where also the observations
$\{y_\mu\}_{\mu=1,\dots,M}$ are independent and each of them depends
on a subset of variables denoted $\partial \mu$. The likelihood of the
teacher's model is then written as 
\be
        P_{\rm tm}(\by|\bx) = \prod_{\mu=1}^M   P_{\rm tm}(y_\mu| \{x_i\}_{i\in \partial
        \mu}) \, .
\ee

Given the information available to the student, she then 
follows the strategy of Bayesian statistics. The prior information she
has from the teacher is denoted as $P(\bx)$, where $\bx$ are the
variables to be inferred. The statistical model is represented by its
likelihood $P(\by|\bx)$. Again we assume that both these can be
written in a product form as above. She then considers Bayes'
theorem~(\ref{posterior}) and writes the posterior probability which
contains all the information available to her
\be
       P(\bx | \by) =   \frac{1}{Z(\by)} \prod_{i=1}^N   P(x_i) \prod_{\mu=1}^M  P(y_\mu|\{x_i\}_{i\in \partial  \mu})  \, . \label{post_fact} 
\ee
A very useful quantity to define that derives directly from the posterior
distribution is the marginal probability of one variable
 \be \mu_i(x_i) = \int P(\bx
| \by) \prod_{j\neq i} {\rm d} x_j\, . \label{marg} 
\ee

Examples:
\begin{itemize}
\item Let us set the problem of simple linear regression that we used
  as an example in section \ref{sec:example} in the teacher-student
  setting. Consider the teacher generated a number $g^*$ from some
  probability distribution $P_{\rm tp}(g)$, this $g^*$ is the
  ground-truth value we aim to find back as precisely as possible from the noisy measurements. Then the teacher generated
  heights $y_i$ independently from a Gaussian probability distribution
  of mean $-g^*t_i^2/2$ and variance $\Delta^*$, where $t_i$
  are the known measurement times, $i=1,\dots,N$. He then handed
  the heights $\by$ and the distribution $P_{\rm tp}(g)$, $P_{\rm
    tm}(y_i|g)$ to the student. Set in this way the Bayesian solution
  presented in sec.~\ref{sec:example} has no ambiguity in terms of
  assuming the noise was Gaussian and independent, nor in terms of the
  choice of the prior distribution for $g$.  
\item Historically the teacher-student setting was introduced for the
  problem of perceptron, called model B in
  \cite{gardner1989three}. In perceptron the aim is to learn {\it weights}
  $J_i$, $i=1,\dots,N$ in such a way that the scalar product with
  each of $P$ {\it patterns} $\xi_i^\mu$, $\mu=1,\dots,P$ is
  constraint to be either smaller ($y_\mu=-1$) of larger ($y_\mu=1$) than some threshold
  $\kappa$. Mathematically stated one requires for all $\mu=1,\dots,M$
  \be
      y_\mu  =  {\rm sign}( \sum_{i=1}^N J_i \xi_i^\mu -  \kappa)
      \, .   \label{perceptron}
  \ee 
  This perceptron setting models classification of patterns into two
  classes. One way to set a solvable version of the model is
  to consider both $\xi_i^\mu$ and $y_\mu$ to be independent
  identically distributed (iid) random
  variables. The goal is then to find a set of weights $J_i$ such that
  the constraints (\ref{perceptron}) are satisfied. Another way to set the
  problem that corresponds to the teacher-student scenario is to
  consider some teacher weights $J_i^*$,
  generated from some probability distribution, e.g. $\pm 1$ uniformly
  at random. The teacher perceptron then
  generates the output as $y_\mu= {\rm sign}( \sum_{i=1}^N J^*_i
  \xi_i^\mu -  \kappa)$, where $\xi_i^\mu$ are some known iid patterns. The goal is then to recover the teacher
  weights $J_i^*$ from the knowledge of $\xi_i^\mu$ and  $y_\mu$. A very comprehensive presentation and solution of the perceptron that uses this teacher-student terminology appears in \cite{NishimoriBook01}. 
\end{itemize}

\subsubsection{Bayes optimal versus mismatched}
\label{sec:Bayes_optimal}
Throughout this manuscript we will be distinguishing two main cases
depending on what information about the distributions $P_{\rm tp}(\bx)$ and $P_{\rm tm}(\by|\bx)$ the teacher gave to the student
\begin{itemize}
   \item Bayes optimal case: In this case the teacher hands to the
     student the full and correct form of both the prior distribution
     $P_{\rm tp}(\bx)$ and the likelihood $P_{\rm tm}(\by|\bx)$. 
   \item Mismatching prior and/or model: In this case the teacher hands no or
     only partial information about the
     prior distribution $P_{\rm tp}(\bx)$ and/or the
     statistical model $P_{\rm tm}(\by|\bx)$. 
    In the present manuscript we will assume that the teacher handed the correct
     functional form of the prior distribution and of the statistical model, but not the values of
     the parameters $\theta$ and $\Delta$, respectively. 
\end{itemize}

Let us now consider the following exercise. Take $\bx^*$ to be the 
ground truth values of the variables generated by the teacher and take
$\bx,\bx_1,\bx_2$ to be three independent samples from the posterior
probability distribution $P(\bx|\by)$. We then consider some
function $f({\bf a},{\bf b})$ of two configurations of the variables
${\bf a},{\bf b}$. Consider the following two expectations 
\bea
    {\mathbb E}\left[f(\bx_1,\bx_2)\right] &= &\int  f(\bx_1,\bx_2) P(\by) 
    P(\bx_1|\by) P(\bx_2|\by)  {\rm d}\bx_1 \,  {\rm d}\bx_2 \,  {\rm d}\by\, ,  \\
 {\mathbb E}\left[f(\bx^*,\bx)\right] &=& \int  f(\bx^*,\bx)   P(\bx^*,\bx)  {\rm
   d}\bx\,  {\rm d}\bx^* = \int  f(\bx^*,\bx)   P(\bx^*,\bx,\by)  {\rm
   d}\bx\,  {\rm d}\bx^* {\rm d}\by \nonumber \\ &=& \int  f(\bx^*,\bx)   P(\bx|\by,\bx^*) P_{\rm tm}(\by | \bx^*)  P_{\rm tp}(\bx^*) {\rm
   d}\bx\,  {\rm d}\bx^* {\rm d}\by \, .
\eea 
where we used the Bayes formula. We further observe that $P(\bx|\by,\bx^*)=P(\bx|\by)$ because~$\bx$ is
independent of $\bx^*$ when conditioned on~$\by$. Remarkably, in the
Bayes optimal case, i.e. when $P(\bx)= P_{\rm tp}(\bx)$ and
$P(\by|\bx)= P_{\rm tm}(\by|\bx)$, we then obtain 
\be 
   {\rm Bayes \, \,  optimal:} \quad \quad {\mathbb E}\left[f(\bx_1,\bx_2)\right]=   {\mathbb E}\left[f(\bx^*,\bx)\right]\, , \label{Nish_gen}
\ee
meaning that under
expectations there is no statistical difference between the ground
truth assignment of variables $\bx^*$ and an assignment sampled
uniformly at random from the posterior probability distribution. This
is a simple yet immensely important property that will lead to numerous
simplifications in the Bayes optimal case and it will be used in
several places of this manuscript, mostly under the name {\it
  Nishimori condition}. 

In the case of mismatching prior
or model the equality (\ref{Nish_gen}) typically does not hold. 

\subsubsection{Estimators}
\label{sec:estimators}

What is the optimal estimator $\mathbf{\hat{x}}$ for the variables
$\bx$? The answer naturally depends on what the quantity we aim to
optimize is. The most commonly considered estimators are:

\paragraph*{\it Maximum a posteriori (MAP) estimator.} MAP simply maximizes the
     posterior distribution and is given by 
   \be
        \hat {\bx}^{\rm MAP}  = \underset{\bx}{\rm argmax} \,   P(\bx | \by)\, .
   \ee
    The main disadvantage of the MAP estimator is the lack of
    confidence interval and related overfitting issues. 
\paragraph*{\it Minimum mean squared error (MMSE).} Ideally we would like
to minimize the squared error between
    $\hat \bx$ and $\bx^*$ 
\be
{\rm SE}(\hat \bx,\bx^*) = \frac{1}{N}\sum_{i=1}^N (\hat x_i -  x^*_i)^2 \, . 
\ee
In general, however, we do not know the ground truth
    values of the variables $\bx^*$.
    The best we can do within Bayesian statistics is to assume that  $\bx^*$ is distributed
    according to the posterior probability distribution. We then 
    want to find an estimator $\hat \bx$ that minimizes the squared
    error averaged over the posterior
\be
    {\rm MSE}(\hat \bx) = \frac{1}{N}\int   P(\bx | \by) \sum_{i=1}^N (\hat x_i - x_i)^2
    {\rm d}\bx \label{MSE}\, .
\ee
By a simple derivative with respect to $\hat x_i$ we see that the minimum of
this quadratic function is achieved when 
\be
     \hat x^{\rm MMSE}_i  = \int   P(\bx | \by) x_i {\rm d}\bx = \int
     \mu_i(x_i)\, x_i {\rm d}x_i\, , \label{MMSE_est}
\ee
where the right hand side is actually the mean of the marginal on the $i$-th
variable defined by eq.~(\ref{marg}).

The value of the MMSE is then computed as the MSE (\ref{MSE})
evaluated at $\hat x_i$ (\ref{MMSE_est}). 
\paragraph*{\it Maximum mean overlap (MMO).} In cases where the support of the
variables is discrete and the variables represent indices of
types rather than continuous values, the mean-squared error is not very
meaningful. In that case we would rather count how many of the $N$
positions obtained the correct type. We define the overlap $O$ as the
fraction of variables for which the estimator $\hat \bx$ agrees with the ground truth
\be
 O(\hat \bx,\bx^*) =\frac{1}{N} \sum_i \delta_{x^*_i,\hat x_i} \, .
\ee 
As in the case of MMSE, when we do not have the knowledge of the
ground truth, the best Bayesian estimate of the overlap is its average
over the posterior distribution 
\be
  {\rm  MO}(\hat \bx) = \frac{1}{N} \int  P(\bx | \by)   \sum_i
  \delta_{x_i,\hat x_i}  {\rm d}\bx \, .
\ee
The mean overlap ${\rm MO}$ maximized over the estimator $\hat \bx$ leads
to the maximum mean overlap estimator 
\be
      \hat x^{\rm MMO}_i = \underset{x_i}{\rm argmax} \, \mu_i(x_i) \, , \label{eq:MMO}
\ee 
where $\mu(x_i)$ is the marginal probability of variable $i$ having
type $x_i$, defined by eq. (\ref{marg}). We should note that the MMO is not an
estimator very commonly considered in statistics, but given that
statistical physics often deals with discrete variables, and
anticipating the close relation, the MMO will turn out instrumental. 

Let us compare the MAP estimator with the ones based on marginals. In
most cases the MAP estimator is easier to approach algorithmically
(from the computational complexity point of view optimization is
generically easier than enumeration). Moreover, in many traditional
settings, where the number of samples is much higher than the number
of variables, there is usually very little difference between the two
since the marginal probabilities are basically peaked around the value
corresponding to the MAP estimator.  For instance, let us consider
the teacher-student scenario with unknown parameters $\theta$ of the teacher's prior, or
$\Delta$ of the teacher's model.
In situations where the number of samples $M$ is large, but the
parameters $\theta$ and $\Delta$ are scalar (or low-dimensional), the
posterior probability of these parameters is so closely concentrated
around its expectation that there is no practical difference between the MAP and
the marginals-based estimators for $\theta$ and~$\Delta$. When it
comes to estimating the variables $\bx$ then in the setting of
high-dimensional statistics (see Sec. \ref{sec:high}), where the
number of samples $M$ is comparable to the number of variables $N$ the
MAP estimator for $\bx$ will generically provide crude overfitting and
we will prefer estimators based on marginals.
Another reason to favor marginalization based
estimators is because they come directly with a notion of
statistical significance of the estimate.  Yet another reason is that
from the perspective of spin glass theory
there might even be computational advantages (ground states
might be glassy, whereas Bayes-optimal computations are not, see
sec~\ref{RSB_Nish}).

Let us now investigate the MMSE (and the MMO) estimator in the Bayes
optimal case in the view of equality (\ref{Nish_gen}). In section
\ref{sec:Bayes_optimal} we concluded that in the Bayes optimal case (i.e. when
the teacher handed to the student the precise form of the prior
distribution and of the model likelihood) and in expectation, a random
sample $\bx$ from the posterior distribution can be replaced by the
ground truth assignment $\bx^*$ without changing the values of the
quantity under consideration. It hence follows that 
\be
       {\rm Bayes \, \,  optimal:} \quad \quad   {\rm MMSE} = {\mathbb
         E}[  {\rm SE}(\hat \bx, \bx^*) ] \, . \label{Nish_MMSE}
\ee
In words, the MMSE computed solely using the posterior distribution is
on average equal to the squared error between the MMSE estimator and
the ground truth value of the variables $\bx^*$. This is a very nice
property. An analogous identity holds between the MMO and the average of
the overlap between the MMO estimator and the ground truth
assignment
\be
       {\rm Bayes \, \,  optimal:} \quad \quad   {\rm MMO} = {\mathbb
         E}[  {\rm O}(\hat \bx, \bx^*) ] \, . \label{Nish_MMO}
\ee

\subsubsection{The high-dimensional limit}
\label{sec:high}

Let us denote by $N$ the number of variables
$\bx=\{x_i\}_{i=1,\dots,N}$ (referred to as dimensionality
in statistics), and by $M$ the number of observations (or samples)
$\by=\{y_i\}_{i=1,\dots,M}$.  Traditionally, statistical theory would
consider the case of a finite number $N$ of variables/parameters to
estimate and a diverging number of samples $M$. 
Think for instance of the two 
examples (decaying particles and linear regression) from sec.~\ref{sec:example}. 

In today's data deluge more than ever, it is crucial to extract as
much information as possible from available data. From the information
theoretic perspective, we would expect that useful information is
extractable even when the number of variables (dimensionality) $N$ is
comparably large (or even somewhat larger) than the number of
observations (samples) $M$. In that case, separating the useful
information about $\bx$ from noise is much more challenging. This is
referred to as the {\it high-dimensional statistical theory}. It is, in
particular, in this high-dimensional framework, where the amount of
data to analyze is huge, that we need to pay attention not only to
solvability but also to computational efficiency, keeping in mind both
questions from section \ref{sec:inference}.

\begin{figure}[!ht]
\begin{center}
  \resizebox{12cm}{!}{\includegraphics{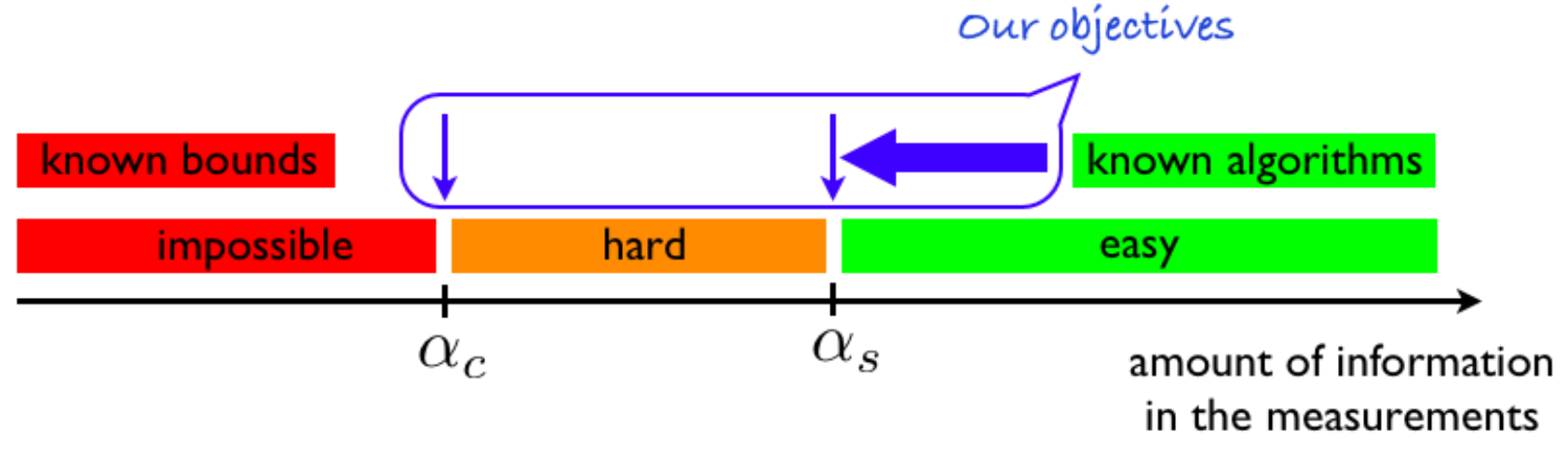}}
\end{center}
  \caption{ \label{fig:schema} Schematic representation of a typical
    high-dimensional inference problem. 
  }
\end{figure}

Typically, we will consider the limit where $\alpha=M/N$ is fixed, whereas $N\to
\infty$. In this limit the following scenario, illustrated 
schematically in Fig.~\ref{fig:schema}, very often applies:  
For low values of $\alpha<\alpha_c$, successful inference of the variables
is not possible for any algorithm, the corresponding observed
information is simply insufficient. For high values of
$\alpha>\alpha_s$, computationally efficient inference algorithms do
exist. In an intermediate regime $\alpha_c < \alpha< \alpha_s$, successful inference is in principal possible but algorithmically
considerably harder than above $\alpha_s$. In many settings, the values $\alpha_c$ and $\alpha_s$ can be defined in a precise
mathematical way as values at which some measure of performance (or its first derivative) presents a discontinuity
in the limit of large size. In such a case, we talk about thresholds or
phase transitions (see section \ref{sec:concepts} for a more precise
meaning of the second). In
many problems the intermediate hard phase is missing and $\alpha_c=\alpha_s$. On the other hand, settings and examples
where the hard phase exists ($\alpha_c<\alpha_s$) are theoretically very interesting and challenging many researchers
to prove some formal results about this hardness. For a general inference problem, the values of $\alpha_c$ and $\alpha_s$
are not known, and neither are efficient algorithms that would work down to $\alpha_s$. The state-of-the-art results
usually only give a lower bound on $\alpha_c$, and an algorithmic
upper bound on $\alpha_s$. 

The objective of a large part of the statistical physicists interested
in these directions and included in this manuscript could be
summarized as follows (see also the schema Fig.~\ref{fig:schema}):
Determine the threshold $\alpha_c$ between the {\it
  inference-impossible} and {\it inference-possible} region, and the
threshold $\alpha_s$ between the {\it inference-hard} and {\it
  inference-easy} region.  And develop efficient algorithms that would be
able to infer the variables successfully with as little information
contained in the observations as corresponds to $\alpha_s$.

\subsection{Statistical physics and inference}
\label{phys_inf}
The connection between statistical physics and statistical inference
has a long history. Arguably, the most influential early works are due
to E. T. Jaynes, starting with his {\it Statistical mechanics and
  information theory} in 1957 \cite{jaynes1957information} where he
develops the maximum entropy principle, thus building upon the
pioneering work of C. Shannon \cite{Shannon48}, who named the amount of
information as entropy inspired by mathematically analogous formulas
in thermodynamics. Jaynes actually suggests that statistical mechanics
does not have to be regarded as a physical theory dependent on its
assumptions, but simply as a special case of a more general theory of
Bayesian inference.  An inspiring reference is a collection of lecture
notes {\it Statistical Physics to Statistical Inference and Back} from
a meeting held in Carg\`ese in 1992 \cite{grassberger1994statistical}. The goal of that meeting was precisely
to revise the foundational connections between the two fields.  This
close connection leads to a continued cross-fertilization between the
two fields. Important methodological developments in one field find
numerous applications in the other, and a new applicational area in one
field profits greatly from the methodology more common to the other.

The present manuscript is mainly concerned with recent applications of
methods from spin glasses and glasses \cite{MezardParisi87b} to
inference problems and associated algorithms. This direction was
perhaps most influenced by the work on simulated annealing by
Kirkpatrick, Gelatt, and Vecchi
\cite{KirkpatrickGelatt83}. Connections between spin glasses and error
correcting codes \cite{Sourlas89}, or neural networks such as the
perceptron \cite{seung1992statistical,watkin1993statistical} belong
among the early very influential works.
This line of research is relatively well explored and illustrated on a
number of problems as seen from classical books such as the one by
H. Nishimori {\it Statistical Physics of Spin Glasses and Information
  Processing An Introduction} \cite{NishimoriBook01}, or more recently
by M. M\'ezard and A. Montanari, {\it Information, Physics, and Computation} \cite{MezardMontanari07}.

Data processing is, however, a very fast developing field with many
emerging yet far reaching applications. On top of that, there was an
important development in the theory of spin glasses on sparse random
graphs in the early 2000s \cite{MezardParisi01,MezardParisi03}. This
manuscript aims at exposing some of this recent development.

\subsubsection{Basic dictionary}

Let us first establish a dictionary between statistical inference and
statistical physics. 
The posterior probability (\ref{post_fact}) can be written as
\be
      P(\bx|\by) = \frac{1}{Z(\by)}  \exp{\left\{ \beta \sum_{\mu=1}^M  \log{ P(y_\mu| \{x_i\}_{i\in \partial
        \mu})  } + \beta \sum_{i=1}^N \log{ P(x_i)}\right\}}  =
\frac{1}{Z(\by)} \exp{ -\beta {\cal H}(\bx,\by)}\, .  \label{Boltz}
\ee
We have used the exponential form to note that it takes the form of
the Boltzmann probability distribution on variables
(discrete or continuous) with $Z$ being the partition function,
$\log{ P(x_i)}$ the local magnetic field, and $\log{ P(y_\mu| \{x_i\}_{i\in \partial \mu})  } $
being the interaction between variables, ${\cal H}(\bx,\by)$ being
the Hamiltonian. We also introduced the auxiliary
parameter $\beta$, that was previously $\beta=1$, and plays the role
of the
inverse temperature. The whole exponent divided by the inverse
temperature is then minus the Hamiltonian ${\cal H}$ of the physical
system. 

Note that the distribution (\ref{Boltz}) encompasses a wide range of models,
the variables $\bx$ can be discrete (e.g. spins) or continuous, the
interactions can depend on few or many variables (e.g. in the planted
spin glass of section \ref{planted_Ising} the interactions are
pair-wise, and in the linear estimation of sec.~\ref{chap:CS} each of the
interactions involves all of the variables). 

From a physics point of view the marginal probabilities (\ref{marg}) are nothing else than
local magnetizations obtained by summing the Boltzmann distribution
over all but one variable. The MMSE (or the MMO) estimator is then computed
using these local magnetizations (at $\beta=1$).  On the other hand
the MAP estimator can  be viewed as the ground state of the physical
system, i.e. the minimum of the Hamiltonian. Another way to think
about the MAP estimator is as the zero temperature limit, $\beta \to \infty$,  of MMSE estimator.

\subsubsection{The teacher-student scenario defines the planted ensemble}
\label{planted_intro}

Looking at the Boltzmann distribution (\ref{Boltz}) it could represent
quite a generic statistical physics model where the observations $\by$
play the role of a quenched disorder. Keeping in mind that we want to
study Bayesian inference in the teacher-student scenario of
sec.~\ref{teacher_student} we have to emphasize that the disorder
$\by$ is generated in a very particular way (by the teacher) related
to the form of the Boltzmann distribution itself.  This becomes
particularly striking in the Bayes-optimal setting in which the
teacher handed to the student all the information about how he
generated the data $\by$ except the realization $\bx^*$.  

In the rest of this manuscript we call models created in the
teacher-student scenario of sec.~\ref{teacher_student} the {\it planted models}.
We call a disorder $\by$ that results from the teacher-student scenario the {\it
  planted disorder}. We contrast it with the most commonly
considered case of randomly-quenched disorder where the component
$y_\mu$ are independent identically distributed variables. In the
planted disorder the components $y_\mu$ are not independent, they are
correlated because they all depend on the ground-truth $\bx^*$ that
the teacher used to generate $\by$. We call $\bx^*$ the {\it planted
  configuration}. This is the same difference between solving a random
system of linear equations $\by={\bf F} \bx$ (which will have no solution
if ${\bf F}$ is a $M \times N$ matrix and $M>N$) and solving one where
the left hand side was generated as $\by={\bf F} \bx^*$, instead of just being random. In the second case, we
have planted a solution $\bx^*$ in the problem.

A generic strategy to define the planted statistical ensemble is to
take an inference problem in the teacher-student setting. The
Bayes-optimal posterior probability distribution of that inference
problem then corresponds to the Boltzmann distribution of the planted
model at temperature $T=1/\beta=1$. What makes the planted model
particular is the form of the quenched disorder that is not
independently random but instead correlated in a very particular way.  

Perturbing a random ensemble by hiding a solution in it is
often a natural thing to do. For instance, protein folding is often
modeled, in statistical physics, by a random energy model (to take
into account its glassiness) in which one special configuration (the
so-called native configuration) is artificially created with a very
low energy~\cite{bryngelson1987spin}.  Thinking of Bayesian
inference in terms of the planted ensemble of statistical physics
models is implicitly done in a number of works. For instance the
pioneering articles of Sourlas on finite-temperature decoding in error
correcting codes \cite{Sourlas89,Sourlas94-2,Sourlas94} and the
relation to the Nishimori line, summarized excellently in
\cite{NishimoriBook01}, were very influential. In mathematics, planting
was introduced by Jerrum \cite{jerrum1992large} as a model to study
hardness of finding a large clique planted in a graph, and also by
Jerrum and Sorkin \cite{jerrum1993simulated} as a model to study graph
bisection. Planting was used in many mathematical works either as a
useful model or as a proof technique. It is the use of planting in
proofs for constraint satisfaction problems
\cite{achlioptas2008algorithmic} that brought the authors' attention
to this concept.  The authors find that making the connection
between planting and inference very explicit brings a lot of insight
into a generic inference problem and for this reason we build the
present review around this connection.

To clarify further what is exactly the planted ensemble we give several
examples and counter-examples:
\begin{itemize}
   \item Planted coloring. Graph coloring is a famous mathematical problem where nodes
     of a graph are assigned colors with a constraint that two
     connected nodes should not have the same color. Coloring of
     random graphs is a constraint satisfaction problem, well
     studied in statistical physics
     \cite{MuletPagnani02,ZdeborovaKrzakala07}. In planted coloring,
     the teacher first generates a random assignment of colors to
     nodes $\bx^*$, and then he generates edges of the 
     the graph at random but in a way that two nodes
     of the same color are not connected.  The edges of the graph play
     the role of the planted disorder $\by$.  The graph (i.e. its set
     of edges) is then handed to
     the student, but not the planted configuration $\bx^*$. The role of the
     student is to find a configuration as correlated to the planted
     one as possible. Physics-study of the planted coloring appeared in \cite{krzakala2009hiding},
     and we will see that it is a special case of the stochastic block
     model to which we devote section \ref{chap:clustering}. 
   \item Planted perceptron. In sec.~\ref{teacher_student} we
     presented the perceptron as the problem for which the
     teacher-student terminology was established. Formulated as a
     planted problem, the set of weights $\bJ^*$ the teacher-perceptron
     uses is the planted configuration. The patterns $\xi_i^\mu$ are
     in the physics works considered as iid, they play the role
     of the ordinary randomly-quenched disorder. The teacher generates
     the outputs $y_\mu$ using the patterns and the planted weights
     $\bJ^*$ (\ref{perceptron}). The output $\by$ then plays the role
     of the planted disorder. The planted weights $\bJ^*$ are then
     hidden from the student and her goal is to recover them as
     precisely as possible. The associated posterior probability
     distribution is the Boltzmann measure of the planted
     perceptron. On the other hand, when asking how many random patterns
     a perceptron can classify (the so-called capacity
     \cite{gardner1989three}), one is studying a non-planted problem.
   \item Hopfield model \cite{Hopfield82} is a well
     known model studied in statistical physics. The goal in the
     Hopfield model is to construct an Ising spin model for which a
     certain pre-defined set of randomly generated patterns
     $\xi_i^\mu\in \{\pm 1\}$, $i=1,\dots,N$ and $\mu=1,\dots,M$,
     serve as configurations attracting local dynamics. It is, in
     fact, a model of {\it associative memory} where, when starting
     close to one of the patterns, the dynamics would converge back to
     it. The Ising model studied by Hopfield has Hebb-like
     interactions $J_{ij}=\sum_{\mu=1}^M \xi_i^\mu \xi_j^\mu$. 

     We could study the inference problem for recovering the patterns
     $\xi_i^\mu$ from the knowledge on the $J_{ij}$. For a single
     pattern, $M=1$, the posterior distribution of this inference
     problem would indeed correspond to the Hopfield model, i.e. Ising
     spin model with interactions $J_{ij}$. However, for more than one
     pattern $M>1$ the posterior of the inference problem will be a
     function of $NM$ variables, and will be very different from the
     Boltzmann distribution of the Hopfield model (which is just a
     function of only $N$ spins). This shows that inferring or remembering are very
     different tasks.  Clearly, the associative memory studied in the
     Hopfield model, where one start the dynamics close to one of the
     patterns, is different from inference of patterns from the $J_{ij}$. The
     Hopfield model is {\it not} a planted model in the sense to which
     we adhere in this review.
\end{itemize}

Since this is important we reiterate that the planted model is a
statistical physics model where the Boltzmann distribution is at the same time the
posterior probability distribution of the corresponding inference
problem. There are many models considered in the literature where the
quenched disorder is created using some auxiliary variables, but the
Boltzmann distribution is not the posterior probability for inference
of these auxiliary variables. Then we do not talk about a planted
model, an example is given above using the Hopfield model with more than one
pattern.

\subsubsection{Inference via Monte Carlo and variational mean-field}
\label{sec:MC_VMF}

When it comes to solving an inference problem, i.e. evaluating the
corresponding MMO of MMSE estimators, see sec.~\ref{sec:estimators},
there are many algorithmic methods studied and used in computer
science, statistics and machine learning. For nice and comprehensive
textbooks see e.g. \cite{MacKay03,wasserman2013all}. Two of the classical
inference algorithms are particularly interesting in the present
context because they originate in statistical physics. These are the
Gibbs sampling of Monte Carlo Markov chain (MCMC) based methods \cite{geman1984stochastic,gelfand1990sampling,neal1993probabilistic} and the
variational mean-field inference methods \cite{PetersonAnderson87,jordan1999introduction,opper2001advanced,wainwright2008graphical}. Both
these methods and their applications are well covered in the existing
literature and in this review we will not focus on them. We only
describe in the present section their basic concepts and ideas and 
relation to the message-passing-based algorithms on which we focus in
the manuscript. 

The development of Monte Carlo goes back to the well know paper of
Metropolis et al. \cite{MetropolisRosenbluth53} and was widely used
since in physics and other scientific fields. The Metropolis-Hastings
algorithm \cite{MetropolisRosenbluth53,hastings1970monte}, most
commonly used in physics, is based on the comparison of
energy cost in one state of a system and a small perturbation of the state (typically the
value of one variable is changed). When the cost decreases the change
is accepted, when the cost increases the change is accepted with a
probability exponentially proportional to the cost difference, $p =
e^{-\beta \Delta E}$, where $\beta$ is a inverse-temperature-like
parameter.  Another perturbation is introduced and the procedure is
repeated. 

Gibbs sampling \cite{geman1984stochastic,gelfand1990sampling}, known as heat-bath in physics, is a variant on
the Metropolis-Hastings algorithm where given a fixed state of all but one
variable, the probability distribution of the one variable is
evaluated conditioning on the fixed state of the other variables. The
value of the variables is then sampled from this probability
distribution. The process is then repeated. 
Both Metropolis-Hastings and Gibbs sampling are designed to sample the Boltzmann
probability measure. But they can also be famously turned into a generic
purpose optimization algorithms called simulated annealing
\cite{KirkpatrickGelatt83}. 

The advantage of Monte Carlo based methods it that if iterated for
sufficiently long time they provide the exact answer for a very
generic and broad class of systems (only ergodicity and balance
condition are required). The disadvantage is that the time needed to
obtain the exact result can be as large as exponential in the size of
the system and it is in general very hard to decide what is the
running time necessary for satisfactory precision. Nevertheless, Monte
Carlo based methods are very useful in many cases, in particular when
the domain of variables is discrete, and throughout this
manuscript we will be comparing to them. 

Tracing the precise origins of the use of variational mean-field for
inference from data is a little harder. Ideas related to variational
mean-field were used in statistical physics since Gibbs (together with
Maxwell and Boltzmann) created the field, and are certainly present in
the works of Van der Waals \cite{vdW} and Pierre Weiss \cite{PW}
arround the 1900's. The modern view on variational mean-field for
inference and learning is very nicely summarized in
\cite{MacKay03,jordan1999introduction}.

The basic idea of variational inference is to approximate the
posterior probability distribution $P(\bx|\by)$ (\ref{Boltz}) by
another probability distribution $Q(\bx|\by)$ such that (a) for
$Q(\bx|\by)$ the Bayesian estimators (MMO or MMSE) can be evaluated
tractably, and (b) the Kullback-Leibler divergence 
\be
D_{\rm KL}(Q||P) = \sum_\bx  Q(\bx|\by) \log{\frac{Q(\bx|\by)}{P(\bx|\by)}}
\ee
between $P$ and $Q$ is the smallest possible. The variational
inference methods are very fast and in many settings and practical
problems they provide precise approximations. However, in some settings they provide crude approximations and sometimes
misleading results (see an example in sec.~\ref{sec:MF}). 

The present manuscript focuses on message passing inference
algorithms, such as belief propagation (BP). In physics the origins of
these methods trace back to the Bethe approximation
\cite{bethe1935statistical}, the work of Peierls \cite{peierls1936statistical} and the Thouless-Anderson-Palmer equations
\cite{ThoulessAnderson77}. Interestingly, these equations can be seen
as ``improved'' variational mean-field equations, and this is at the roots of many
perturbation approaches (such as
Plefka/Yedidia-Georges~\cite{plefka1982convergence,georges1991expand}
or Kikuchi \cite{Kikuchi51}).

In information theory belief propagation was invented by Gallager
\cite{Gallager62} and in Bayesian inference by Pearl \cite{Pearl82}. In an excellent article
Yedidia, Freeman and Weiss \cite{YedidiaFreeman03} relate belief
propagation and the variational mean field methods. However, strictly
speaking belief propagation is not a variational method, because the
related Bethe free energy does not have to be larger than the true
free energy. Compared to variational mean field methods belief
propagation is as fast, and generally more precise. An important property on
which we focus in this manuscript, is that on a class of models, that
can be called mean-field spin glasses, belief propagation method gives
asymptotically exact results (whereas variational mean field does not)
and it is amenable to exact analysis that enables detailed
understanding of many systems of interest.

\subsubsection{Useful statistical physics concepts}
\label{sec:concepts}

Traditional physical systems are composed of a number of spins or
molecules that is comparable to the Avogadro number $N_A\approx
10^{23}$. It is hence almost always relevant to investigate the limit
of large system size that is in statistical physics referred to
as the {\it thermodynamic limit}.

\paragraph*{\it Self-averaging.}

In the analogy between inference and statistical physics of
disordered systems, the
observations~$\by$ play the role of the so-called {\it quenched
  disorder} in the Hamiltonian in eq. (\ref{Boltz}). Generically,
quantities of interest, e.g. the magnetization of variable number $10$, depend on the realization of this quenched
disorder. 

Statistical physics, however, mostly focuses on quantities that have the so-called {\it self-averaging} property, i.e. in
the thermodynamic limit their value does not depend on the realization
of the disorder $\by$ (but only on its statistical properties). More formally, a
quantity $A(\by)$ is self-averaging if for every $\epsilon>0$
\be
         \lim_{N\to \infty} {\rm Prob} \left\{ |A(\by)-  {\mathbb
             E}\left[ A(\by)\right]| > \epsilon  \right\}  = 0 \, .
\ee 
One of the most important quantities that we expect to be self-averaging is the
free energy density 
\be
      f(\beta,\by,N) =-\frac{1}{N\beta} \log{Z(\by)}\,   
\ee 
that is assumed to have a thermodynamic limit 
\be
     f(\beta) = \lim_{N\to \infty}  f(\beta,\by,N)   \label{free_energy}
\ee 
that does not
depend on the realization of $\by$. 

Among the quantities discussed previously in this manuscript, the ones
for which the self-averaging property is particularly interesting are the
MMSE and the MMO. Under self-averaging, the equalities (\ref{Nish_MMSE}-\ref{Nish_MMO}) hold not only on
average, but also when the averages on the r.h.s. are removed. This
means that the typical squared distance between the ground truth
and the MMSE (MMO) estimator is equal, in the thermodynamic limit, to
the one evaluated without the knowledge of the ground truth. Clearly,
focusing on the analysis of self-averaging quantities greatly
simplifies the description of the behavior in the thermodynamic
limit and enables a large part of the results obtained in statistical physics. 

It should be noted, however, that from a rigorous point of view, proving that the above quantities indeed are
self-averaging poses a non-trivial technical challenge. This is a recurring discrepancy between
theoretical physics and rigorous approaches, in physics we make
assumptions that are correct and enable progress and understanding of
the problems, but that are not always easy to prove.

\paragraph*{\it Phase transitions.} Another concept borrowed from
physics that will turn out very influential in the analysis of
high-dimensional inference are phase transitions. Intuitively, a phase
transition is an abrupt change in behavior as a parameter
(commonly the temperature) is tuned.  Since not everything that
is termed {\it phase transition} in computer science or engineering
literature is a genuine phase transition from a physics point of view,
it is useful to make some reminders.

First of all, true phase transitions do not exist at
finite system size. In mathematical physics a phase transition is
defined as a non-analyticity in the free energy density $f(\beta)$~(\ref{free_energy}). Since the finite $N$ free energy density is a
logarithm of a sum of exponentials, it is always an analytical
function. Only when the limit $N\to \infty$ is taken can the
non-analyticities appear. Therefore when we talk about a phase
transition, we should always make sure that it indeed corresponds to a
non-analyticity and not to a smooth (even though perhaps very rapid)
change.  

Secondly, a traditional phase transition is always associated to either a {\it critical
slowing down} or to {\it
metastability}. According to the Ehrenfest's classification of phase
transitions, there are two main types
\begin{itemize}
   \item {\it 1st order phase transitions} are associated to a discontinuity in
     the first derivative of the free energy. The first derivative of
     the free energy is usually related to some measurable quantity
     that consequently has a discontinuity. Since a number of
     quantities, including e.g. the MMSE or the magnetization, are related to this first
     derivative, they also display a discontinuity at a first order
     phase transition. The mechanism behind a 1st order phase
     transition is a competition between two stable phases of the system. For
     the sake of vocabulary let us call the two phases  ``paramagnet''
     and ``ferromagnet''. Around the phase transition there is a
     region of phase coexistence in which the properties of the system are ruled by
     properties of both the phases. In systems that cannot be embedded
     into a (finite dimensional) Euclidean space we can define sharp
     boundaries of this parameter-region, called the {\it spinodal}
     points. The phase with lower free energy
     (i.e. higher log-likelihood) is said to be thermodynamically stable,
     the other one is called metastable. Metastability has profound consequence on the behavior
     of dynamics and algorithms. We will come back to this in detail. 
   \item {\it 2nd order phase transitions} are associated to a discontinuity in
     the second derivative of the free energy. This type of phase
     transition is associated to a so-called {\it critical slowing
       down}, meaning that dynamical processes become increasingly slow
     next to the phase transition. Diverging timescales are closely
     linked to diverging length-scales and diverging strength
     of fluctuations. These are described by the theory of criticality and critical
     exponents that ruled statistical physics in the second half of
     last century.    
\end{itemize}

Genuine phase transitions are quite peculiar creatures, and this is
largely why they fascinate many physicists. To make a distinction in
vocabulary in physics we use {\it cross-over} for a more generic
rapid change in behavior as a parameter is tuned. We will keep the
term {\it phase transition} or {\it threshold} for what is described
above.  Going back to Figure~\ref{fig:schema}, a physicist's mind
becomes very intrigued when for a given inference problem we find that,
as the amount of information in the measurements is increased,
there exist genuine phase transitions. For a
physicist, a phase transition is something intrinsic, so we
immediately expect a profound meaning and non-trivial consequences for
the system under investigation.

\section{Planted spin glass as a paradigm of statistical inference}
\label{planted_Ising}


As biology has its drosophila, or computational complexity its K-SAT problems, in
statistical physics the canonical example is the Ising model 
\be
   {\cal H}(\bS) = - \sum_{(ij)\in E} J_{ij} S_i S_j
   - \sum_{i=1}^N h_i S_i\, , \label{Ham_Ising}
\ee
where $\bS=\{S_i\}_{i=1,\dots,N}$ and the spins are $S_i \in \{-1,1\}$, $E$ is the set of interacting pairs,
$J_{ij}$ is the interaction strength between spins and $h_i$ is the
magnetic field. Statistical physics is the study of the Boltzmann probability
measure on spin configurations 
\be
      \mu(\bS)=\frac{1}{Z(\beta,\bJ,\bh)} e^{-\beta {\cal
          H}(\bS)} \, .  \label{Boltz_Ising}
\ee

In this section we study an inference problem, that we call {\it
  the planted spin glass}, for which, in the
teacher-student setting of sec.~\ref{teacher_student}, the posterior probability distribution is the
Boltzmann distribution of the Ising model. The purpose of this
exercise is simply that we will be able to use a large part of the
knowledge existing in the literature on spin glasses to describe in
detail the properties of this inference problem. Moreover, the
physical picture that arises in this section generalizes to other
inference problems in the teacher-student setting. Once we understood in detail the
planted spin glass problem, we can basically deduce without calculation what
will qualitatively happen in a wide range of other inference problems.  
Whereas the problem presented in this section may appear as a
very special and relatively simple case, all the concepts that we illustrate on this
example are very generic and apply to many other highly non-trivial
inference problems.  Sections
\ref{chap:clustering} and \ref{chap:CS} then give two recently studied examples.

The Ising model with the Hamiltonian (\ref{Ham_Ising}) is presented in
numerous textbooks. The most commonly considered geometries of
interactions are finite dimensional lattices, fully connected graphs,
random graphs, or for instance social networks. Interactions $J_{ij}$
are chosen such that $J_{ij}\ge 0$ in the ferromagnetic case (most
commonly $J_{ij}=J$ for all $(ij)\in E$) while in the antiferromagnetic
case $J_{ij}\le 0$. In the random field Ising model the fields $h_i$
are random. Perhaps the most intriguing case of an Ising model is the
spin glass, in which the interactions $J_{ij}$ are taken independently at
random from some distribution including both negative and positive
numbers. In most of the situations considered in physics literature, the interaction couplings $\bJ$ are
chosen independently of any specific configuration $\bS$.

\subsection{Definition of the planted spin glass}
\label{sec:planted_def}

To motivate the planted spin glass we consider the following problem: We have a
(large) number $N$ of people and a pack of cards with the number $+1$
or $-1$ on each of them. There is about the same number of each. We
deal one card to every person.
In the next step we choose pairs of people and
ask them whether they have the same card or not. Their instructions
are to roll a dice and according to the result answer truthfully with
probability $\rho$ and falsely with probability $(1-\rho)$. We collect
the answers into a matrix $J_{ij}$, $J_{ij}=1$ if the pair $(ij)$
answered that they had the same card, $J_{ij}=-1$ if the answer was
that they did not have the same card, and $J_{ij}=0$ if we did not ask
this pair. We want to model this system, and one of the questions we have is whether and
how we can estimate which person has which card.
 
Thinking back about the teacher-student scenario of
sec.~\ref{teacher_student} the teacher distributes the cards, i.e. generates a latent variable
$S^*_i\!=\!\pm 1,\ 1\!\leq\!i\!\leq\!N$ for each of the nodes/people
independently at random.  We then select the graph of
observations. Let us assume that the pairs to whom we ask questions
are chosen at
random, and that we ask $M$ of them. The set of queried pairs $E$ can then
be represented by an Erd\H{o}s-R\'enyi (ER) random graph
\cite{ErdosRenyi59,ErdosRenyi60} on $N$ nodes having average degree
$c=2M/N$, as defined in section \ref{sec:RandomGraphs}.
Graphs created in this manner are at the core of many results
in this review. The teacher collects the answers into a matrix
$J_{ij}\!=\!\pm1,\ (ij)\in E$ taken from the following distribution:
\begin{equation}
\label{edgeweights}
P(J_{ij}  \lvert S^*_i,S^*_j) \!=\! \rho \delta(J_{ij}-S^*_i S^*_j) +
(1-\rho) \delta(J_{ij}+S^*_i S^*_j)\,,
\end{equation}
where $\rho$ is a bias of the dice (i.e. the probability of
getting the correct answer), and $\delta(\cdot)$ is the Dirac delta function.
The latent values $S_i^*$ are then hidden from the student and the
goal is to infer from the values of $\bJ$ as precisely as possible which node had which
card. Depending on the situation, the teacher did or did not reveal the
value $\rho$ he used. For the moment we will assume $\rho$ is
known. 

Let us look at the simplest cases: In the noiseless case,
$\rho\!=\!1$, each connected component of the graph has only two
possible configurations (related by global flip symmetry). 
When the graph is densely connected, and hence has a giant component, one can therefore easily recover the two groups. When
$\rho\!=\!1/2$, it is as if we chose the couplings independently at
random just as in the most commonly considered spin glass model. The values~$\bJ$ do not contain any information about the latent variables
$S^*_i$, and recovery is clearly impossible. What is happening in
between, for $1/2<\rho<1$? (The behavior for $0 < \rho < 1/2$ is
then deduced by symmetry.)

The above model is simple to state, but its solution is very
interesting, and rather non-trivial. This problem appeared in several
different contexts in the literature, and is called 
{\it censored block model} in \cite{abbe2014decoding,abbe2013conditional,saade2015spectral}.
From a rigorous point of view this problem is still a subject of
current research. In the physics literature a closely related model
is known as the {\it Mattis model} \cite{mattis1976solvable}. Here, we will
call the above problem the {\it planted spin-glass model} to
differentiate it from the standard spin-glass model, where we chose all
the interactions $\bJ$ randomly and uncorrelated. It should be noted,
that ``spin glass'' here is used as a name for the model, it does not
necessarily imply that this model must exhibit glassy behavior. The
phase diagram on the planted spin glass is discussed in detail in
Sec.~\ref{sec:planted_phase}. We stress that the main goal in inference is
to retrieve the planted configuration, whereas the usual goal in
physics of spin glasses is to understand the glassy behavior of the
model. These two goals are of course related in some aspects, but also
complementary in others. 

More formally defined, to create an instance, that is a triple $(\bS^*,G,\bJ)$, of the planted spin
glass we proceed as follows
\begin{itemize}
\item We consider that the hidden assignment $S^*_i$, called the {\it planted configuration}
  of spins, is chosen uniformly at random from all the
  $2^N$ different possibilities.
\item We then choose a graph of interactions $G=(V,E)$, but not
  yet the associated values of $J_{ij}$, independently of the planted
  configuration $\bS^*$.
\item For each edge $(ij)\in E$, the value of $J_{ij}$ is chosen to be
  either $+1$ or $-1$ with a probability taken from (\ref{edgeweights}). The
  signs of the couplings then carry information about the planted
  configuration.  
\end{itemize}

So far the relation with the Ising model may not be very clear. To
see it very explicitly we write eq.~(\ref{edgeweights}) differently. Without losing
generality, we can define a parameter $\beta^*$ by
\be
\rho = \frac{e^{\beta^*}}{2\cosh{\beta^*}}~~\text{and}~~~1-\rho =
\frac{e^{-\beta^* }}{2\cosh{\beta^*}} ~~\text{with}~~~
\beta^* =\frac 12 \log{\left( \frac{\rho}{1-\rho} \right)}\, ,
\ee
so that eq.~(\ref{edgeweights}) reads
\begin{equation}
\label{edgeweights2}
P(J_{ij}  \lvert S^*_i,S^*_j) \!=\! \frac{e^{\beta^* J_{ij} S_i^* S_j^*}}{2\cosh{\beta^*}}\,.
\end{equation}
Since we want to know the values of the spins $\bS$ given the knowledge
of the couplings $\bJ$, we write the Bayes formula
\be
P({\bf S}|{\bf J}) = \frac{P({\bf J}|{\bf S}) P({\bf S})}{P({\bf J}) }
= \frac{ \prod_{(ij)\in E} e^{\beta^* J_{ij} S_i S_j } }{2^N
  (2\cosh{\beta^*})^M P(\bJ)} \label{planting_posterior}
\ee 
where we used (\ref{edgeweights2}). The posterior probability
distribution has to be normalized, therefore
\be
    P(\bJ) = \frac{ Z(\beta^*,\bJ) }{  2^N (2\cosh{\beta^*})^M  }\, , \label{PJ}
\ee
where $ Z(\beta^*,\bJ)$ is the partition function of the Ising model
(\ref{Boltz_Ising}) at inverse temperature $\beta^*$ (in zero magnetic field). 
In other words, the posterior distribution
is simply the standard Boltzmann measure of the Ising model at a
particular  inverse
temperature $\beta^*$.  In general we shall consider this Ising model at a
generic inverse temperature $\beta \neq \beta^*$. This is because a priori we
do not know the proportion of errors. In the teacher-student setting,
the teacher may not have handed to
the student the value $\beta^*$.

The special case $\beta=\beta^*$ corresponds to the Bayes optimal
inference as discussed in section \ref{sec:Bayes_optimal}. One of the
most physically important properties of the Bayes-optimal case is that
the planted configuration can be exchanged (under averages, or in the
thermodynamic limit) for a random configuration taken uniformly from the
posterior probability measure, which is in physics simply called an {\it equilibrium
configuration}. All equilibrium configurations with high probability
share their macroscopic properties, e.g. their energy, magnetization,
various overlaps etc. This is directly implied by the Nishimori condition
(\ref{Nish_gen}) derived in Sec.~\ref{sec:Bayes_optimal} for the Bayes-optimal
inference. This property will turn out to be crucial in the present
context. 

As discussed in Sec.~\ref{sec:estimators}, since we are dealing with a
problem with discrete variables (spins), it is desirable to minimize the number of mismatches in the
estimation of the planted assignment. We hence aim to evaluate the 
maximum mean overlap (MMO) estimator (\ref{eq:MMO}) which in the case
of an Ising model is simply 
\be \hat{S}_i = \text{sign}~(m_i) \, ,
\label{rules}
\ee
where $m_i$ is the equilibrium value of the magnetization of node $i$ in the Ising
model. We reduced the inference of the planted configuration to the
problem of computing the magnetizations in a disordered Ising model as
a function of temperature. 

An off-the-shelf method for this that every physicist has in her
handbag is a Monte-Carlo simulation, for instance with the Metropolis
algorithm. In Sec.~\ref{sec:cavity} we shall see that for the
mean-field setting (i.e. the underlying graph being random or fully
connected) there exists an exact solution of this problem via the
cavity method. However, before turning to the mean-field setting we
want to discuss concepts that are more generic and we will hence keep
Monte-Carlo in our mind as a simple and universal method to evaluate
phase diagrams of Hamiltonians with discrete degrees of freedom.

\subsection{Nishimori's mapping via gauge transformation}
\label{sec:Nish}

For the planted spin glass there is an alternative way that provides 
understanding of its phase diagram and quite interesting additional
insights
\cite{nishimori1980exact,nishimori1981internal,NishimoriBook01}. One
takes advantage of the
so-called gauge transformation
\be
       S_i S_i^*\to \tilde S_i \, , \quad \quad 
       J_{ij} S_i^* S_j^* \to \tilde J_{ij}  \, .
\ee 
If this transformation is applied, the Ising Hamiltonian is conserved
since all $S_i=\pm 1$.
In the new variables the planted configuration becomes ferromagnetic:
$\tilde S_i^* =(S_i^* )^2=1$. What happened to the couplings? The
energy of each edge in the planted configuration has not
been changed by the gauge transform. Since the frustrated interactions
were chosen independently at random in the planting, after the gauge
trasformation we end up with a standard spin glass with a fraction 
\be
    \rho = \frac{e^{\beta^*}}{2\cosh{\beta^*}}
\ee
of positive $J_{ij}=+1$ couplings and the rest $-1$, where the $-1$'s
are chosen independently at random.

After the gauge transformation, the planted assignment
has been transformed to the ferromagnetic one where all spins are
positive. The overlap between the planted and another randomly chosen equilibrium
configuration becomes simply the magnetization. 
The system is now a standard spin glass with iid interactions,
albeit with a ferromagnetic bias $\rho$. The question of the identification
of the hidden assignment is now simply mapped to the question of
finding the ferromagnetic ``all spins up'' configuration. This
is a step that from the physics point of view simplifies the
discussion. It should be stressed, however, that the theory of
Bayes-optimal inference, that is the main subject of the present
manuscript, applies to a class of planted models for which a similar gauge
trasformation does not always exist.  

The phase diagram of the spin glass models, with a ferromagnetic bias
$\rho$, versus a temperature $1/\beta$, has been determined and one has
simply to look it up in the literature
\cite{KwonThouless88,carlson1990bethea,carlson1990betheb}. The
non-zero value of magnetization in the ferromagnetically biased model then translates into the ability to be
able to infer the planted configuration better than by a random
guessing. The value of the local magnetization of spin $i$ is then related to the MMO estimator
via (\ref{rules}). 
In a remarkable contribution, Nishimori realized that
the physics on the line $\rho= e^{\beta} / (2\cosh{\beta})$ is
particularly simple and quite different from a generic~$\beta$. This line is now called the {\it Nishimori line} and
corresponds to nothing else but the condition for Bayes optimality
$\beta=\beta^*$ in the planted model. 

Consequently, a number of interesting properties, that do not hold in
general, hold on the Nishimori line. These are called {\it Nishimori
conditions}. In general, Nishimori conditions are properties that hold in the Bayes
optimal case, such as (\ref{Nish_MMSE}) and (\ref{Nish_MMO}). Another
useful Nishimori identity concerns the average energy $E = \langle
{\cal H} \rangle/N$
(\ref{Ham_Ising}). For more Nishimori conditions see \cite{NishimoriBook01,krzakala2011melting2}. Using again the property that the planted
configuration can replace an equilibrium configuration
we obtain immediately for the planted spin glass on Erd\H{o}s-R\'enyi
random graph (as defined in section \ref{sec:RandomGraphs}) of average degree $c$
\be
E = - \frac{c}{2}\tanh{\beta^*}\, .  \label{E_Nish}
\ee

\begin{figure}[!ht]
\begin{center}
 \resizebox{10cm}{!}{\includegraphics{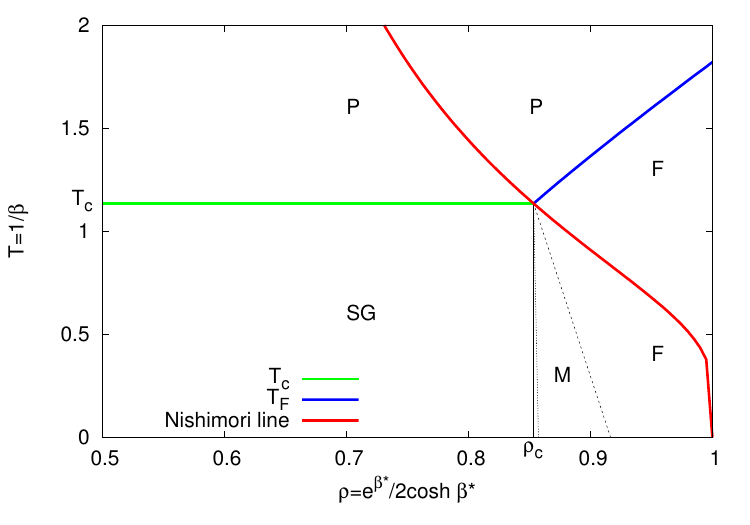}}
\end{center}
\caption{\label{fig:phase_planted} Phase diagram of the planted spin glass as
  a function of temperature $T=1/\beta$ and density $\rho$ of edge
  interactions that agree with product of spins in the planted configuration. This phase diagram is for random
  regular graphs of degree $c=3$. The Nishimori line, 
  $\rho = e^{\beta}/(2\cosh{\beta})$ (red online), corresponds to the Bayes optimal
  inference. There is the spin glass (SG) phase where
  the system is in a glassy state, the ferromagnetic (F) phase, the
  paramagnetic (P) phase and the mixed (M) one where the system is
  both SG and F. The boundary of the mixed phase (black dashed) is
  only a guide for the eye between the multi-critical point and the
  zero-temperature point from \cite{castellani2005spin}. The same for
  the boundary of the SG phase (black dotted) that is obtained from
  the inexact 1RSB approximation and is hardly distinguishable from the
  vertical line (black full).}
\end{figure}

In Fig.~\ref{fig:phase_planted} we give the phase diagram of the spin
glass at temperature $T=1/\beta$ with ferromagnetic bias $\rho$. As an
example we use the spin glass on a random graph in which case an exact
solution exists (see sec.~\ref{sec:cavity}). However, on a generic geometry
the phase diagram could be obtained using Monte-Carlo simulation and
would be qualitatively similar. Due to the Nishimori mapping this is
also the phase diagram of the planted spin glass, where the
ferromagnetic phase is understood as correlated to the planted
configuration. The phenomenology of a rather generic Bayesian inference problem can be directly read off from this
phase diagram.

\subsection{Bayes-optimality on the Nishimori line}
\label{sec:second}

First, let us assume that we know (or that we learn from data, see next section)
the true value of $\beta^*$ (or equivalently $\rho$) for which the
couplings $\bJ$ have been generated. Then, our consideration is
restricted to the Nishimori
line. If the temperature is large enough, i.e. $\beta^*$ is low enough, we are in the paramagnetic
phase, and we find that all
marginals (magnetizations) are zero. It would thus be
difficult to find their signs. In fact, this simply means that, in the
paramagnetic phase, inference of the hidden configuration is
impossible. There is just not enough information to say anything about
the hidden/planted assignment.

As the temperature is lowered, i.e. $\beta^*$ is increased, the situation changes. If we
used $\beta^*$ corresponding to an inverse temperature above a certain critical
value $\beta_c$, then a Monte-Carlo
simulation would rapidly discover the ferromagnetic phase where
local magnetizations are (mostly) positive. Applying rule (\ref{rules}) would
give at this point a decent overlap with the true assignment (which
is, after the gauge transformation, all $+1$).

In terms of inference of the planted configuration these two
phases have the following interpretation
\begin{itemize}
   \item Undetectable: For $\beta^* < \beta_c$, $\rho<\rho_c$, the planted spin glass
     system is in the paramagnetic state. Meaning that the set of
     observed variables $\bJ$ does not contain any information about
     the planted configuration and its inference is hence information-theoretically impossible. 
   \item Detectable: For $\beta^* > \beta_c$, $\rho>\rho_c$, the planted spin glass is
     in a ferromagnetic phase where equilibrium configurations are all
     correlated with the planted one and inference is possible, and
     algorithmically tractable e.g. via Monte-Carlo sampling. 
\end{itemize}

For the planted spin glass of a random graph, the critical temperature $T_c=1/\beta_c$ where one goes from {\it impossible
  inference} to {\it easy inference} can be taken from
the literature \cite{Thouless86,carlson1990bethea,carlson1990betheb} (we
will show how to compute it for this case in Sec.~\ref{sec:cavity}) and in the case
of random Erd\H{o}s-R\'enyi graphs (as defined in section \ref{sec:RandomGraphs}) with average degree $c$ it reads 
\begin{equation} 
T^{\rm ER}_{\textrm c}(c) =
\left[\text{atanh}{\frac{1}{\sqrt{c}}}\right]^{-1} ,
\label{sol1_}
\end{equation}
while for regular random graphs it reads
\begin{equation} 
T^{\rm REG}_{\textrm c}(c) =
\left[\text{atanh}{\frac{1}{\sqrt{c-1}}}\right]^{-1} .
\label{sol2_}
\end{equation}
On an Erd\H{o}s-R\'enyi  random graph (as defined in section \ref{sec:RandomGraphs}), this indicates that it is only possible to infer something
about the planted assignment if the number of edges in the graph
$cN/2$ is such that
\begin{align}
\label{exact_transition}
c >c_{\rm detect}=\frac{1}{(2\rho-1)^2}.
\end{align}
If this is not satisfied, the system is in the paramagnetic phase and
nothing can be said about the planted configuration. If, on the other hand, (\ref{exact_transition}) is satisfied, then
the magnetization aligns partly with the planted configuration. This is a
very generic situation in Bayes-optimal inference when we encounter
a second order phase transition. We go from a phase where inference is
impossible, to a phase where it is rather straightforward.
Note that this is not only a statistical physics prediction, but one
that can be proven rigorously. The impossibility result has been
shown in \cite{HLM2012,LMX2013}  and the possibility one (if
(\ref{exact_transition}) is satisfied) in \cite{saade2015spectral}.

\subsection{Parameter mismatch and learning}
\label{sec:learn}

Let us now discuss what happens if one does not know the value
$\beta^*$. One approach that seems reasonable is to use $\beta\to \infty$, so that
we look for the most probable assignment (this would correspond to the
MAP estimator) which is the ground
state of the corresponding Ising model. We observe from the phase
diagram in Fig.~\ref{fig:phase_planted} that for $\rho<\rho_c$, at $T=0$ the system is in the spin
glass phase. In terms of Monte-Carlo, the presence of such a
phase manifests itself via a drastic  growth of the equilibration
time. Computing the corresponding magnetizations becomes very
hard. Moreover, even in the undetectable phase, $\rho<\rho_c$, the
zero-temperature local magnetizations will not be zero. This is a sign of
overfitting that the MAP estimator suffers from. 

In the detectable phase, $\rho>\rho_c$, the ferromagnetic phase
extends only up to a temperature that for an Erd\H{o}s-R\'enyi random
graph corresponds to a certain \be \beta_F= {\rm atanh}\left(
  \frac{1}{c \tanh{\beta^*}} \right)\, .  \ee Hence if we chose too
low value of $\beta$ as a starting point we could miss the
detectability phase. At
low temperatures we might again encounter a spin glass phase that
may or may not be correlated to the planted configuration. In either
case, such a phase causes equilibration problems and
should better be avoided.  Fortunately, there is no such glassy phase
on the Nishimori line as we shall explain in sec. \ref{RSB_Nish}.

The above considerations highlight the necessity of learning parameters
(in this case $\beta^*$). Within the theory described above there is one
particularly natural strategy for learning such parameters, 
closely related to the well known algorithm of expectation
maximization \cite{dempster1977maximum}.  Let us illustrate this
parameter learning on the example of the planted Ising model where we
assume that the parameter $\beta^*$ is not known.

In Bayesian inference, we always aim to write a posterior probability
distribution of what we do not know conditioned to what we do know. In
the present case we need $P(\beta|\bJ)$. Using the Bayes formula we obtain
\be P(\beta | \bJ) = \sum_{\bS} P(\bS,\beta|\bJ) = \frac{P(\beta)}{2^N
  P(\bJ)} \frac{ Z(\beta,\bJ)}{(2\cosh{\beta})^M} \, . \label{P_beta}
\ee We notice that as long as the hyper-prior $P(\beta)$ is
independent of $N$, in the thermodynamic limit this probability
distribution $ P(\beta | \bJ)$ converges to a Dirac delta function on
the maximum of the second fraction on the right hand side.
This maximum is achieved at 
\be
  E = - \left\langle \frac{1}{N} \sum_{(ij)\in E}  J_{ij} S_i S_j \right\rangle =
  -\frac{cN}{2} \tanh{\beta} \, . \label{EM_beta}
\ee 
The corresponding iterative algorithm for learning $\beta$ is as follows. We start with
some initial choice of $\beta$ and compute the average energy to
update the value of $\beta$ according to (\ref{EM_beta}) until a fixed
point is reached. Depending on the initial value of $\beta$, and on
the algorithm used to estimate the average energy, in some
cases we may obtain several different fixed points of
(\ref{EM_beta}). In those cases we choose the fixed point that is the global maximum of
(\ref{P_beta}). 

Notice now a very useful property, the condition for stationarity of
(\ref{P_beta}) is the same as the Nishimori condition (\ref{E_Nish})
and indeed $\beta= \beta^*$ is a fixed point of (\ref{EM_beta}).  This
strategy is very reasonable, in the Bayes-optimal case all the
parameters are the ground truth ones and a number of {\it Nishimori
conditions} hold only in this case. Hence in order to learn the right
values of parameters we iteratively impose the Nishimori conditions to
hold.

Finally, let us note that in the above planted Ising model the
underlying phase transition is of second order, therefore there is no
particular computational issue. Inference is either impossible, or
possible and tractable. The hard phase outlined in section
(\ref{sec:high}) is absent. The picture is richer in problems where in the random case there is a
discontinuous phase transition. Examples of such a case are described in detail in Sec.~\ref{sec:Potts}. 

\subsection{Relations between planted, quenched  and annealed disorders}

From the early works, it sometimes seems necessary to have a gauge
transformation (and consequently a special form of the Hamiltonian) to
be able to speak about the Nishimori conditions or other
properties of the Nishimori line
\cite{nishimori1980exact,nishimori1981internal,georges1985exact}. 
Importantly this is not the case. The properties of
the Nishimori line are more general and apply to every case where it
makes sense to talk about Bayes optimal inference. In fact, a
more general notion of the
Nishimori line can be identified with the notion of
Bayes-optimality.  A nice article on this connection was given
by Iba \cite{iba1999nishimori}.  

In this section, we shall
discuss properties of the planted disorder, and its relation with the more
commonly considered - the randomly-quenched and the annealed disorders. This clarifies the generic connection
between statistical physics of spin glasses and Bayesian inference. We stress again that none of this requires a gauge symmetry of the underlying Hamiltonian. 

Let us consider back the Hamiltonian of the Ising system
(\ref{Ham_Ising}). For a given system, e.g. the magnet on the fridge in
your house, the set of couplings $J_{ij}$ is a given fixed finite set
of numbers. In the planted spin glass above, this set was the values
of interactions $\bJ$ we
observed. 

In statistical physics, where in the vast majority of cases we can take
advantage of self-averaging, see sec.~\ref{sec:concepts}, we often do not think
of a given set of numbers $J_{ij}$ but instead about a probability
distribution over such sets. For instance randomly-quenched disorder
is generated following
\be P(J_{ij})=\frac 12 \delta(J_{ij}-1) + \frac 12 \delta(J_{ij}+1) \label{J_uni}
\ee 
for all $(ij)\in E$, where $E$ are the edges of the graph of
interactions. In this way we do not have a given Hamiltonian, but
rather its probability of occurrence, or equivalently an ensemble
where instances appear proportionally to this probability. Properties of the system then have to be averaged over the
disorder (\ref{J_uni}). This view is adopted in statistical physics
because the self-averaging is assumed to hold in most physical systems
of interest and the theory is then
easier to develop. This notion of averaging over disorder is so well
engraved into a physicist's mind that going away from this framework
and starting to use the related (replica and cavity) calculations for a single realization
of the disorder has lead to a major paradigm shift and to the
development of revolutionary algorithms such as the survey propagation \cite{MezardParisi02}.
In this section we do, however, want to stay in the mindset of averaging over
the disorder and explain how to fit in the notion of the {\it planted
  ensemble}, i.e. ensemble of instances where we average over the
disorder that was planted.

\subsubsection{The quenched and annealed average}
How to take the average over disorder of $J_{ij}$ meaningfully is a problem that was solved a long time ago by Edwards
in his work on spin glasses and vulcanization (see the
wonderful book \textit{Stealing the gold}
\cite{goldbart2005stealing}). If one takes a large enough system, he
suggested, then the system becomes \textit{self-averaging}: all
extensive thermodynamic quantities have the same values (in densities)
for almost all realizations of the Hamiltonian. Therefore, one needs to compute the average free energy
\beq
f_{\textrm{quenched}}=\[[\frac FN \]] = \lim_{N \to \infty} -\frac
{1}{\beta N} \[[\log{Z}\]],
\eeq
where here $\[[\cdot\]] $ denotes the average over the disorder. This is called the quenched average, and the corresponding
computations are the quenched computations. In fact, the
self-averaging hypothesis for the free energy has now been proven
rigorously in many cases, in particular for all lattices in finite
dimension \cite{WehrAizenman90} and for some mean-field models
\cite{GuerraToninelli02}.  Edwards
did not stop here and also suggested (and gave credit to Mark Kac for
the original idea) a way to compute the average of the 
logarithm of $Z$, known today as the \textit {replica trick}, using
the identity
\beq \log{Z}=\lim_{n\to 0} \frac {Z^n-1}n . \eeq
The idea here is that
while averaging the logarithm of $Z$ is usually a difficult task,
taking the average of $Z^n$ might be feasible for any integer value of $n$.
Performing a (risky) analytic continuation to
$n=0$, one might compute the averaged free energy over the disorder as
\beq f_{\textrm{{quenched}}}= - \lim_{N\to \infty} \frac{1}{N\beta} \lim_{n \to 0}
\frac{[Z^n]-1}{n}. \eeq

The computation of the quenched free energy is in
general still very difficult and remains a problem at the core of studies in statistical physics of
disordered systems.
It is \textit {much easier} to consider the so-called annealed
average, one simply averages the partition sum and only \textit{then}
takes the logarithm
\beq
f_{\textrm {annealed}}= -\frac{1}{N \beta} \log{[Z]}.
\eeq
This is a very different computation, and of course there is no reason for it to be equal to the quenched computation. 

For the example of the Ising model with binary uniformly distributed
couplings (\ref{J_uni}) we obtain (independently of the dimension or geometry)
\be
   - \beta  f^{\rm Ising}_{\textrm {annealed}} = \log{2} + \frac{c}{2}
   \log(\cosh{\beta})\, ,  \label{f_annealed}
\ee
where $c$ is the average coordination number (degree) of the interaction graphs. 

It is important to notice that the annealed average is \textit {wrong} if one wants to do
physics. The point is that the free energy is an extensive quantity,
so that the free energy per variable should be a quantity of $O(1)$
with fluctuations going in most situations as $\propto (1/\sqrt{N})$ (the
exponent can be different, but the idea is that fluctuations are
going to zero as $N \to \infty$). The partition sum $Z$, however, is
exponentially large in $N$, and so its fluctuations can be huge. The
average can be easily dominated by rare, but large, fluctuations.

Consider for instance the situation where $Z$ is $\exp{(-\beta N)}$
with probability $1/N$ and $\exp{(-2 \beta N)}$ with probability
$1-1/N$. With high probability, if one picks up a large system, its
free energy should be $f_{\rm quenched}=2$, however in the annealed computation one
finds
 \beq 
[Z]= \frac 1N \exp{(-\beta N)} + \((1-
\frac 1N\)) \exp{(-2\beta N)} 
\eeq 
and to the leading order, the annealed free energy turns out to be
$f_{\textrm{annealed}}=1$.

One should not throw away the annealed computation right away, as we
shall see, it might be a good approximation in some cases. Moreover, it turns out to
be very convenient to prove theorems. Indeed, since the logarithm is a
concave function, the average of the logarithm is always smaller or
equal to the logarithm of the average, so that
\beq
f_{\textrm {annealed}} \le f_{\textrm {quenched}}.
\eeq
This is in fact a crucial property in the demonstrations of many
results in the physics of disordered systems, and in computer science
this is the inequality behind the  ``first moment method" \cite{moore2011nature}.

Furthermore, there is a reason why, in physics, one should sometimes
consider the annealed average instead of the quenched one. When the
disorder is changing quickly in time, on timescales similar to those
of configuration changes, then we indeed need to average both over
configurations and disorder and the annealed average is the correct
physical one. This is actually the origin of the name {\it annealed} and
{\it quenched} averages.

\subsubsection{Properties of the planted ensemble}
In sec.~\ref{sec:planted_def} we described the planted spin glass, where we
first considered the planted configuration $\bS^*$ and a graph
$G=(V,E)$, and then we
considered the probability over interactions $P(\bJ|\bS^*)$
eq.~(\ref{edgeweights2}). The ensemble of instances $(\bS^*,G,\bJ)$ generated by the
planting, i.e. by the three steps of Sec.~\ref{sec:planted_def}, is called the planted ensemble. 

From a physics point of view the two most important properties of the
planted ensemble, the two golden rules to remember, are: 
\begin{itemize}
\item The planted configuration is an equilibrium configuration of the
  Hamiltonian derived from the posterior distribution. By this we mean
  that the planted configuration has exactly the same macroscopic
  properties (e.g. its energy or magnetization) as any other
  configuration that is uniformly sampled from the posterior
  distribution.  Indeed, when we view planting as an inference
  problem, we can write the corresponding posterior probability
  distribution as in eq. (\ref{planting_posterior}). In section
  \ref{sec:Bayes_optimal} we derived that in the Bayes optimal
  inference the planted configuration behaves exactly in the same way
  as a configuration sampled from the posterior.

  Given that generating an equilibrium configuration from a
  statistical physics model is in general a difficult task (this is,
  after all, why the Monte-Carlo method has been invented!) this is
  certainly a very impressive property. It actually means that, by
  creating a planted instance, we are generating at the same time an
  equilibrium configuration for the problem, something that would
  normally require a long monte-carlo simulation.
   \item   The realization of the disorder of the planted problem is not
  chosen uniformly, as in the randomly-quenched ensemble (\ref{J_uni}), but instead each
  planted instance appears with a probability proportional to its
  partition sum. This can be seen from eq. (\ref{PJ}) where the
  probability distribution over $\bJ$ in the planted ensemble is
  related to the partition function of the planted system $Z(\bJ)$. A
  more explicit formula can be obtained using the annealed partition
  function (\ref{f_annealed}) 
\be
 P_{\rm planted}(\bJ) = \frac{Z(\bJ)}{\Lambda Z_{\rm annealed}}\, , \label{P_pl}
\ee
 where $\Lambda$ is the number of all possible realizations of the
 disorder, $\Lambda=2^M$ for the Ising model. Note the $1/\Lambda$ can
 also be interpreted as the probability to generate a given set of
 couplings in the randomly-quenched ensemble where the couplings are 
 chosen uniformly at random. 
\end{itemize}

\begin{figure}[!ht]
\begin{center}
  \resizebox{10cm}{!}{\includegraphics{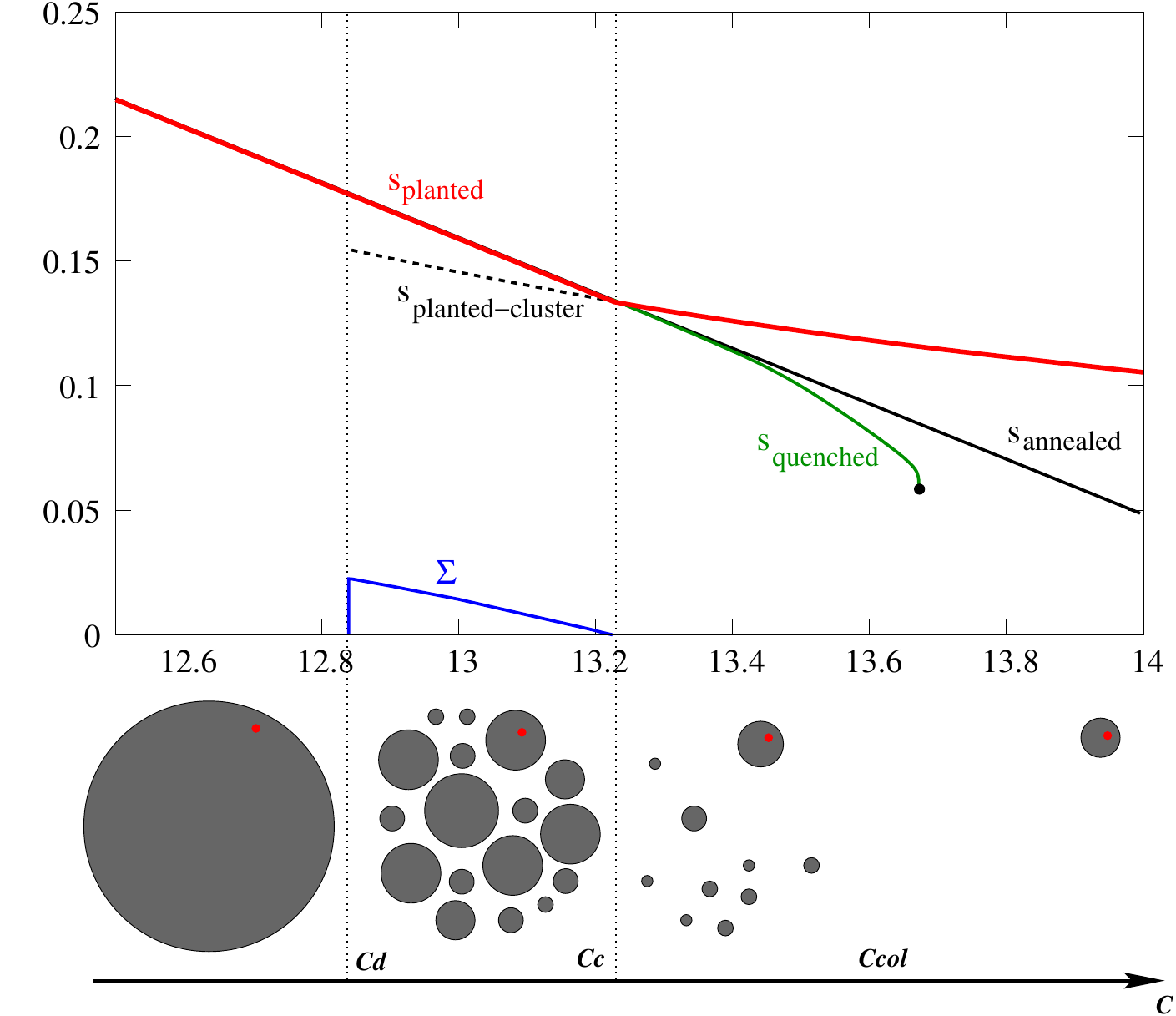}}
\end{center}
\caption{ \label{col_entropies} 
  Entropy density (which is equal to the free energy at zero temperature) for the
  5-coloring problem on an
  Erd\H{o}s-R\'enyi random graph (as defined in section \ref{sec:RandomGraphs}) of average degree $c$. 
  The planted entropy (in red) is compared to the quenched one (in
  green) and the annealed one (in black). For $c<c_c$ the three curves
  agree. The quenched entropy goes to $-\infty$ at the colorability
  threshold $c_{\rm col}$. The region between $c_d$ and $c_c$ is
  relevant only for cases with a first order phase transition as
  discussed in Sec.~\ref{sec:Potts}. In mean field theory of spin
  glasses this region is called the d1RSB phase \cite{BouchaudCugliandolo98}. We also show the
  entropy of the planted cluster below $c_c$ (black dashed), which is less than the
  total entropy, the difference is called the ``complexity"
  $\Sigma$ (in blue) and is the logarithm of the number of
  pure states. 
  Bottom: Sketch of the shape of the space of all valid colorings in
  the planted case. At an average
  degree $c_d$ the space of solutions shatters into exponentially many
  clusters, the planted cluster being one of them. Beyond $c_c$ the
  planted cluster contains more solutions than all the others
  together. At $c_s$ the last non-planted cluster disappears (this
  is the coloring threshold in a non planted, purely random
  model). 
  Figure taken from \cite{krzakala2009hiding}.}
\end{figure}

The average over the planted ensemble might seem related to the annealed average
because for instance the planted energy (\ref{E_Nish}) is always equal to
the annealed average energy, which is easily derived from the annealed
free energy (\ref{f_annealed}). However, the free energy of the
planted ensemble is in general different from the annealed free energy
(\ref{f_annealed}), as illustrated e.g. in Fig.~\ref{col_entropies}.

\subsubsection{Quiet planting}
\label{quiet_planting}

Let us now imagine we have found a system where the (quenched) free
energy is exactly equal to the annealed free energy. Then according to
eq.~(\ref{P_pl}) we see that for such a system
$P_{\textrm {planted}}(\bJ)=1/\Lambda=P_{\textrm {quenched}}(\bJ)$,
meaning that generating instances from the planted ensemble is the
same thing as generating from the randomly-quenched ensemble. Such
planting is denoted as {\it quiet}.  In fact, we do not even need the
free energies to be exactly equal, it is sufficient that the free
energy densities are the same. This is because atypical instances are
usually exponentially rare and hence any difference in the free
energies that does not show up in the leading order will not generate
atypical instances.

It so happens that in mean-field systems the (self-averaging) free energies of
paramagnets are indeed equal to the annealed free energies. The
equilibrium in these systems can hence be studied using the planted
ensemble and this is greatly advantageous as we will see in section
\ref{equlibration}. 

The idea of quiet planting comes from a rigorous work
\cite{achlioptas2008algorithmic} where the above notion was proven
rigorously. In mathematics this property is formalized and the two ensembles
termed {\it contiguous} \cite{achlioptas2008algorithmic,mossel2012stochastic}. 
The paper that pioneered the usage of quiet planting in physics, and
that also established the name ``quiet'', is 
 {\it Hiding  quiet solutions in random constraint satisfaction problems}
\cite{krzakala2009hiding}.  

\subsection{Replica symmetry breaking} 
\label{sec:RSB}

The replica and cavity method, that are among the main assets that
statistical physics offers for analysis of inference problems, were
originally developed to understand behavior of glasses and spin
glasses via the so-called replica symmetry breaking (RSB). We will
explain in the next section, that for Bayes-optimal inference RSB is
not needed. But to appreciate fully the results exposed in this
manuscript it is useful to have a very basic idea of what RSB is and
what it implies.

RSB is usually explained together with the related (relatively
involved) calculations, see
e.g. \cite{NishimoriBook01,MezardMontanari07}. In this section we
try to convey the basic ideas behind RSB without going into the
computations. We rely on the descriptions of what would happen to
a Monte Carlo simulation in a system that
exhibits various versions of RSB.

Let us start with the concept of {\it replica symmetry}. For the
purpose of the present manuscript we can say that a probability
measure $P(\bx|\by)$ is replica symmetric if a MCMC that is
initialized at equilibrium, i.e. in a configuration sampled uniformly
at random from $P(\bx|\by)$, is able to sample $P(\bx|\by)$ (close to) uniformly
in a number of iterations that is linear in the size of the system.  A
probability measure $P(\bx|\by)$ exhibits dynamical one-step of
replica symmetry breaking (d1RSB) if MCMC initialized at equilibrium
explores in linear time only configurations that belong into an
exponentially small (in $N$) fraction of the equilibrium
configurations. In both RS and d1RSB the distance between two
configurations sampled from the Boltzmann probability measure is the same number
with probability going to one (as $N\to\infty$).

We speak about static one-step replica symmetry breaking (1RSB) if
MCMC initialized at equilibrium is able to explore configurations
belonging to a finite fraction of all equilibrium
configurations. Moreover the distance between two randomly sampled
configurations can take (with probability going to one) one of two
possible values.

A probability measure $P(\bx|\by)$ corresponds to full-step replica
symmetry breaking (FRSB) if the distance between two randomly chosen
configurations is distributed according to some probability
distribution with continuous non-trivial support. In terms of behavior
of MCMC, in a FRSB regime the time needed to sample configurations
uniformly starting from equilibrium has a more complex behavior:
relaxations are critical and a power-law behavior is
expected~\cite{PhysRevB.25.6860}. 

\subsection{No replica symmetry breaking in Bayes-optimal inference.} 
\label{RSB_Nish}
From the methodological point of view, the Nishimori line has one absolutely
crucial property: the absence of glass phase at equilibrium. In the
words of the mean-field theory of spin glasses, there is no static
replica symmetry breaking on the Nishimori line. Note that there might
be, and indeed is in all the cases where the underlying transition is of
the 1st order, the dynamical one-step RSB phase (d1RSB) \cite{BouchaudCugliandolo98}. The marginals in the
d1RSB phase are, however, still exactly described by the belief
propagation algorithm (as described in section \ref{sec:cavity}), and this is what matters to us in the Bayes-optimal
inference.  

A replica symmetry breaking phase (or equivalently the static spin glass phase) can be
defined by the non-self-averaging of the overlap $q$ between two
configurations randomly sampled from the Boltzmann
distribution. Depending on the realization of the sampling, the
Hamming distance between the two configurations is with nonzero
probability different from its average. 

In section \ref{sec:Bayes_optimal} we derived that, on the Nishimori line, the overlap $m$ between a typical configuration
and the planted one is equal to the overlap between two typical
configurations $q$. With similar arguments one can derive equality of the distribution (over
realizations of the sampling) $P(q)$ of the overlaps between two
configurations randomly sampled from the posterior, and the
distribution of the magnetizations $P(m)$, which is the overlap between the
planted configuration and a random sample from the posterior
\cite{NishimoriBook01}.  In physics, the magnetization is always argued to
be self-averaging (the same is true for the free energy density, and
the magnetization is its derivative with respect to the magnetic field),
and hence on the Nishimori line the overlap is also
self-averaging. From this, one argues that $P(q)$ is trivial,
which in mean-field spin glasses indicates the absence of a spin glass
phase. However, from a rigorous point of view, self-averaging of the
magnetization remains an open problem.

Another way to see the problem is to remark that, given that the spin glass
susceptibility is equal to the ferromagnetic one for planted systems
(see for instance \cite{krzakala2011melting1,krzakala2011melting2}),
then it is hard to reconcile the presence of an equilibrium
ferromagnet phase (where the susceptibility is finite) with an equilibrium
spin glass phase (where the susceptibility diverges).

A complementary (rigorous) argument was presented by Montanari
\cite{Montanari08} who showed that in sparse systems (i.e. each
observations depends on a bounded number of variables) in the
Bayes-optimal setting, two point correlations decay. In the static spin
glass phase these correlations could not decay.  

This means that we can exclude the presence of a static
equilibrium transition to a glass phase on the Nishimori line. Note,
however, that the dynamics can still be complicated, and in fact as we
mentioned, a
dynamic spin glass (d1RSB) phase can appear. As soon as we
deviate from the Bayes optimal setting and mismatch the prior, the
model or their parameters, the stage for glassiness to arrive is
open. And indeed as we see in Fig.~\ref{fig:phase_planted} there are
regions with RSB when $\beta \neq \beta^*$, typically when we take the
temperature too small. When studying
inference with a mismatched prior or model we thus always have to keep this
in mind before making claims about exactness.

\subsection{Applications of planting for studies of glasses}
\label{sec:glasses}
Now that we have explored the concept of planting and quiet planting,
and its connection to inference, we will show that it also has a
number of consequences for the study of structural glasses in
physics. While these are not directly relevant to inference, it
illustrates in many ways the fascinating properties of the planted
ensemble and the deep connection between inference, optimization
problems and the statistical physics of glasses, and we thus summarize
these properties in this section.

\subsubsection{Equilibration for free in glasses}
\label{equlibration}

Let us discuss the physical consequences of the fact that in quietly
planted systems the planted configuration behaves exactly the same way as
any other equilibrium configuration. We can hence literarily
equilibrate for free. This is particularly interesting in the study of
glassy models. As a matter of fact, the very concept of glassiness can be defined as super-fast growth of
the equilibration time as the glass transition temperature is
approached~\cite{BouchaudCugliandolo98}.

The core of the interest in glasses is concerned by finite (usually three) dimensional
models. In those cases the annealed and quenched free energies are
known not to coincide. A class of theories of glasses is, however, based on mean-field
(i.e. systems living on random geometries of fully connected graphs)
disordered systems that present a discontinuous phase transition. The
analog of $c_d$ in these systems is called the dynamical glass
transition, and the analog of $c_c$ is called the Kauzmann glass
temperature \cite{berthier2011theoretical}. Properties of the equilibration time are well studied and
it turns out that the equilibration time (i.e. number of Monte Carlo steps
per variable in the thermodynamic limit) diverges as the dynamical
temperature is approached \cite{MontanariSemerjian06b} and is
exponentially large in the system size between the dynamical and
Kauzmann temperature. This in practice means that numerical
simulations of these models are very cumbersome. 

In most of the mean-field models used for the studies of glasses, the
quiet planting works. Numerical simulations studying the dynamics from
equilibrium can hence be speeded up considerably by initializing in
the planted configuration. A particular version from of this trick was first
used in \cite{MontanariSemerjian06} where it enabled for the first
time precise simulation of the dynamics starting from equilibrium. It
was later popularized by \cite{krzakala2009hiding} and has now been
used in a considerable number of works. Without being exhaustive let
us mention
Refs. \cite{mari2011dynamical,foini2012relation,bapst2013effect,charbonneau2014hopping,mariani2015calorimetric,charbonneau2015numerical}.

Even if one does not care about the equivalence to the
randomly-quenched ensemble one can use planting to obtain equilibrium
configurations even in general finite dimensional systems. Indeed the
property of planted configuration being the equilibrium one for a
corresponding Boltzmann measure is general. This has been used in
numerical studies of so-called pinning in glasses, see
e.g.~\cite{hocky2014equilibrium}.

\subsubsection{Following states and analyzing slow annealing}
\label{sec:foll}
In a very influential work, Franz and Parisi \cite{FranzParisi97}
considered the so called ``potential'' method, that aimed to analyse
the properties of a single pure state in glasses.  Perhaps more
interestingly, this allows to describe the adiabatic evolution of
these glassy Gibbs states as an external parameter, such as the
temperature, is slowly tuned. Again, there is a deep connection with
the planted ensemble which we shall now
discuss. Ref. \cite{krzakala2009hiding} studied the state that is
created in the planted coloring around the planted
configuration. Adding to this the concept of quiet planting, this
framework allows us to study the {\it shape} of states in mean-field
glassy systems and is equivalent to the Franz-Parisi construction.

When a macroscopic system is in a given phase, and if one tunes a
parameter, say the temperature, very slowly then all observables, such
as the energy or the magnetization in a magnet, will be given by the
equilibrium equation of state. In a glassy system, where there could
be many such states, it is very interesting to see how they evolve
upon cooling, as they tend to fall off equilibrium very fast. This is
for instance what is happening in the so-called 3-XOR-SAT problem,
also called 3-spins
spin-glass~\cite{GrossMezard84,RicciWeigt01,MezardRicci03}. This is
nothing but an Ising spin glass where the two body interaction of
Hamiltonian~(\ref{Ham_Ising}) is replaced by a three spin interaction
(i.e. each term is a triplet $J_{ijk} S_i S_j S_k$).

\begin{figure}[!ht]
\begin{center}
  \resizebox{10cm}{!}{\includegraphics{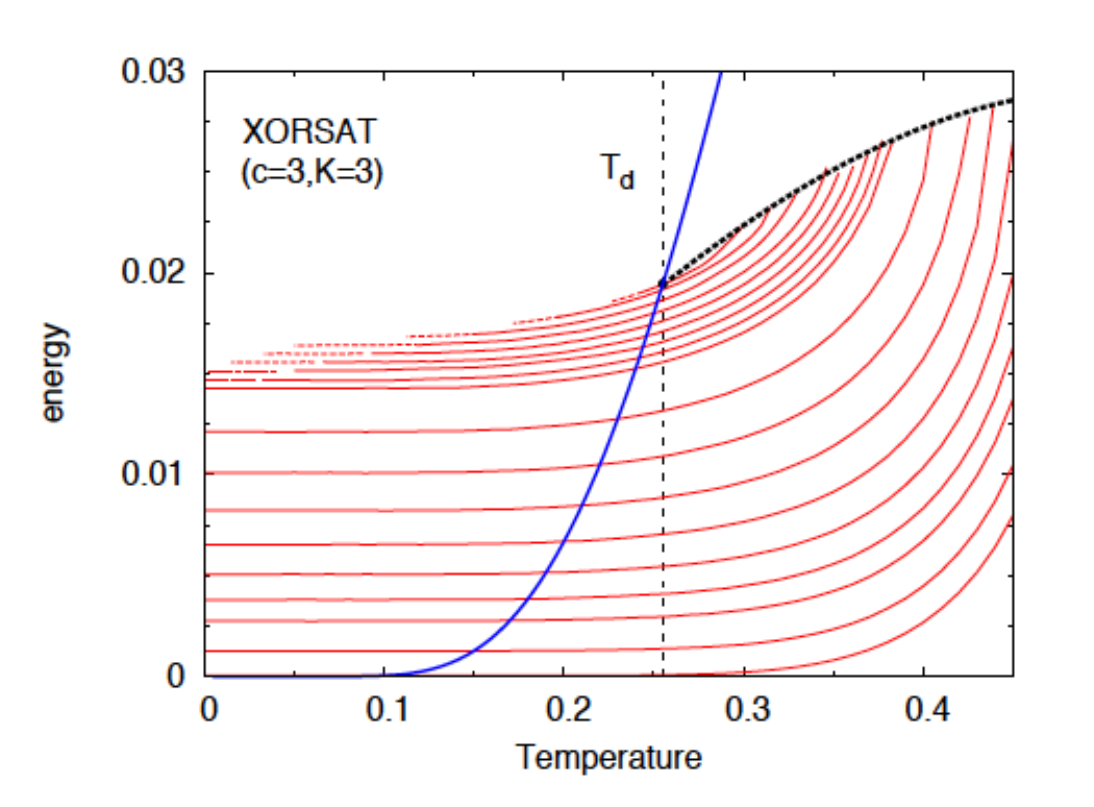}}
\end{center}
  \caption{ \label{fig:following} State following in regular
    3-XOR-SAT. Figure taken from \cite{zdeborova2010generalization}.}
\end{figure}

A plot that illustrates this phenomena is given in
Fig.~\ref{fig:following}. The blue line is the equilibrium (in
technical terms, the one-step replica symmetry breaking solution)
energy of the 3-regular hyper-graph (this means that every variables
has exactly $3$ triplets of interactions, chosen randomly) 3-XOR-SAT
problem as a function of temperature~$T$.  For this system quiet
planting works at {\it all} temperatures.  The red lines are energies
that we obtain when we plant quietly (at temperature where the red
lines crosses the blue one) and {\it then} lower or increase the
temperature and monitor the corresponding energy given by the
dynamics. This procedure has been called {\it state following} in
\cite{krzakala2010following,zdeborova2010generalization}, because
physically this is a study of the corresponding state at temperature
at which the state is no longer among the equilibrium ones.

Note that such state following was often described in physics of
glasses on the level of a thought-experiment. Methods for quantitative
analysis existed only for the particular case of spherical $p$-spin
model, via an exact solution of the dynamics
\cite{CugliandoloKurchan93,crisanti1993sphericalp,BouchaudCugliandolo98}. For
other systems, in particular diluted ones, state following was
considered to be an open problem, see e.g. the concept suggested but
not analyzed in \cite{KrzakalaKurchan07}. The method for state
following is mathematically very much related to the concept of
Franz-Parisi potential at two different temperatures
\cite{FranzParisi97,BarratFranz97}. 

From an algorithmic point of view, the method of state following opens
the exciting possibility of analyzing limiting energy of a very slow
annealing. It is well known that in a finite size system annealing
always finds the ground state, in a time that is in general
exponential in the system size. A crucially interesting question is:
What is the lowest achievable energy when the annealing rate scales
linearly with the size of the system? In mean-field models of glasses
equilibration is possible in a time that is linear in the system size
down to the dynamical temperature~$T_d$ (or equivalently up to $c_d$),
after that the dynamics gets blocked in one of the exponentially many
states. If we take an equilibrium state at $T_d-\epsilon$ and follow
it down to zero temperature the resulting energy is a very good
candidate for the behavior and for the limiting energy of very slow
simulated annealing.  Very slow annealing for the Ising $p$-spin model
was compared to the
theoretical prediction from state following in \cite{krzakala2013performance} .

\paragraph*{\it Paradox in state following.}
There is, however, a problem that appears when one tries to carry out
the calculation for following states that are at equilibrium close
to~$T_d$. When following the state down in temperature, the replica
symmetry gets broken at some point (the boundary between full and
dashed lines in Fig.~\ref{fig:following}). Moreover, at yet lower
temperature the solution correlated to the planted configuration
disappears (in a mathematically analogous way as when a spinodal line
is encountered). Even using the 1RSB computation to describe the
internal structure of the state we were not able to follow it down to
zero temperature.  A possible explanation is that the correct
description of this region is the full-step replica symmetry breaking
(FRSB). To test this hypothesis Ref. \cite{sun2012following} analyzed
the mixed spherical $p$-spin model where the FRSB is tractable.

The findings of \cite{sun2012following}, however, raised even more
questions. It is found by the Nishimori mapping to the model with a
ferromagnetic bias (as described in section \ref{sec:Nish}) that there
is no FRSB solution to the problem, yet there is no magnetized 1RSB
solution at very low temperature either.  Yet in MCMC simulation one
always confirms the expectation that as temperature is lowered the
bottom of the state is correlated to the reference configuration. In
other words, using the Nishimori mapping, the non-magnetized spin
glass phase that is assumed to exist at low temperatures for
$\rho>\rho_c$ in Fig.~\ref{fig:phase_planted} is not physical in the
context of state following.  These observations are hence an
inconsistency in the theory. Loose ends such as this one are always
interesting and basically always eventually lead to new theoretical
development. This state following paradox should hence not be
forgotten and should be revisited as often as new possibilities open.

One such possibility might be related to the calculation involving a
chain of reference (planted) configurations at lower and lower
temperatures as suggested in \cite{franz2013quasi}. 

\subsubsection{Melting and glass transition}

Using the planted models and extended form of the Nishimori conditions
the analogy between glassy dynamics and melting was studied in 
\cite{krzakala2011melting1,krzakala2011melting2}. Perhaps the
theoretically most intriguing and very fundamental question in the theory
of glasses is: Is there a true glass transition in a
finite-dimensional system (i.e. system where the graph of interactions
can be embedded into finite-dimensional Euclidean space)? Such an ideal glass transition has to come
with diverging time
and length scales when approaching the transition. A definitive
answer to this question is difficult to obtain as both
simulations and experiments are faced with the extremely
slow dynamics. According to the random first order theory
of the glass transition
\cite{kirkpatrick1987connections,kirkpatrick1987stable}, there exist
systems with time and length scales with a genuine divergence
at the ideal glass transition temperature. Moreover this theory is
able to predict in which systems such an ideal glass transition exists
and in which it does not. However, 
the random first order theory is not free from criticisms, see for instance
\cite{biroli2012random} and references therein.

The question of existence of the ideal glass transition remains open, but in 
\cite{krzakala2011melting2} it was derived that if there is a
finite-dimensional model with a first order
phase transition on the Nishimori line (or equivalently in the Bayes
optimal inference) then there is an ideal glass transition as well. This result was reached by using the Nishimori mapping from a planted
model to a ferromagnet, sec.~\ref{sec:Nish}, and comparing their dynamical
properties. In particular, if there is a first order
phase transition on the Nishimori line then it comes with two quite
uncommon properties - absent latent heat, and the melting dynamics being equivalent to the stationary
equilibrium dynamics. Arguably, first order phase transitions are easier
to describe or exclude than the tricky glass transition. This is hence an interesting path towards the
fundamental question of existence of the ideal glass transition. 

\newpage

\section{Phase diagram from the cavity method}
\label{sec:cavity}
Now we move to special cases of inference problems where the phase
diagram and phase transitions can be computed analytically exactly,
and for which related algorithms more efficient than Monte-Carlo can
be derived.  We shall achieve this goal using extended mean-field
methods.  Slightly abusively we will call mean-field systems those
where these methods lead to exact results.

Mean-field methods are often the first step in the understanding of
systems in physics, and disordered systems are no exception.  In spin
glasses, in particular, the mean-field model proposed in $1975$ by
D. Sherrington and S. Kirkpatrick \cite{SherringtonKirkpatrick75} has
been particularly important. We refer to the classical literature
\cite{MezardParisi87b} for the replica solution, involving the
so-called replica symmetry breaking, devised by Parisi. It is not that
we are not interested in the replica method, quite the contrary, the
replica method is at the roots of many results in this review. We
shall, however, focus here in the alternative cavity method
\cite{MezardParisi85c,MezardParisi01,MezardParisi03} since this leads
more naturally to specific algorithms and algorithmic ideas. The
development of more efficient algorithms is the goal in inference
problems.

There are two main types of lattices for which the mean-field theory
is exact. The first type is the fully connected lattice underlying the
canonical Sherrington--Kirkpatrick model.  Similar solutions have been
derived for the p-spin model \cite{GrossMezard84} and for the Potts
glass model \cite{GrossKanter85}, and these played a major role in the
development of the mean-field theory of the structural glass
transition
\cite{KirkpatrickThirumalai87a,KirkpatrickThirumalai87b,KirkpatrickThirumalai89}.
The linear estimation problem that we treat in Sec.~\ref{chap:CS} is
an example of a problem on a fully connected lattice.

The second type of lattice for which the mean-field theory is exact is
given by large random graphs with constant average degree, a case commonly refereed to as ``Bethe lattice'' in the
physics literature~\cite{VianaBray85}.  In 2001 M\'ezard and Parisi
\cite{MezardParisi01,MezardParisi03}, using the cavity method, adapted
the replica symmetry breaking scheme to solve models on such sparse
lattices. There is a number of reasons why Bethe lattices are
interesting, the two major ones are: (a) because of the finite
degree, they provide a better approximation of the finite
dimensional case and because the notions of distance and neighboring
can be naturally defined, and (b) the related random graphs are
fundamental in many interdisciplinary cases, for instance in computer
science problems, which is our main motivation here.

The cavity method is a generalization of the Bethe and Onsager ideas
\cite{Bethe35,Onsager44} for disordered systems. It was initially
developed by M\'ezard, Parisi and Virasoro \cite{MezardParisi85c} as a
tool to recover the solution of the Sherrington-Kirkpatrick model
without the use of replicas. It was subsequently developed by M\'ezard
and Parisi to deal with the statistical physics of disordered systems
on random graphs \cite{MezardParisi01}.  There is a deep global
connection between the Bethe approximation and the so-called belief
propagation (BP) approach in computer science (in error
correction~\cite{Gallager68} and Bayesian networks~\cite{Pearl82}) as
was realized in connection to error correcting codes by Kabashima and
Saad \cite{kabashima1998belief}, and was put in a more general setting
by \cite{YedidiaFreeman03}. For a more recent textbook that covers in
detail the cavity method and belief propagation see
\cite{MezardMontanari07}, for a very comprehensive explanation of
belief propagation see \cite{YedidiaFreeman03}. 


\subsection{Random graphs}
\label{sec:RandomGraphs}
We will remind basic properties of sparse {\it Erd\H{o}s-R\'enyi} (ER)
\cite{ErdosRenyi59,ErdosRenyi60} random graphs, that we used briefly already in
sec.~\ref{sec:planted_def}. At the core of the cavity method is the
fact that such random graphs locally look like trees, i.e. there are
no short cycles going trough a typical node. 

An ER random graph is taken uniformly at random from the ensemble,
denoted ${\cal G}(N,M)$, of graphs that have $N$ vertices and $M$
edges. To create such a graph, one has simply to add random $M$ edges to an
empty graph. Alternatively, one can also define the so called
${\cal G}(N,p)$ ensemble where an edge exists independently for each
pair of nodes with a given probability $0<c/N<1$. The two ensembles are
asymptotically equivalent in the large $N$ limit, when
$M=c(N-1)/2$. The constant $c$ is called the average degree. We denote by 
$c_i$ the degree of a node $i$, i.e. the number of nodes to which $i$
is connected. The degrees are distributed according to
Poisson distribution, with average $c$.

Alternatively, one can also construct the so-called {\it regular} random
graphs from the ensemble ${\cal R}(N,c)$ with $N$ vertices but where
the degree of each vertex is fixed to be exactly $c$. This
means that the number of edges is also fixed to $M=cN/2$.

We will mainly be interested in the sparse graph limit, where 
$N\to\infty$, $c\sim O(1)$, $M\sim N$. The key point is that, in this
limit, such random graphs can be considered locally as trees \cite{JansonLuczak00,Bollobas01}.
The intuitive argument for this result is the following one: starting from
a random site, and moving following the edges, in $\ell$ steps $c^{\ell}$
sites will be reached. In order to have a loop, we thus need $c^{\ell}\sim
N$ to be able to come back on the initial site, and this gives
$\ell \sim \log(N)$. 


\subsection{Belief propagation and the cavity method}
\label{sec:BPexplained}
Now that we have shown that random graphs are locally tree-like, we
are ready to discuss the cavity method. We will consider a system with
Hamiltonian
\be
{\mathcal H}=-\sum_{(ij) \in {E}} J_{ij} S_i S_j -\sum_{i} h_{i}
S_i
\ee
and derive the thermodynamics of this model on a tree. We have to deal with
$N$ variables $S_i$ with $i=1,\ldots,N$, and $M$ interactions, that we
denote $\psi_{ij}(S_i,S_j)$, on each edge $(ij)$ of the graph
$G(V,E)$, with $\psi_{ij}(S_i,S_j)=\exp(\beta J_{ij}S_iS_j)$.
Further denoting $\psi_{i}(S_i)=\exp(\beta h_{i}S_i)$ the ``prior''
information the Hamiltonian gives about $S_i$, the total
partition sum reads
\be
Z=\sum_{\{{S}_{i=1,\dots,N}\}}\[[\prod_i \psi_{i}(S_i)
\prod_{(ij)\in E}\psi_{ij}(S_i,S_j)\]] \ee

To compute $Z$ on a tree, the trick is to consider instead the
variable $Z_{i \to j}(S_i)$, for each two adjacent sites $i$ and $j$,
defined as the {\it partial} partition function for the sub-tree
rooted at $i$, when excluding the branch directed towards $j$, with a
fixed value $S_i$ of the spin variable on the site $i$. We also need
to introduce $Z_i(S_i)$, the partition function of the entire complete
tree when, again, the variable $i$ is fixed to a value $S_i$. On a
tree, these intermediate variables can be exactly computed according to
the following recursion
\bea Z_{i \to j}(S_i) &=&
\psi_{i}(S_i) \prod_{k \in \dimj} \left(\sum_{S_k} Z_{k \to i}(S_k)
  \psi_{ik}(S_i,S_k) \right) \ , \label{eq:Zitoj}\\ \qquad Z_i(S_i) &=&
\psi_{i}(S_i)\prod_{j \in \di} \left( \sum_{S_j} Z_{j \to i}(S_j)
  \psi_{ij}(S_i,S_j) \right) \ ,
\label{eq:Zi}
\eea
where $\dimj$ denotes the set of all the neighbors of $i$, except spin
$j$. In order to write these equations, the only assumption that
has been made was that, for all $k\neq k' \in \dimj$, $Z_{k \to i}(S_k)$
and $Z_{k' \to i}(S_{k'})$ are independent. On a tree, this is
obviously true: since there are no loops, the sites $k$ and $k'$ are
connected only through $i$ and we have ``cut" this interaction when
considering the partial quantities. This recursion is very similar, in
spirit, to the standard transfer matrix method for a one-dimensional
chain.

In practice, however, it turns out that working with partition
functions (that is, numbers that can be exponentially large in the
system size) is somehow impractical, and we can thus normalize
eq. (\ref{eq:Zitoj}) and rewrite these recursions in terms of
probabilities. Denoting $\eta_{i\to j}(S_i)$ as the marginal
probability distribution of the variable $S_i$ when the edge $(ij)$ has been
removed, we have 
\beq \eta_{i \to j}(S_i)=\frac{Z_{i \to j}(S_i)}{\sum_{S'}Z_{i \to
    j}(S')} \, ,~~~~~ \eta_i(S_i)=\frac{Z_i(S_i)}{\sum_{S'}Z_i(S')} \,
.\eeq
So that the recursions (\ref{eq:Zitoj}-\ref{eq:Zi})  now read
\bea 
\eta_{i \to j}(S_i) &=&
\frac{\psi_{i}(S_i)}{z_{i \to j}} \prod_{k \in \dimj} \left(\sum_{S_k}
  \eta_{k \to i}(S_k) \psi_{ik}(S_i,S_k) \right) \, ,   \label{eq:msg1}
\\
\eta_i(S_i) &=& \frac{\psi_{i}(S_i)}{z_i} \prod_{j \in \di} \left(
  \sum_{S_j} \eta_{j \to i}(S_j) \psi_{ij}(S_i,S_j) \right) \ ,
\label{eq:msg2}
\eea
where the $z_{i \to j}$ and $z_i$ are normalization constants
defined by:
\bea
z_{i \to j} &=& \sum_{S_i}\psi_{i}(S_i)
\prod_{k \in \dimj} \left(\sum_{S_k} \eta_{k \to i}(S_k) 
 \psi_{ik}(S_i,S_k)  \right) \ ,  \label{eq:norm1} \\
z_i &=& \sum_{S_i}  \psi_{i}(S_i)\prod_{j \in \di} \left( 
\sum_{S_j} \eta_{j \to i}(S_j)  \psi_{ij}(S_i,S_j)  \right) \ .
\label{eq:norm2}
\eea

The iterative equations (\ref{eq:msg1},\ref{eq:msg2}), and their
normalization (\ref{eq:norm1},\ref{eq:norm2}), are called the belief propagation
equations. Indeed, since $\eta_{i \to j}(S_i)$ is the 
distribution of the variable $S_i$ when the edge to variable $j$ is
absent, it is convenient to interpret it as the ``belief'' of the
probability of $S_i$ in absence of $j$.  It is also called a ``cavity"
probability since it is derived by removing one node from the graph. 
The belief propagation equations are used to define the belief propagation algorithm
\begin{itemize}
\item Initialize the cavity messages (or ``beliefs'')
  $\eta_{i \to j}(S_i)$ randomly or following a prior information $\psi_i(S_i)$ if
  we have one.
\item Update the messages in a random order following the belief propagation recursion
  eq. (\ref{eq:msg1},\ref{eq:msg2}) until their convergence to their fixed
  point. 
\item After convergence, use the beliefs to compute the complete
  marginal probability distribution $\eta_{i}(S_i)$ for each
  variable. This is the belief propagation estimate on the marginal
  probability distribution for variable $i$. 
\end{itemize}
Using the resulting marginal distributions, one can compute, for instance, the
equilibrium local magnetization via $m_i = \langle S_i \rangle =
\sum_{S_i} \eta_i(S_i) S_i$, or basically any other local quantity of
interest. 

At this point, since we have switched from partial partition sums to
partial marginals, the astute reader could complain that it seems that
we have lost out prime objective: the computation of the partition
function. Fortunately, one can compute it from the knowledge of the
marginal distributions.  To do so, it is first useful to define the
following quantity for every edge $(ij)$:
\beq z_{ij} = \sum_{S_i,S_j} \eta_{j \to i}(S_j)
\eta_{i \to j}(S_i) \psi_{ij}(S_i,S_j) = \frac{z_j}{z_{j \to i}} =
\frac{z_i}{z_{i \to j}} \ ,
\label{eq_norm_Ising2}
\eeq where the last two equalities are obtained by plugging 
(\ref{eq:msg1}) into the first equality and realizing that it almost
gives eq.~(\ref{eq:norm2}).  Using again
eqs.~(\ref{eq:msg1}-\ref{eq:norm2}), we obtain \bea \nonumber z_i
&=&\sum_{S_i} \psi_{i}(S_i) \prod_{j \in \di} \left(
  \sum_{S_j} \eta_{j \to i}(S_j)  \psi_{ij}(S_i,S_j)  \right) \\
&=& \sum_{S_i} \psi_{i}(S_i) \prod_{j \in \di} \left( \sum_{S_j}
  \frac{Z_{j \to i}(S_j)}{\sum_{S'} Z_{j \to i}(S')}
  \psi_{ij}(S_i,S_j) \right) =\frac{\sum_{S_i} Z_i(S_i) }{\prod_{j \in
    \di} \sum_{S_j} Z_{j\to i}(S_j) } \ , \eea and along the same
steps \beq z_{j\to i} = \frac{ \sum_{S_j} Z_{j\to i}(S_j) }{\prod_{k
    \in \partial j \setminus i} \sum_{S_k} Z_{k\to j}(S_k) } \ .  \eeq

For any spin $S_i$, the total partition function can be obtained using
$Z = \sum_{S_i} Z_i(S_i)$.  We can thus start from an arbitrary spin
$i$ \beq Z = \sum_{S_i} Z_i(S_i) = z_i \prod_{j \in \di} \left(
  \sum_{S_j} Z_{j\to i}(S_j) \right)= z_i \prod_{j \in \di} \left(
  z_{j\to i} \prod_{k \in \partial j \setminus i} \sum_{S_k} Z_{k\to
    j}(S_k) \right) \ , \eeq and we continue to iterate this
relation until we reach the leaves of the tree.  Using
Eq.~(\ref{eq_norm_Ising2}), we obtain
\beq Z = z_i \prod_{j \in \di} \left( z_{j \to i} \prod_{k
    \in \partial j \setminus i } z_{k\to j} \cdots \right) = z_i
\prod_{j \in \di} \left( \frac{z_j}{z_{ij}} \prod_{k \in \partial j
    \setminus i} \frac{z_k}{z_{jk}} \cdots \right) = \frac{\prod_{i}
  z_i}{\prod_{(ij)} z_{ij}} \, .\eeq

We thus obtain the expression of the free energy in a
convenient form, that can be computed directly from the knowledge of the cavity
messages, often called the {\it Bethe free energy} \beq\label{f_Bethe}
\begin{split}
  &f N = -T\log Z =  \sum_i f_i - \sum_{(ij)} f_{ij} \ , \\
  &f_i = -T \log z_i \ , ~~~f_{ij} = -T \log z_{ij} \ ,
\end{split}\eeq
where $f_i$ is a ``site term'' coming from the normalization of the marginal distribution of site $i$, and is related to the
change in $Z$ when the site $i$ (and the corresponding edges) is added
to the system. $f_{ij}$ is an ``edge'' term that can be interpreted as
the change in $Z$ when the edge $(ij)$ is added. This provides a
convenient interpretation of the Bethe free energy
eq.~(\ref{f_Bethe}): it is the sum of the free energy $f_i$ for all
sites but, since we have counted each edge twice we correct this by
subtracting $f_{ij}$ (see for instance \cite{MezardParisi03}).

We have now entirely solved the problem on a tree. There is, however,
nothing that prevents us from applying the same strategy on any
graph. Indeed the algorithm we have described is well defined on any
graph, but we are not assured that it gives exact results nor that it
will converge. Using these equations on graphs with loops is
sometimes referred to as ``loopy belief propagation'' in Bayesian
inference literature~\cite{YedidiaFreeman03}.

When do we expect belief propagation to be correct? As we have have discussed,
random graphs are locally tree-like: they are trees up to any
finite distance. Further assuming that we are in a pure thermodynamic
state, we expect that we have short range correlations, so that the
large $O(\log{N})$ loops should not matter in a large enough system,
and these equations should provide a correct description of the
model. 

Clearly, this must be a good approach for describing a system in a
paramagnetic phase, or even a system with a ferromagnetic transition
(where we should expect to have two different fixed points of the
iterations). It could be, however, that there exists a huge number of
fixed points for these equations: how to deal with this situation?
Should they all correspond to a given pure state? Fortunately, we do
not have such worries, as the situation we just described is the one
arising when there is a glass transition. In this case, one needs to
use the cavity method in conjunction with the so-called ``replica
symmetry breaking'' approach as was done by M\'ezard and Parisi
\cite{MezardParisi01,MezardParisi02,MezardParisi03}.  In this review,
however, we discuss optimal Bayesian inference and we are guaranteed
from the Nishimori conditions discussed in sec.~\ref{RSB_Nish} that no
such replica symmetry breaking occurs. We can assume safely that
in the Bayes-optimal setting the simple belief propagation approach is always providing the correct
description (unless there are other sources of correlations in the system, e.g. the
graph is not random but a lattice or the quenched disorder is
correlated in a non-local manner).

\subsection{Properties of the randomly-quenched ensemble}
\label{sec:random}
The first system on which we will illustrate how to use the cavity
method and the related belief propagation to compute the phase diagram
is the Viana-Bray (V-B) model for the so-called diluted spin glass
\cite{VianaBray85}.  The reason we discuss first the solution for this well known
model is that, using the theory developed in sec.~\ref{planted_Ising}, we will be
able to read off the solution for the planted spin glass model from
the phase diagram of the randomly-quenched model, see Fig.~\ref{fig:phase_planted}.  Viana and Bray studied the Ising spin glass
Hamiltonian (\ref{Ham_Ising}) with the lattice being an
Erd\H{o}s-R\'enyi random graph with average degree $c$. Interactions
$J_{ij}$ are chosen from $\pm 1$ independently and uniformly at random and, for
simplicity, we consider the absence of a magnetic field. We call the
ensemble of instances generated this way the randomly-quenched
ensemble. 

The description of the V-B model within this approach was originally
done by \cite{bowman1982spin}.  Instead of working with the iteration
(\ref{eq:msg1}), it is convenient to rewrite it in terms of an
``effective'' local cavity fields $u_{i \to j}$ defined as
\be \eta_{i \to j}(S_i) = \frac{ e^{u_{i \to j} S_i} }{2\cosh{
    e^{u_{i \to j} }}}\, .\ee
With a bit of algebra, the belief propagation equation (\ref{eq:msg1})
can be written as
\be u^{i\to j}_t = f(J_{ij},\{{ u^{k\to i}_{t-1}\}}_{k\in \partial i
  \setminus j} )= \frac{1}{\beta} {\rm atanh}\left[ \tanh(\beta
  J_{ij}) \tanh\left(\beta \sum_{k \in \partial i \setminus j} u^{k\to
      i}_{t-1} \right)\right] \, ,\label{VB_update} \ee
where we have added an index~$t$ that stands for a ``time'' 
iteration. 
As stated, iterating this procedure allows to actually solve the
problem on a tree, which is at the root of the cavity method. Here,
however, we apply this procedure on the graph, iteratively, hence the
time indices in eq.~(\ref{VB_update}). The local magnetization is then
computed from the BP fixed point using the hyperbolic tangent of the
local field acting on the spin, and thus reads \be m_i = \tanh{( \beta
  \sum_{j \in \partial i} u^{j\to i} )} \, .  \ee

Applying the same notation to the Bethe free energy per site (\ref{f_Bethe}) leads to 
\be 
\beta N f = \sum_i (c_i-1) \log\left[ 2 \cosh(\beta h^i)\right]
-\sum_{(ij)\in E}
\log\left[\sum_{S=\pm1} 2e^{S\beta J_{ij}} \cosh(\beta h^{j \to i} +
  S\beta h^{i \to j})\right] \, , \ee
where $h^i=\sum_{k \in \partial i} u^{k \to i}$ and
$h^{i \to j}= \sum_{k \in \partial i \setminus j} u^{k\to i}$, and
$c_i$ is the degree of node $i$. Physical properties of the system are
computed from the fixed points of the BP equations
(\ref{VB_update}). At this point, there are two different strategies to
solve the problem. The first one is to investigate numerically what is
happening on a given realization of the problem, i.e. to generate a
large random graph and to actually run the BP iteration on it. This is
actually very instructive (as we shall see) but no matter how large
our graph will be, it will not be at the thermodynamic limit.

The second strategy is to work on an infinite graph directly via a so
called {\it population dynamics} \cite{MezardParisi01}. We
attempt to describe the model by the knowledge of the probability
distribution ${\cal P}(u)$ of the messages $u$ on an edge with value
$J$ over the (infinite) graph.  Consider for instance the case of a
random graph with fixed degree $c_i=c$. Clearly, within a
pure Gibbs states, ${\cal P}(u)$ satisfies the BP equations (\ref{VB_update}) so that
\be
\label{pop_dyn} {\cal P}(u) = \frac 1{\cal Z} \int P(J){\rm d}J
\delta[u-f(J,\{{ u_d \}}_{d=1,\dots,c-1})] \prod_{d=1}^{c-1} {\cal P}(u_d) {\rm d}{u_d}  \ee
where $f(.)$ is given by eq.~(\ref{VB_update}). While this appears to
be a difficult equation to solve explicitly, it can be conveniently
approximated to arbitrary precision with the so-called population
dynamics method \cite{MezardParisi01}. In this case, one models the
distribution $ {\cal P}(u)$ by a population of $\cal{M}$
values
\be {\cal P}_{\rm pop}(u) = \frac 1{\cal {M}} \sum_{l=1}^{{\cal M}}
\delta(u-u_l) \, . \ee
Plugging this expression in eq.~(\ref{pop_dyn}) leads to the population
dynamics algorithm:
\begin{itemize}
\item Initialize all values $u_l$ randomly.
\item For a given element $l$ of the population, draw $c-1$ values of $u$ from the population,
  and a value $J$ of the coupling from $P(J)$, compute a new value
  $u_{\rm new}=f(J,\{{ u \}})$, and replace the element $u_l$ by $u_{\rm new}$.
\item Repeat the last step until ${\cal P}_{\rm pop} (u)$ has converged to
  a good approximation of ${\cal P} (u)$.
\end{itemize} 
This method is very simple and, given the population dynamics
distribution, allows to compute the value of the free energy per site
via (\ref{f_Bethe}).

Either way, depending on the inverse temperature $\beta$ in the
Viana-Bray model, one observes one of the two following possibilities:
At low $\beta$ there is a stable {\it paramagnetic} fixed point
$u^{i\to j}=0$ for all $(ij)\in E$. The corresponding paramagnetic
free energy reads \be -\beta f(\beta,\bJ) = \log{2}+ \frac{c}{2}
\left[ \log{(2\cosh{\beta})} -\log{2} \right] \, . \label{free_para}
\ee We do notice that the same expression is obtained for the annealed
free energy for this model.

At high $\beta$ (low temperature) the iterations of (\ref{VB_update})
do not converge on a single graph. The last statement, of course, can
not be checked simply when performing the population dynamics
algorithm. A similar effect can, however, be observed by performing two-populations dynamics with the same random numbers starting from
infinitesimally close initial populations. At high $\beta$ the two populations will diverge to entirely different values
\cite{PagnaniParisi03}. This is a clear sign of long range
correlations and of the appearance of a spin glass phase. We
are, of course, interested to know the precise location of the
phase transition.

Thouless \cite{Thouless86} analyzed the threshold value of $\beta_c$
by linearizing the update (\ref{VB_update}) around the uniform fixed
point, thereby obtaining \be u^{i \to j}_t = \tanh(\beta J_{ij})
\sum_{k\in\partial i \setminus j} u^{k \to i}_{t-1} \,
.\label{Bethe_lin} \ee Thouless then argues that averaged over the
disorder $J_{ij}$ the above equation gives zero, and when its square
is averaged we get \be \langle u^2_t \rangle= c \tanh^2{\beta} \,
\langle u^2_{t-1} \rangle\, , \label{SG_stab} \ee where we remind that
$c$ is the average degree of the ER graph, that is also equal to the
average branching factor of the tree that approximates locally the
random graph. We thus expect that the paramagnetic phase becomes
unstable when $c \tanh^2{\beta} \ge 1$.

In fact, by analyzing such perturbations, one is investigating the
behavior of the susceptibility. Indeed the spin glass susceptibility
can be written as
\begin{equation} 
\chi_{\textrm{SG}}\equiv \beta \sum_i \langle S_0 S_i \rangle_{\textrm c}^2 
\approx \beta \sum_{r=0}^{\infty} c^r \overline{\<S_0 S_r \>_{\textrm c}^2} .
\label{eq_susc}
\end{equation} 
In eq.~(\ref{eq_susc}), $\overline{\cdots}$ represents a spatial
average over the whole graph, $\<\cdots\>$ represents a thermal
average, $S_r$ are all spins at distance $r$ from $S_0$, and $X_{\textrm c}$ is the connected version of the correlation
function $X$:
$\langle S_0 S_i \rangle_{\textrm c}=\langle S_0 S_i \rangle-\langle
S_0\rangle \langle S_i \rangle$.
The correlation can be computed using derivatives with respect to an
external magnetic field on the spin $S_0$
\begin{equation} 
\beta \overline{\<S_0 S_i \>_{\textrm c}^2} = \overline {\(( \frac {\partial
    \<S_i\>} {\partial \<h_0\>} \))^2} .
\end{equation}
Since the physical field is a function of the cavity fields, we can
monitor the propagation of the response using the chain rule
\begin{equation}
\frac {\partial
  \<S_r\>} {\partial \<h_0^{\textrm c}\>}=\frac {\partial \<S_r\>} {\partial
  \<h_{r}\>} \frac {\partial \<h_{r}\>}{\partial \<h_{r-1}\>} 
 \frac {\partial \<h_{r-1}\>}{\partial \<h_{r-2}\>} 
\cdots \frac
{\partial \<h_1\>}{\partial \<h_0\>} 
\end{equation}
where we indicate with $h_i$ the cavity field going from spin $i$ to
spin $i+1$ for simplicity and $h_{r}, h_{r-1},..., h_0$ is the
sequence of cavity fields on the shortest path from $S_0$ to $S_r$ (we
are sure that there exists just one shortest path because we are
locally on a tree).  To check whether or not the susceptibility
defined in eq.~(\ref{eq_susc}) diverges, one should thus simply
analyze the square of the derivative of the belief propagation equations,
as we have done in eq.~(\ref{SG_stab}). The conclusion is that the
critical temperature is given by the condition that a
perturbation would grow, so that
\be \beta^{\rm
  ER}_c = {\rm atanh}{\left( \frac{1}{\sqrt{c}}\right)}\,
.  \label{SG_beta} \ee
Given the branching ratio is $c-1$ for a random regular graph, we get
in this case:
\be
     \beta^{\rm reg}_c = {\rm atanh}{\left( \frac{1}{\sqrt{c-1}}\right)}\, .  \label{SG_beta_2}
\ee
It follows that for $\beta<\beta_c$ the state of the Viana-Bray model
is correctly described by the paramagnetic fixed point. For
$\beta>\beta_c$ the Bethe approximation fails and for the correct
physical description the so-called replica symmetry breaking framework
needs to be applied \cite{MezardParisi01}. Since, in this review, we are
concerned by the Bayes-optimal inference that effectively lives on the
Nishimori line and does not need the concept of replica symmetric
breaking, we refer the reader interested in this to the textbook
\cite{MezardMontanari07}.

\subsection{Phase diagram of the planted spin glass}
\label{sec:planted_phase}
Now we compute the phase diagram of the planted spin glass, where the
interactions $\bJ$ are chosen conditionally on the planted configuration $\bS^*$,
following eq.~(\ref{edgeweights2}). In this case the Bethe approximation and its linearization are still described by (\ref{VB_update})
and (\ref{Bethe_lin}). In the planted model the stability analysis of
the paramagnetic fixed point is slightly different, since the
distribution from which the interactions~$J_{ij}$ are taken is not
random but depends on the planted configuration.
This is where our mapping to the Nishimori ensemble is making the
problem rather easy to analyze anyway. Indeed, according to our
analysis in the last section, we simply need to ask whenever, on the
Nishimori line with a bias
$\rho = \frac{e^{\beta^*}}{2\cosh{\beta^*}}$, the system acquires a
spontaneous magnetization. Or more simply, we just need to compute the
phase diagram when a ferromagnetic bias is present. Taking the average
of (\ref{Bethe_lin}) and counting properly the probability of $J_{ij}$
given $S_i^*$ and $S_j^*$ we immediately see that, while the spin
glass transition is unchanged, the ferromagnetic one is now given by
\be
    \langle u_t \rangle= c  (\tanh{\beta^*}) (\tanh{\beta})
    \, \langle u_{t-1} \rangle\, .    
\ee
According to this computation, the system thus starts to polarize towards
the ferromagnetic configuration when
\be \beta > \beta_F = {\rm atanh}\left( \frac{1}{c \tanh{\beta^*} }
\right) \, .  \label{ferro_stab} \ee 
For the Bayes-optimal case (or equivalently, on the Nishimori line)
when $\beta^*=\beta$, this immediately gives the transition from the
no-detectability phase to the recovery phase that we gave in eq.~(\ref{exact_transition}). We see now how easy it
was to compute it using the cavity method and the Nishimori mapping.

A couple of subtle but interesting points can be made about the phase
diagram, always keeping in mind the fact that the inference problem
can be seen as looking for the ferromagnetic state on the Nishimori
line. 
\begin{itemize}
\item As $\beta^*\to \infty$ ($\rho=1$) we recover the ordinary
ferromagnetic Ising model with its well known critical temperature. 
\item It is interesting to observe that there a multi-critical point
  at $\beta={\rm atanh}(1/\sqrt{c})$, where all the three conditions
  (\ref{SG_beta}), (\ref{ferro_stab}) and Bayes-optimality,
  $\beta=\beta^*$, meet. This point, of course, is on the Nishimori
  line, and is precisely the transition point
  eq.~(\ref{exact_transition}).
\item For the Bayes-optimal case, $\beta^*=\beta$, for all
  temperature higher than the transition point $\beta^*<\beta_c$, the
  paramagnetic fixed point is the {\it only fixed point} of the
  recursion~(\ref{VB_update}). Since the paramagnetic and annealed
  free energies are equal, this is a case where the planting is quiet,
  as defined in section \ref{quiet_planting}. This means that in the
  whole region $\beta^*<\beta_c$ the planted ensemble behaves exactly
  the same as the randomly-quenched ensemble. This is indeed reflected
  in the phase diagram in Fig.~\ref{fig:phase_planted}, where
  everything on the left of the multi-critical point is independent of
  the value of $\rho$.  
\item Out of the Bayes-optimal case, i.e. when $\beta^*\neq \beta$,
  there exists at very low temperature (large $\beta$) a so-called
  mixed phase where equilibrium configurations are correlated to the
  planted one (the ferromagnetic after our mapping) but where the
  equilibrium phase is actually a spin glass one. 
\end{itemize}
The phase diagram of the planted spin glass is summarized in
Fig.~\ref{fig:phase_planted} where the boundary between the paramagnet
(P) and the spin glass (SG) is plotted in green, between the
paramagnet (P) and ferromagnet (F) in blue, and the Nishimori line
corresponding to the Bayes-optimal inference in red. We also sketch
the position of the mixed phase (M).

\subsection{Systems with first order phase transitions}
\label{sec:Potts}

In the above planted Ising spin glass model the underlying phase
transition is of second order. In that case there is no particular
computational issue. Either inference is impossible, or it is possible and
tractable. The hard phase outlined in section (\ref{sec:high}) is
absent.

The picture is richer in problems when a discontinuous phase
transition is found in the randomly-quenched ensemble.  Famously, this
is the case in the low density parity check error correcting codes as
described e.g. in \cite{NishimoriBook01,MezardMontanari07}, and has
been analyzed also in details in the pioneering work of Nishimori and
Wong \cite{nishimori1999statistical}. The examples of clustering of
networks and compressed sensing (to which we dedicate sections
\ref{chap:clustering} and \ref{chap:CS}) also fall in this
category. 

\begin{figure}[!ht]
\begin{center}
 \resizebox{10cm}{!}{\includegraphics{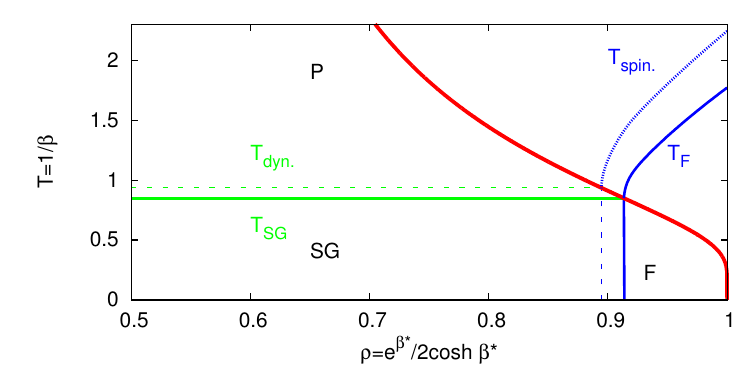}}
\end{center}
\caption{ \label{fig:phase_planted_pspin} Phase diagram of the 3-XOR-SAT
  (or $3$-spin glass) problem as a function of temperature
  $T=1/\beta$ and density of ferromagnetic bonds $\rho$. This
  phase diagram is for random regular hyper-graphs of degree $L=5$. The
  Nishimori line, $\rho = e^{\beta}/(2\cosh{\beta})$, corresponds to
  the Bayes-optimal inference of a planted configuration. As in
  Fig.~\ref{fig:phase_planted} there is the spin glass (SG) phase
  where the system is in a glassy state, the ferromagnetic (F) phase,
  and the paramagnetic (P) phase (for simplicity, the mixed phase is
  not shown). A major difference with respect to
  Fig.~\ref{fig:phase_planted}, however, is the fact that the
  ferromagnetic transition is now a first order one, with a spinodal
  line. The spin glass transition is also a first-order one, with the
  so-called dynamic transition happening {\it before} the static
  transition.}
\end{figure}

In order to observe such a phenomenon, we have to change the
Hamiltonian eq.~(\ref{Ham_Ising}) and move from two-body interactions to
three-body interactions. This leads to the so-called 3-spin spin glass
model, which we already mentioned in sec.~\ref{sec:foll} under the
name 3-XOR-SAT~\cite{GrossMezard84,RicciWeigt01,MezardRicci03}. The
Hamiltonian now reads
\be
   {\cal H}(\bS) = - \sum_{(ijk)\in E} J_{ijk} S_i S_j S_k \, , \label{Ham_XORSAT}
\ee
Again, we can create this model on a random hyper-graph. For instance, we can consider a $L$ regular
hyper-graph where each variable belongs to exactly $L$ triplets,
chosen randomly at random. Again we can study this model under a
planted setting, or equivalently, on the Nishimori line. What is
changing with respect to the preceding section?
First of all, we need to rewrite the belief propagation equation. It
is an exercise (see for instance \cite{RicciWeigt01,MezardRicci03}
or the appendix A in \cite{zdeborova2010generalization}) to see that
they now read
\be m^{i\to a} = \tanh{\left( \beta \sum_{b\in\partial i \setminus a}
    J_b \prod_{j\in \partial b\setminus i} m^{j\to b} \right)} \, 
\label{cavity_ap}
\ee 
where $a,b$ denote the triplets of
interactions. Repeating the analysis of the previous section, one can
obtain the phase diagram which we show in Fig.~\ref{fig:phase_planted_pspin}.

At first sight, this phase diagram looks really similar to the one
from Fig.~\ref{fig:phase_planted}. What has
changed now is the fact that the transition is a first order one. Let
us first concentrate on the Nishimori line. As soon as $T<T_{\rm spin}$ there
are two fixed points of BP, a paramagnetic and a ferromagnetic one, and
the one dominating the measure is the one with lower free energy. Only
when $T<T_{F}$ does the free energy of the ferromagnetic fixed point
become lower than the paramagnetic one. This means that, for
$T<T_{F}$, a perfect sampling would give a configuration with a
positive magnetization. Using the inference-Nishimori connection,
this means that inference is possible for $T<T_{F}$, and impossible
for $T>T_{F}$: the equilibrium phase transition marks the information
theoretic transition.

Now, let us try to actually perform inference in practice, for
instance by running belief propagation or a Monte-Carlo
algorithm. Unfortunately, we have no idea about what is the planted
configuration so at best we can initialize randomly. The
problem is that this will put us in the paramagnetic state, which
exists and is locally stable at all positive temperatures. Both BP or
MCMC will simply explore this paramagnetic state, and, in any
sub-exponential time, will miss the existence of the
ferromagnetic one. This is the classical picture of a first-order phase
transition in a mean-field model. Because of the first
order transition, we are trapped into a high free energy minima that
is stable at all temperatures, and therefore, we cannot perform
inference. While we {\it should} be able
to find a configuration correlated with the planted one (in the sense
that it is information-theoretically possible), we cannot do it within
the class of existing polynomial algorithms.
Related ideas, in the context of the glass transition, were originally
present in \cite{FranzMezard01}. This illustrates the difference
between a ``possible'' inference and an ``easy'' one in the sense of
section~\ref{sec:high}.

XOR-SAT is actually a little bit special in the sense that the easy
detectable phase (see below) is missing. There is a spinodal
transition ($T_{\rm spin}$) where the ferromagnetic phase becomes
unstable, but not a second one where the paramagnet become
unstable. In most models, however, if the temperature is lowered a
second spinodal is reached, the paramagnetic phase becomes eventually
unstable and inference is possible.

Let us thus discuss another example where such a transition
arises. For this, we turn to the planted coloring problem~\cite{krzakala2009hiding}.
\begin{figure}[!ht]
\begin{center}
  \resizebox{10cm}{!}{\includegraphics{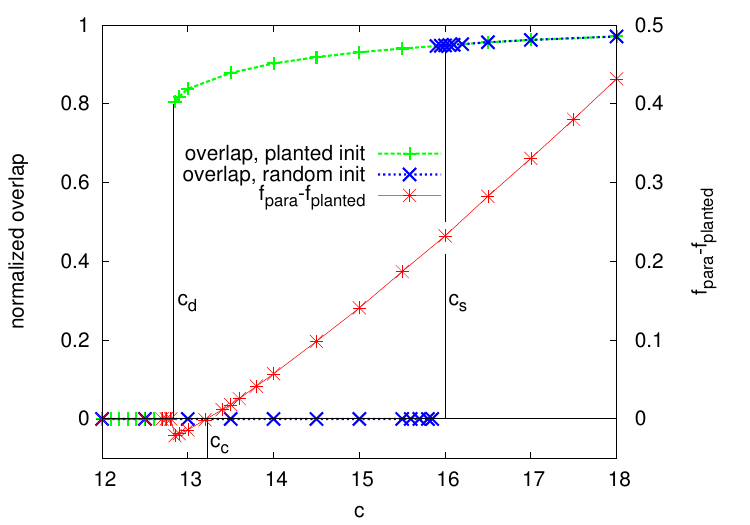}}
\end{center}
  \caption{ \label{fig:planted_col} The phase transitions for
    planted 5-coloring. The overlap (normalized in such a way that
    for a random configuration it is zero) achieved by BP initialized
    in the random and planted configuration. The difference between
    the corresponding free energies is also plotted.  Figure taken from \cite{decelle2011asymptotic}.}
\end{figure}
In planted coloring one first chooses a random assignment of $q$
colors out of the $q^N$ possible ones. Then, one puts $M=cN/2$ edges
at random among all the edges that do not have the same color on
their two adjacent nodes. The Bayes optimal inference in this setting
evaluates marginals over the set of all valid colorings of the created
graph. Computation of these marginals builds on previous works
\cite{MuletPagnani02,ZdeborovaKrzakala07}, where the problem of
coloring random graphs was studied, and goes as follows: One considers
belief propagation initialized in two different ways. In the random
initialization the messages are chosen at random but close to the
paramagnetic ones. In the planted initialization the initial values of
the messages are set to point in the direction of the planted
configurations. A fixed point is reached from both these
initializations and its free energy is evaluated. One obtains the
following picture, illustrated in Figs.~\ref{col_entropies} and
\ref{fig:planted_col} for planted 5-coloring
\begin{itemize}
    \item Undetectable paramagnetic phase, for average degree
      $c<c_d$. The planting is quiet and belief propagation converges
      to the uniform fixed point both from random and planted
      initialization. We will call $c_d$ the {\it dynamical} or the
        {\it clustering} phase transition \cite{KrzakalaMontanari06,krzakala2009hiding}. 
    \item Undetectable clustered  phase, or dynamic one-step replica
      symmetry breaking (d1RSB) phase, for average degree $c_d < c <
      c_c$. The planting is still quiet, but the set of solutions is
      split into exponentially many clusters that are uncorrelated and all look the same as
      the one containing the planted solutions.  We will refer to
      $c_c$ as the {\it detectability} phase transition. 
     In the randomly-quenched ensemble $c_c$ is the condensation transition \cite{KrzakalaMontanari06,krzakala2009hiding}.
    \item Hard detectable phase, for average degree $c_c <c <
      c_s$. This is arguably the most peculiar phase of the four. The
      Bayes optimal marginals are correlated to the planted
      configuration, however, there is a metastable paramagnetic phase
      to which belief propagation converges from a random
      initialization. A ghost of the quiet planting is visible even for
      $c>c_c$ in the sense that the space of configurations looks exactly
      like the one of the randomly-quenched ensemble except for the planted
      cluster. 
    \item Easy detectable phase, for average degree $c>c_s$. Belief
      propagation finds a configuration correlated to the planted
      coloring. The phase transition $c_s$ can be located doing a local stability
      analysis of the uniform BP fixed point, along the same lines as
      done for the planted spin glass in section \ref{sec:planted_phase}. We
      will refer to $c_s$ as the {\it spinodal} or the {\it hard-easy}
      phase transition. In the randomly-quenched ensemble, $c_s$ is the point starting from which
    belief propagation stops converging, related to the local
    instability towards the spin glass phase \cite{ZdeborovaKrzakala07,krzakala2009hiding}. 
\end{itemize} 
A careful  reader will recognize that this corresponds to the
situation we have described in the introduction when we draw
figure~\ref{fig:schema}. The situation in this figure is that of a
problem with a first order transition, where the possible inference is
defined by the equilibrium transition, and the easy/hard transition is
defined by the presence of a spinodal marking the instability of the
paramagnet towards the planted state. This is arising in learning
problems as diverses as the planted perceptron, community detection
with $4$ or more communities, compressed sensing, error correction,
etc.
If, however, the transition is a second order one, then these two
transition coincides as in the case of the planted spin glass (or, as
we shall see, community detection with two equal sizes communities).
We find it remarkable that the century old classification of phase
transition in physics, introduced by Paul Ehrenfest in
$1933$~\cite{Ehrenfest}, turns out to be the right tool to discuss a
deep phenomena of statistical inference and computational complexity.

\subsection{Comments on the phase transitions in various problems}
\label{phase_various}
One more question we want to clarify is whether the concept of {\it
  quiet planting} works for every inference problem. Here the answer
is both yes and no. First of all, for quiet planting to work, we need
to be in a setting where belief propagation has the uniform (sometimes
confusingly called factorized) fixed point, where the messages in the
fixed point do not depend on the indices of the edge. If such a fixed
point exists then the annealed free energy density equals the quenched
one \cite{Mora07}, and consequently the argument about planting being
quiet works. There are many problems where BP on the randomly-quenched
ensemble does not have a uniform fixed point, the most studied
examples are perhaps the problem of $K$-satisfiability (K-SAT), or the Ising
model in a random external magnetic field $h_i$.

Even in problems such as $K$-SAT, where canonical BP does not have a
uniform fixed point \cite{MonassonZecchina96}, we can introduce a
reweighted Boltzmann measure, as done in
\cite{krzakala2014reweighted}, in such a way that the corresponding BP
fixed point is uniform. Planting according to this reweighted measure
then leads to instances that are contiguous to the random ones. Note
that this reweighting and corresponding planted instances were used in
the proof of \cite{AchlioptasMoore02,achlioptas2004threshold} and
studied also in \cite{barthel2002hiding,jia2005generating}. The
planted solution is not an equilibrium solution with respect to the
canonical Boltzmann measure but with respect to the reweighted
measure.

Another comment is related to the
following natural question: How to distinguish between problems where the transition is continuous or
discontinuous? And when it is discontinuous, how wide is the hard
phase? Is there some criteria simpler than performing the
whole analysis? To distinguish between the first and second order transition, we do not know of any
other way. We only have a couple of rules of thumb - in Potts model with sufficient number of
states continuous transitions become discontinuous, and in models
where more than two variables interact, e.g. the $p$-spin model or
$K$-SAT, the continuous transition becomes discontinuous for
sufficiently large $K$ and $p$ \cite{KrzakalaMontanari06}. But there are binary pairwise problems
such as the independent set problem where at sufficiently large average degree
the continuous transition also becomes discontinuous \cite{barbier2013hard}. 

Interestingly, one can identify a
class of constraint satisfaction problems (CSP) where the hard phase is very wide. In
\cite{krzakala2014reweighted} these problems were identified as those
with $c_s\to \infty$. Previously, only XOR-SAT was known to have
this property \cite{FranzMezard01}, but in
\cite{feldman2015complexity} a whole hierarchy
of such problems was identified and is related to $k$-transitivity of
the constraints. A constraint is $k$-transitive if and
only if the number of satisfying assignments compatible with any given
assignment of any $k$ variables is the same. For all CSPs that are at
least 2-transitive the hard phase extends over the whole region of
linearly many constraints, equivalently $c_s\to \infty$. Calling~$r$ the smallest
number such that a constraint is not $r$-transitive, the number of
constraints predicted to render the problem tractable scales as
$N^{r/2}$ \cite{feldman2015complexity}. Interestingly, this
classification makes the $K$-XOR-SAT the hardest of all CSPs (despite
the fact that it can be solved by Gaussian elimination, which is not
robust to any kind of noise). The notion of 2-transitivity, and
consequently infinitely large hard phase, extends also to continuous
variables, and as it was recently pointed out, renders tensor
factorization and completion very difficult
\cite{richard2014statistical}. Physically, the corresponding systems are such
when the paramagnetic phase is locally stable down to zero
temperature \cite{FranzMezard01}. Examples of such systems indeed include the Ising p-spin
(related to $K$-XOR-SAT) or the spherical $p$-spin model related to
tensor factorization. 

Having examples where the hard phase is
huge is very useful when thinking about formal reasons
of why this phase is hard. In \cite{feldman2015complexity} the authors proved
hardness for a class of the so-called {\it statistical algorithms} using
precisely this kind of constraint satisfaction problems. Arguably, the
class of statistical algorithms is too restrictive, but we believe
that the instances of CSP with such a large hard phase should be
instrumental for other attempts in this direction. We shall summarize
in section \ref{sec:hard} arguments for why we think there is
something fundamentally hard about this phase. 

In \cite{zdeborova2011quiet} we discussed the example of planted locked constraint satisfaction
problems (CSP). These are particularly interesting for a couple of
reasons. First of all, the valid solutions are separated by extensive
Hamming distance from each other and therefore BP either converges to a
paramagnet or points exactly towards the planted configuration. Note
that low-density parity check (LDPC) error correcting codes have precisely
this property. Locked CSP can be thought of as non-linear
generalization of LDPC. A second reason why locked CSPs are interesting
is that in some cases the hard phase is very wide and moreover the
planted assignment is with high probability the only one (perhaps up to
a global flip symmetry). The planted locked CSP are hence attractive
candidates for a one-way function that could find applications in
cryptography. Having hard CSP instances that have exactly one valid
assignment was used in studies of quantum annealing where having
non-degenerate ground state is greatly advantageous, see
e.g. \cite{hen2011exponential}.

 \newpage

\section{From physics insight to new algorithmic ideas}
\label{sec:algs}
In the introduction, sec.~\ref{sec:inference}, we presented two main
questions about inference that we
aim to answer. One more fundamental, about phase diagrams, phase
transitions and related insights, the second more practical, about
the development of algorithms. This section is about algorithmic
contributions obtained using the insight resulting from the statistical
physics analysis. For clarity, all is again illustrated on the planted spin glass model.

With the work of M\'ezard, Parisi and Zecchina on survey propagation
\cite{MezardParisi02}, the statistical physics community learned that
the replica and cavity calculations are as powerful when used as
algorithms as they are when used as tools for analyzing the phase diagram.  The
connection between the Bethe approximation and belief propagation was
noticed earlier in connection to error correcting
codes by Kabashima and Saad \cite{kabashima1998belief}, and in a more
general setting by \cite{YedidiaFreeman03}. But looking at the volume
of related literature, it is only after survey propagation that the
whole statistical physics community started to look for explicit
algorithmic applications of the Bethe approximation. 

Here we review some recent physics-related contributions to algorithmic development for
Bayes-optimal inference. 

\subsection{Non-backtracking spectral method}
\label{sec:NB_planted}

Notice, within the historical perspective, that the Bethe solution for
the spin glass on a random graph was written by Bowman and Levin
\cite{bowman1982spin}, i.e. about 16 years before realizing that the
same equation is called belief propagation and that it can be used as an
iterative algorithm in a number of applications.

The paper by Thouless \cite{Thouless86}, who analyzed the spin glass
critical temperature, implicitly includes another very interesting
algorithm, that can be summarized by equation (\ref{Bethe_lin})
where BP is linearized. The fact that this equation has
far-reaching implications when viewed as a basis for a spectral algorithm waited for its discovery even
longer. Notably until 2013 when the methodologically same algorithm was
suggested for clustering of networks \cite{krzakala2013spectral}. This
work will be discussed in sec.~\ref{sec:NB}. We now turn back
to our pedagogical example, the $\pm J$ planted spin glass, and related
spectral algorithm as described and analyzed in~\cite{saade2015spectral}. Related ideas also appeared in
\cite{zhang2014non}. 

First note that a large class of inference algorithms is based on a
spectral method of some kind. Probably the most widely used is the
so-called principal component analysis (PCA). Its
operation can be thought of as revealing the internal structure of the
data in a way that best explains the variance in the data. In PCA the
information is contained in singular vectors (or eigenvectors for
symmetric matrices) corresponding to the largest singular values. Resulting
spectral algorithms are used everyday to solve numerous real world
problems. The same strategy of computing leading singular (or eigen-) vectors of some
matrix related to the data is applied in many settings including the
planted partitioning problem closely related to the planted spin glass
model \cite{coja2010graph,coja2009finding}, or in understanding
financial data \cite{laloux1999noise,bouchaud2003theory}.

A natural question that arises is whether a spectral algorithm can
find a configuration correlated to the planted assignment in the
planted spin glass in the whole region $\beta^*>\beta_c$ where belief
propagation can asymptotically do so.  
Facing this question, many experts would probably suggest to take the matrix $J_{ij}$ (with $J_{ij}=0$
when $(ij)\notin E$) or its associated Laplacian (perhaps normalized
in some way) and compute the leading eigenvector. This is indeed a
good strategy, but on sparse Erd\H{o}s-R\'enyi random graphs it does not work down to the threshold 
$\beta_c$. The problem is that for an Ising model living on a large sparse
Erd\H{o}s-R\'enyi graph the tail of the spectra of the commonly considered
matrices is dominated by eigenvectors localized on some small subgraphs
(e.g. around high-degree nodes or on some kind of hanging
sub-trees). 

The strategy in designing spectral method that does not have the
problem with localized eigenvectors (at least on a much larger class
of graphs) is to try to mimic what BP is doing when departing from the
paramagnetic fixed point. Eq.~(\ref{Bethe_lin}) is precisely the linearization
of BP around that fixed point and suggests a definition of a so-called
{\it non-backtracking matrix} $B$ defined on directed edges as
\be
      B_{i\to j, k\to l} = J_{ij} \delta_{il} (1-\delta_{jk}) \, .\label{NB_matrix}
\ee
Denoting $M$ the number of edges in the non-directed graph, this is a $2M \times 2M$ matrix.
The term $(1-\delta_{jk})$ means that a walk that follows non-zero
elements of the matrix is not going back on its steps, this motivates
the name of the matrix.  
Building on \cite{krzakala2013spectral,bordenave2015non}, it was proven in \cite{saade2015spectral} that the
planted configuration is indeed correlated with the
signs of elements of the eigenvector of $B$ corresponding to the
largest (in module) eigenvalue as soon as $\beta^*>\beta_c$. 

\begin{figure}[!ht]
\begin{center}
  \resizebox{11cm}{!}{\includegraphics{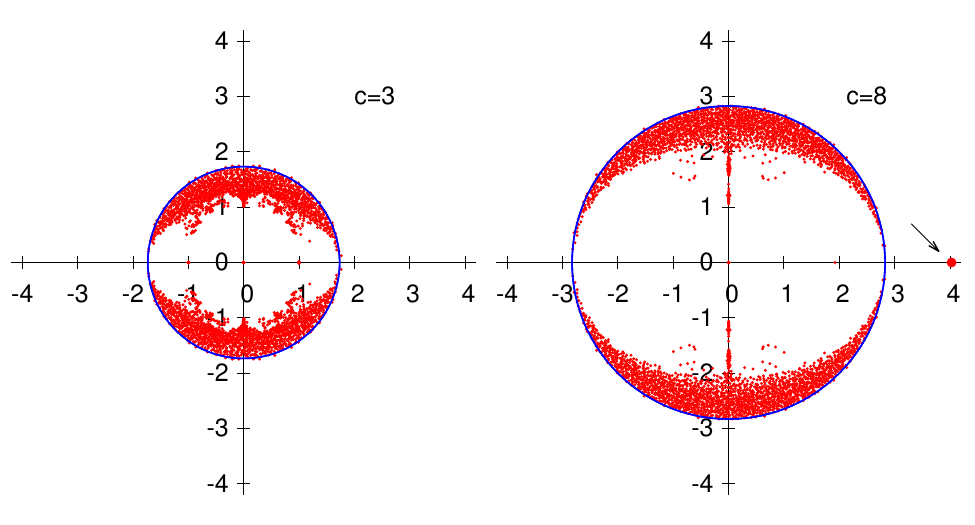}}
\end{center}
  \caption{ \label{fig:NB_spectrum} Spectrum of the non-backtracking
    operator for the $\pm J$ planted spin glass model. Right: In the undetectable
    phase where all eigenvalues are inside a circle of radius
    $\sqrt{c}$. Left: In the detectable phase where one eigenvalue is
    outside the circle on the real axe, its value is $c\tanh{\beta^*}$. Figure taken from \cite{saade2015spectral}.}
\end{figure}

Fig.~\ref{fig:NB_spectrum} depicts the spectrum on the matrix $B$
(\ref{NB_matrix}) for a planted spin glass model on an ER graph of average
degree $c$ and having $N=2000$ nodes. The $J_{ij}=\pm 1$ were generated following the procedure
described in sec.~\ref{sec:planted_def} with $\beta^*=0.55$. We
observe (the proof is found in \cite{saade2015spectral}) that on
sparse random graphs
\begin{itemize}
    \item{For $\beta^*<\beta_c= {\rm atanh}(1/\sqrt{c})$, corresponding to the
        paramagnetic phase of the planted spin glass model: The spectrum of $B$ is
      the same as it would be for $\beta^*=0$, bounded by a circle of
      radius $\sqrt{c}$.}
    \item{For $\beta^*>\beta_c=  {\rm atanh}(1/\sqrt{c})$, corresponding to the
        ferromagnetic phase of the planted spin glass model: The spectrum of
      $B$ has the same property as in the previous case, except for
      two eigenvalues on the real axes. One of them is out of the
      circle and has value $c \tanh \beta^*$. Its corresponding
      eigenvector is positively correlated with the planted
      assignment.}
\end{itemize}

The performance of the non-backtracking spectral algorithms will be
further discussed and compared to other algorithms for the example of
clustering of networks in sec.~\ref{sec:NB}. 
In sec.~\ref{real_nets} we will also illustrate how this method performs on
some non-random graphs. Just as all the other spectral algorithms, this method is not
robust to adversarial changes in a small part of the graph (such as
planting a small dense sub-graph). This can hence be a disadvantage
with respect to methods that optimize some global cost function that is
not affected by small local changes. We, however, illustrate in Fig.~\ref{fig:real_spectrum}
that spectrum of the non-backtracking matrix of many real graphs does
not deviate excessively from its behavior on random graphs.

\subsection{On computational hardness of the hard phase}
\label{sec:hard}

In section \ref{sec:Potts} we described the four phases that we
observe in the
Bayes optimal setting of inference problems. The most theoretically
intriguing is the hard detectable phase that appears between the detectability
transition $c_c$, and the spinodal transition~$c_s$. In this phase
both Monte Carlo and belief propagation, that aim to sample the
Bayes-optimal posterior distribution, fail and get blocked in
the metastable paramagnetic phase. The spinodal transition $c_s$ is
also a barrier for all known spectral algorithms. Some of the spectral methods
including the one based on the non-backtracking matrix saturate this
threshold. 

There are many other problems where a phase of the same physical
origin was discovered. Among them the best known are the LDPC error
correcting codes \cite{MezardMontanari07} and the planted clique problem
\cite{jerrum1992large,alon1998finding,juels2000hiding,deshpande2015finding}. Some
recent works prove
hardness in other problems assuming the hardness of planted clique (in the corresponding
region)
\cite{berthet2013computational,hajek2014computational,ma2015computational,cai2015computational}. There
is a lot  of current interest in this kind of results.

When a physicist sees a phenomenon as the one above she immediately
thinks about universality and tends to conjecture that the spinodal
transition in inference problems will be a barrier for all polynomial
algorithms. There is, however, the case of planted XOR-SAT that
presents the same phenomenology as described in section
\ref{sec:Potts}. Moreover, the hard phase in planted XOR-SAT extends
to infinity (the paramagnetic state is always locally stable). The
work of \cite{feldman2015complexity} actually suggests that the right
scaling for the number of clauses at which planted K-XOR-SAT becomes
easy is when the number of clauses is at least $O(N^{K/2})$. Yet,
thanks to its linearity planted K-XOR-SAT can be solved in polynomial
time by Gaussian elimination. Hence the above conjecture about the
hard phase is false. But the physicist's intuition is rarely entirely
wrong, and in those rare cases when it is indeed wrong it is
for absolutely fundamental reasons. In our opinion, the question:
\begin{center}
          In which exact sense is the hard phase hard? 
\end{center}
is the most fundamental question related to the interface between
statistical physics, statistical inference and average computational
complexity. 

A line of research that should be explored, and seems quite accesible,
is proving or disproving that there is no way to use belief
propagation or Monte-Carlo sampling with a mismatched prior or model in
order to improve the position of the spinodal transition. 

Let us also note here that the spinodals of first order phase
transitions in Bayes-optimal inference problems create a much cleaner
algorithmic barrier than phase transitions described in optimization
or in constraint satisfaction problems such as $K$-satisfiability. In
$K$-SAT the idea of hard instances being close to the satisfiability
threshold goes back to the works by
\cite{CheesemanKanefsky91,SelmanMitchell96,monasson1999determining}. However,
even after more than 20 years of study it is still not clear what is
the limiting constraint density up to which polynomial algorithms are
able to find satisfiable solutions in random $K$-SAT (even at very large $K$). For a related discussion see \cite{zdeborova2009statistical}.

\subsection{Spatial coupling}
\label{sec:coupling}

Spatial coupling is, from a physics point of view, a very natural strategy
to beat the hard-easy spinodal transition in systems where the
structure of the graphical model can be designed. It is closely
related to nucleation in physics. An exponentially long living metastability is only possible in mean-field
systems. Let us describe here why it is not possible in systems
embedded in finite Euclidean dimension~$d$. Metastability means that
there are two competing phases. Let us call $\Delta f$ the free
energy difference between them. Let us consider a large system in the
metastable state with a droplet of radius $L$ (small compared to the
size of the whole system) that by random
fluctuation appears to be in the stable state. Creating such a droplet
will cost some energy due to its interface where the two phases do not
match. This is quantified by the surface tension $\Gamma$. The total energy balance can be written as
\be
   E(L)=\Gamma L^{d-1} - \Delta f L^d\, .
\ee
Physical dynamics can be roughly thought of as gradient descent in this energy. Notice that
$E(L)$ is increasing for $L< L^* = \Gamma (d-1)/(d \Delta f)$ and
decreasing for $L>L^*$. This means that if fluctuations created a
droplet larger than the critical size $L^*$ it will continue growing until it invades
the whole system. The above argument works only in geometries where
the surface of a large droplet is
much smaller than its volume. On random sparse graphs or fully
connected geometries it does not work. This is why nucleation does not
exist in mean-field systems. Notice also that nucleation is simpler
for lower dimensions~$d$, the lowest meaningful case being $d=1$. 

We can conclude that in order to avoid long living metastability
we need to work with finite-dimensional systems. But the situation is
not so simple either. In systems where the mean-field geometries are
used, they are used for a good reason. In LDPC codes they assure large
distance from spurious codewords, in compressed sensing random
projections are known to conserve well the available information,
etc. To avoid metastability we need to mix the mean-field geometry on
a local scale with the finite dimensional geometry on the global
scale. This is exactly what spatial coupling is doing. 
Originally developed for LDPC error correcting codes \cite{lentmaier2010iterative,KudekarRichardson10} and closely
related to convolutional LDPC that have longer history \cite{FelstromZigangirov99}, spatial
coupling has been applied to numerous other settings. Again we will
discuss some of them in more details in section \ref{sec:spatial}. To
introduce the concept here, we restrict ourselves to
the spatially coupled Curie-Weiss model as introduced in \cite{Urbankechains} and studied
further in \cite{caltagirone2014dynamics}.  

The Curie-Weiss (C-W) model in a magnetic field is arguably among the
simplest models that present a first
order phase transition. The C-W model is a system of $N$ Ising spins $S_i\in\{-1,+1\}$, 
that are interacting according to the fully connected Hamiltonian 
\begin{equation}
	\mathcal{H}_{C-W}(\bS) = -\frac{J}{N}\sum_{\left\langle i,j\right\rangle} 
	S_i S_j - h \sum_{i=1}^N S_i\, ,
	\label{eq:C-W}
\end{equation}
where the notation $\left\langle\cdot,\cdot\right\rangle$ is used to denote all 
unique pairs, $J>0$ is the ferromagnetic interaction strength and $h\in \mathbb{R}$ is the external 
magnetic field. 

Let us briefly remind what is obvious to everybody trained in
statistical physics. For positive magnetic field $h>0$ the equilibrium magnetization $m(J,h)>0$ is 
positive, and vice versa. There exists a critical value of the interaction 
strength $J_c=1$ such that: for $J<J_c$ the $\lim_{h\to 0^+} m(J,h) = 
\lim_{h\to 0^-} m(J,h)  = 0$, and for $J>J_c$ we have $\lim_{h\to 0^+} m(J,h) > 
0 > \lim_{h\to 0^-} m(J,h)$. The latter is a first order phase 
transition, in the ``low temperature'' regime $J>J_c$ the system keeps
a non-zero 
magnetization even at zero magnetic field $h$. There exists a spinodal
value of the magnetic field
\begin{equation}
	h_{\rm s}(J)= \sqrt{J(J-1)} -\atanh\left(\sqrt{\frac{J-1}{J}} \right)   \, ,
	\label{eq:hsp}
\end{equation}
with the following properties: If the magnetizations are initialized to
negative values and the magnetic field is of strength  $0 < h < h_{\rm s}(J)$, 
then both local physical dynamics and local inference algorithms, such as the 
Monte-Carlo sampling, will stay at negative 
magnetization $m^-(J,h)<0$ for times exponentially large in the size 
of the system. The spinodal value of the magnetic field $h_{\rm s}(J)$ 
acts as an algorithmic barrier to equilibration, when initialized at
$S_i=-1$ for all $i=1,\dots,N$, and hence to successful 
inference. For $h > h_{\rm s}(J)$ it is, on the other hand, easy to reach the 
equilibrium magnetization $m^+(J,h)$.

\subsection{The spatially coupled Curie-Weiss model} We introduce here the
construction from \cite{Urbankechains} that might not appear very
transparent at first. But its idea is really simple. We create a
one-dimensional chain of C-W models, put a nucleation seed somewhere
and let the sub-systems interact in order to ensure that the
nucleation proceeds to the whole chain. 

\begin{figure}[t]
	\centering
	\includegraphics[scale = 0.35]{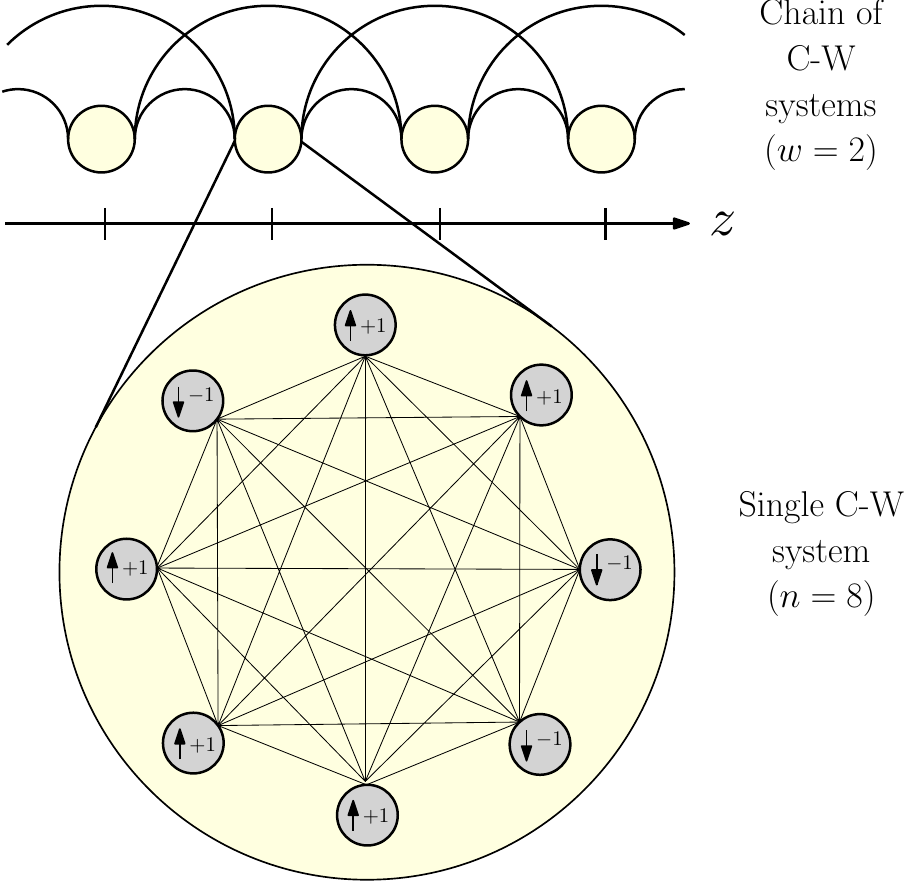}
\includegraphics[scale = 0.6]{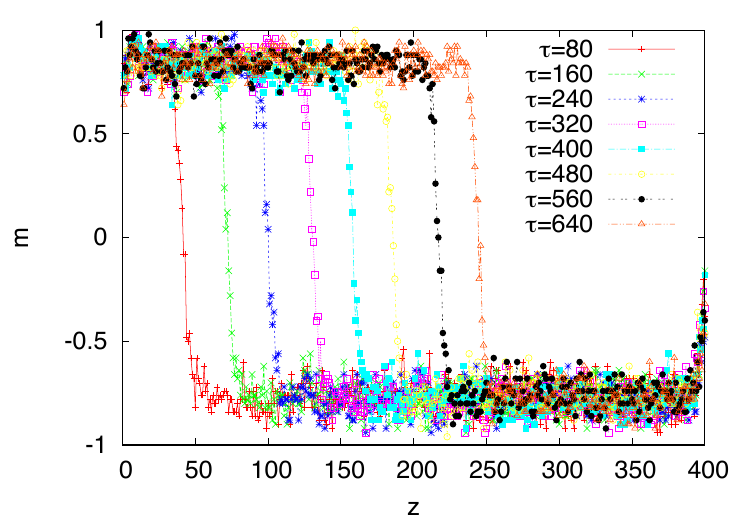}
	\caption{Left: A schematic graphical representation of the
          spatially coupled C-W model.  
		A chain of C-W models interacting within a certain range $w$ ($w=2$ in 
		the figure). In the zoomed part the fully-connected structure of the 
	single C-W model is shown.  A connection between C-W models 
along the chain indicates connections between all spins contained in both 
models. Right: The propagating wave obtained by
		Monte-Carlo heat-bath simulation for the following parameters: $J=1.4$, 
		$L=400$, $n=100$, $h=0.05$, $w=5$, $w_{\mathrm{seed}}=10$ and 
		$h_{\mathrm{seed}}=0.3$.  Simulations were performed using random 
	sequential update and therefore time must be rescaled by
        $\tau=t/N$. The figure is taken from \cite{caltagirone2014dynamics}.}
	\label{fig:model}
\end{figure}

As illustrated in Fig.~\ref{fig:model}, we consider a one-dimensional chain of $\left(2L+1\right)$ C-W systems, where each of the  
C-W system has $n$ spins (referred to as a ``block'') and is labelled by the 
index $z\in\{-L,\ldots,L\}$.  The result is that a configuration $\bS$ of the 
full system is now given by the values of $N=n(2L+1)$ spins, each labelled by a 
compound index:
\begin{equation}
	\bS=\left\{S_{iz}\in\left\{+1,-1\right\}:\ i\in\left\{1,\dots,n\right\}, 
	z\in\left\{-L,\dots,L\right\}\right\}.
	\label{eq:s}
\end{equation}
In a similar way, the uniform external magnetic field $h$ for a single system 
is replaced by an external field profile $h_z$.  As far as the coupling is 
concerned, every spin not only connects to all spins in the same location $z$ 
but also all spins within $w$ blocks from $z$. The corresponding Hamiltonian is 
then
\begin{equation}
	\mathcal{H}_{n,L}\left(\bS\right) =
	-\frac{1}{n}\sum_{\langle iz,jz'\rangle} J_{zz'} S_{iz} S_{jz'} - 
	\sum_{z=-L}^{L} h_z \sum_{i=1}^n S_{iz}.
	\label{eq:H}
\end{equation}
The couplings between spins are $J_{zz'} = J g\left(\vert z-z'\vert/w\right)/w$,
where the function $g$ satisfies the following condition $g\left(\vert
  x\vert\right) = 0 ,\ \forall\ \vert x\vert > 1$
and we choose its normalization to be 
$\frac{1}{w}\sum_{z=-\infty}^{+\infty} g\left(w^{-1}\vert z\vert\right) = 1$.

In order to ensure that the system 
{\it nucleates}, we must increase the field $h_z$ at some point on the chain. We choose 
the magnetic field profile in such a way that $h_z = h_\mathrm{seed} > 
h_\mathrm{sp}$ in some small region given by $z\in 
\{0,\ldots,w_\mathrm{seed}\}$. Everywhere else, $h_z=h \ll h_\mathrm{sp}$, such that the 
average field strength is still small. To illustrate the nucleation in
Fig.~\ref{fig:model} we
initialize to $S_{iz}=-1$ everywhere and let the system evolve under
Monte-Carlo dynamics. We indeed observe a nucleation ``wave'' that
invades the whole system in a time proportional to the number of
blocks. The system can be designed in a such a way that the spinodal
regime entirely disappears \cite{Urbankechains}. 

\subsection{A brief history of time indices in the TAP equations}
\label{sec:TAP}
In physics of disordered system, the historically first form of
iterative procedure that was used to compute marginals (local
magnetizations) in a densely connected spin glass model are the
Thouless-Anderson-Palmer (TAP) equations \cite{ThoulessAnderson77}. In
the early statistical physics works on error correcting codes TAP
equations were used for decoding.  The primary goal of the paper by
Kabashima and Saad \cite{kabashima1998belief}, who pointed out the
connection between belief propagation and the Bethe approximation, was to
compare the performance of TAP and BP. In dense systems the fixed
points of both BP and TAP agree. However, it was reported e.g. in
\cite{kabashima2003propagating,kabashima2003cdma} that TAP equations
did not converge in some cases even when BP does and there is no
replica symmetry breaking.  With the renewed interest in message
passing of dense systems that followed the work of Donoho, Maleki and
Montanari on compressed sensing \cite{donoho2009message}, it is
definitely worth making some up-to-date remarks on the TAP equations
and their convergence.

The fixed point of the TAP equations 
\be
      m_i= \tanh\left[ \beta h+\beta \sum_{j=1}^N J_{ij} m_j - \beta^2 m_i \sum_{j=1}^N J_{ij}^2 (1-m_j^2)\right]
\ee
describes the replica symmetric marginals of the following fully
connected spin glass Hamiltonian 
\be
    {\cal H}_{SK}(\bS) = - \sum_{\langle i,j
      \rangle} J_{ij}S_i S_j  - h \sum_i S_i\, ,
\ee
where $S_i$ are the Ising spins, and $J_{ij}$ are independent random variables of zero mean and variance $\tilde
J/N$. Hence a typical value of $J_{ij} = O(1/\sqrt{N})$.  

Fixed points of the TAP equations are stationary points of the
corresponding TAP free energy \cite{ThoulessAnderson77}. Derived as
such, TAP equations do not include any time-indices and were mostly
iterated in the simplest form of having time $t-1$ on all
magnetizations on the right hand side, and $t$ on the left hand
side. At the same time, as we said above, a number of authors reported
that the TAP equations have problems to converge even when BP does
converge and replica symmetry is not broken
\cite{kabashima2003propagating,kabashima2003cdma}. No satisfactory
explanation for this existed in the literature until very recently
\cite{bolthausen2014iterative}.  The reason behind this
non-convergence was that the time indices were wrong and when done
correctly, convergence is restored, and on sufficiently dense systems
TAP equations behave as nicely as BP. 

For the Ising spin glass this was first remarked by Bolthausen
\cite{bolthausen2014iterative}, who proved the convergence of TAP with the
right time indices in a non-zero magnetic field (and out of the glassy
spin glassy phase). His paper
inspired the proof technique of \cite{bayati2011dynamics} but
otherwise passed quite unnoticed. 
In the context of linear estimation, as discussed in Sec.~\ref{chap:CS},
the analog of the TAP equations is the approximate message passing
algorithm (AMP) \cite{DonohoMaleki09} (but see also \cite{kabashima2004bp}). AMP is often thought of as the large degree limit
of belief propagation and in that way of thinking the correct time indices come up
naturally. Here we re-derive TAP for the Ising spin glass as a
large degree limit of BP to clarify this issue of time indices.

Let us go back to the belief propagation for sparse graphs eq.~(\ref{VB_update}). Define
messages
$m^t_{i\to j} \equiv \tanh{(h+\beta \sum_{k\neq j} u_{t-1}^{k\to i}
  )}$.
Recall that $J_{ij}$, and hence also $u$, is of order $O(1/\sqrt{N})$ and
expand to the leading order. BP then becomes \be m^t_{i\to j}=
\tanh\left( \beta h+\beta \sum_{k\neq j} J_{ki} m^{t-1}_{k\to i}
\right)\, .  \ee In this dense-graph message passing there is not yet
any surprise concerning time indices. The TAP equations are closed on
full magnetizations (not cavity ones) defined as 
\be m^t_{i}=
\tanh\left( \beta h+\beta \sum_{k} J_{ki} m^{t-1}_{k\to i} \right)\,
. \label{TAP_MP} \ee
In order to write the TAP equations we need to write messages in terms
of the marginals up to all orders that contribute to the final
equations. We get by Taylor expansion with respect to the third term

\be m_{i\to j}^t = \tanh\left( \beta h+\beta \sum_{k} J_{ki}
  m^{t-1}_{k\to i} - \beta J_{ij} m^{t-1}_{j\to i}\right) \approx m_i^t
-\beta J_{ij} m^{t-1}_j [1-(m^t_i)^2]\, , \ee 
which leads to the iterative TAP equations
\be m_i^t = \tanh{\left\{ \beta h+ \beta
    \sum_{j} J_{ij} m_j^{t-1} - \beta^2 m_i^{t-2} \sum_j J_{ij}^2
    [1-(m_j^{t-1})^2] \right\}} \, .  \label{TAP:correct} \ee
Observe the time index $(t-2)$ before the last sum, which is not
intuitive on the first sight. These equations are the ones written and
analyzed by Bolthausen \cite{bolthausen2014iterative}.

In his proof, Bolthausen showed that the behavior of these iterative
equations can be followed using what can be called the {\it state
  evolution} of the TAP algorithm, at least outside of the spin glass
phase (the algorithm does {\it not} converge inside the spin glass
phase). It works as follows: we consider the interactions $J_{ij}$ to
have zero mean and variance $\tilde J/N$.  Going back to
eq.~(\ref{TAP_MP}), the argument of the $\tanh$ is a sum of
independent variables (by the assumptions of belief propagation) and
thus follows a Gaussian distribution. Denoting $m$ and $q$ the mean
and the variance of this distribution of messages, two self-consistent
equations can be derived 
\bea m^{t} &=& \frac 1{\sqrt{2 \pi}} \int
e^{-z^2/2}
\tanh{ \(( \beta z \tilde J \sqrt{q^{t-1}}  + \beta  h\))} {\rm d}z  \label{se:tap1}\\
q^t &=& \frac 1{\sqrt{2 \pi}} \int e^{-z^2/2} \tanh^2
{\(( \beta z \tilde J \sqrt{q^{t-1}} + \beta h\))}
{\rm d}z 
\label{se:tap}
\eea 
The fixed point of these equations provides the well known replica
symmetric solution for the magnetization and the spin overlap in the
Sherrington-Kirkpatrick model, as was first
derived in~\cite{SherringtonKirkpatrick75}. Since this is the correct
solution outside the spin glass phase (above the so-called dAT
line~\cite{AlmeidaThouless78}, it is reassuring to see that we have a
convergent algorithm that will find the marginals fast in the densely
connected case.  We will return to this observation in section
\ref{sec:AMP} when we will discuss inference in densely connected
models.

\begin{figure}[!ht]
    \hspace{-8mm}
    \resizebox{16cm}{!}{\includegraphics{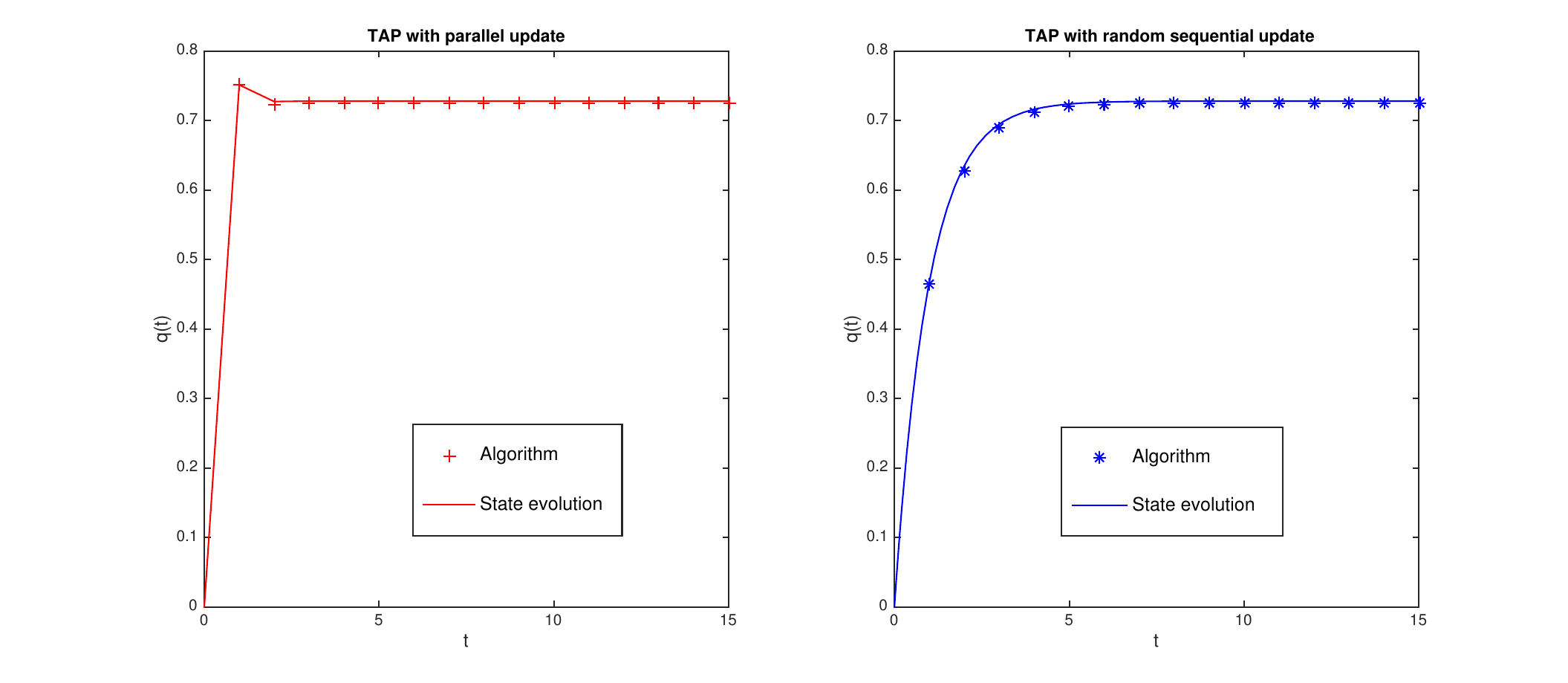}}
  \caption{\label{fig:TAP} State evolution versus actual simulation for
    the TAP equations in the Sherrington-Kirkpatrick with parallel
    update (left) and random sequential one (right), for a spin glass system
    with $N=5000$, $1/\beta=T=3$, $\tilde J=1$ and $h=4$. The variance $q$ of the distribution
    of messages is shown. In both cases, the state evolution equations
    (with the iterative form on the left and the differential one on the
    right) perfectly describe the evolution of the algorithm.}
\end{figure}

Note also that eqs.~(\ref{se:tap1}-\ref{se:tap}) correspond to a parallel update
where all magnetizations are updated at the same time. Very
often, it has been found that such update might be problematic for
convergence and that one should update spins one at a time for better
convergence properties. This, will also be discussed in \ref{sec:AMP}
when discussing the AMP algorithm~\cite{manoel2015swept}. In that case
one chooses randomly at each time step which spin should be updated, then one
can still use eq.~(\ref{TAP:correct}) with a careful book-keeping of the
time-indices. The state evolution eqs.~(\ref{se:tap}) become
in that case differential equations in time. Instead of
$q^{t+1}=f(q^t)$ as in eqs.~(\ref{se:tap}) we have now
${\rm d}q/{\rm d}t=f(q^t) -q^t$ in the continuous time limit. We show an exemple
of the behavior of these two approaches in Fig.~\ref{fig:TAP}. While in
this case the parallel update was faster, we will see that sometimes
the other one could be beneficial because it is smoother and avoids
e.g. the overshoot we see in the first iteration of the parallel update.

These considerations stems from the derivation of TAP equations
starting from the BP equations for random interaction. To finish this
section, we point out that there have been a large body of work in
trying to write TAP equations for interaction that are not random. We refer the interested reader to the book
collection \cite{opper2001advanced} for detailed discussion of the TAP
equations in this context. Along this line of thoughts, the adaptive
TAP approach \cite{opper2001tractable} allows to learn the structure
of the matrix $J$, while the structure of the Onsager term can be
exactly computed depending on the structure of the correlations (see
e.g. \cite{parisi1995mean,shamir2000thouless,opper2001tractable}). 

Since these approaches do not lead to a natural time ordering, it is
worth mentioning that, very recently, these have been revisited
\cite{opper2015theory} in the light of the ``time indices" subtlety,
leading a much more satisfying theory of the dynamics of the TAP
equations on general matrices.

\newpage

\section{Clustering of networks and community detection}
\label{chap:clustering}
Clustering is one of the most commonly studied problems in machine
learning. The goal is to assign data points to $q$ clusters in a way
that the cluster is somehow representative of the properties of the
points inside it. About $15$ years ago, a large number of scientists
got interested in the research on so-called complex networks,
i.e. sets of nodes where some pairs are connected by edges, arising in
a wide range of settings. A natural question followed: How to cluster
networks? That is, how to assign nodes of a graph in groups such that
the largest possible part of the edges can be explain by a set of
affinities between the clusters (instead of between the nodes
themselves)? Typically, one considers a number of clusters much
smaller than the number of nodes. Part of the motivation for studying
this question came from studies of social networks which resulted in
the name for this problem: {\it community detection}.

\subsection{Biased overview of community detection}
\label{sec:community}
The literature on community detection is vast and there are a number
of reviews that make a great effort in putting the existing works in
chronological and merits perspective,
e.g. \cite{fortunato2010community}. Here we give a perspective biased
towards the approach of statistical physics of disordered systems.

In community detection we are given a graph $G(V,E)$ of $N$ nodes and
$M$ edges. We search to assign the nodes into $q$ groups, calling the
assignment $s_i\in \{1,\dots,q \}$, $i=1,\dots,N$. In the most
commonly considered assortative setting, the goal is that there are
many edges between nodes of the same group and few edges among the
groups. Traditionally, in statistics or machine learning, this problem
would be solved by a spectral relaxation. One takes a matrix
associated to the graph, its adjacency, its Laplacian or even
something else, and compute its largest (or smallest) eigenvalues and
associated eigenvectors. The elements of these eigenvectors would then
reveal in a {\it visible} way the community structure
\cite{donath1973lower,fiedler1973algebraic}, for more details see the
nice review \cite{von2007tutorial}.

An influential set of works on community detection is due to Mark
Newman and collaborators who introduced the so-called {\it modularity}
function
\cite{girvan2002community,clauset2004finding,newman2006modularity} 
\be
Q = \frac{1}{2M} \sum_{\langle i,j \rangle} \left(A_{ij} - \frac{d_i
    d_j}{2M }\right) \delta_{s_i,s_j} \, , \ee 
where the sum is over all distinct pairs of nodes, $d_i$ is the degree
of node $i$. Community detection is performed by finding an assignment
of nodes that maximizes this modularity. Optimizing the modularity is
a NP-hard problem. In practice, however, many efficient heuristic
algorithms were suggested for this task, see e.g. 
\cite{fortunato2010community}.

Modularity-based community detection has nice connections to
statistical mechanics and spin glasses. Reichardt and Bornholdt
\cite{reichardt2006statistical} pointed out that modularity
maximization can be seen as a case of finding a ground state of a
particular Potts spin glass. In order to assess the statistical
significance of community structure found by maximizing the
modularity, ref. \cite{reichardt2006statistical} estimated the maximum
value of modularity on a random graph, as the null model. Indeed, on
sparse random graphs the maximum value of modularity can be very large
even when no community structure whatsoever is present. To give an
example, results on random graph partitioning
\cite{zdeborova2010conjecture} teach us that a random 3-regular graph
can be divided into two groups of roughly equal size in such a way
that only about $11\%$ of edges are between the groups. This visually
looks like a very good community structure with high modularity. Yet
there were no communities in the process that created the random
graph. The need to compare the value of modularity to a null model is
not very satisfying. Ideally, one would like to have a method able to
tell that the underlying graph does not contain any communities if it
did not.

Let us assume that we believe there were $q$ communities respectively
taking a fraction~$n_a$ of nodes, $a=1,\dots,q$, $\sum_a n_a=1$, and
the probability that a given random node from community $a$ is connected to
a given random node from community $b$ is $c_{ab}/N$, where $N$ is the
notal number of nodes and $c_{ab}$ is a system size
independent $q\times q$ matrix. There will be roughly $c_{ab} n_a n_b N$ edges between groups $a$ and $b$, and
$c_{aa} n^2_a N/2$ edges inside group $a$. The above very natural assumption is
called the sparse {\it stochastic block model} (SBM)
\cite{white1976social,holland1983stochastic}, and is commonly
considered in this setting.  As Bayesian probability theory teaches
us, the best possible way of inferring the hidden communities is to
enumerate all assignments of nodes into communities that have roughly
the above sizes and respective number of edges between
groups. Subsequently one averages over all such assignments (factoring
away the global permutational symmetry between groups). This way one
ends up with probabilities for a given node to belong to a given
group. For a truly random graph these probabilities will not be
different from the fractions $n_a$.

From a computationally complexity point of view, exactly enumerating
assignments as above is even harder than finding optimal values of a
cost function such as the modularity. However, from a statistical
physics point of view evaluating the above averages is equivalent to
the computation of local magnetizations in the corresponding Potts
glass model. Notice, for instance, that planted coloring
\cite{krzakala2009hiding} is simply a special case of the stochastic
block model ($c_{aa}=0$, and $c_{ab}=c_{cd}$ for all $a\neq b$,
$c\neq d$).  We shall summarize in the rest of this section the
asymptotically exact analysis of the stochastic block
model \cite{decelle2011inference,decelle2011asymptotic}.

\subsection{Asymptotically exact analysis of the SBM}

We explained all the methodology needed for the asymptotic analysis of
the SBM on the example of the planted spin glass in section
\ref{planted_Ising}. In this section, we consider again the teacher-student
scenario where the teacher generates the observed graph from the
stochastic block model. 
We denote $\theta = \{n_a,c_{ab}\}$ the set of parameters of the
SBM. We remind that $n_a$ is the fraction of nodes taken by group
$a$, $a=1,\dots,q$, and $c_{ab}/N$ is the probability that a given node in group
$a$ connects to a given node in group $b$. Again we denote by
$\theta^*$ the actual values that the teacher used to
generate the graph $G$. The graph is represented by its adjacency
matrix $A_{ij}$, with $N$ nodes and $M$ edges. Further $s_i\in \{1,\dots,q\}$ are the Potts
spins denoting the assignments to the respective groups. 

The posterior likelihood of the SBM reads
\be
             P(\bs|G,\theta) = \frac{1}{Z(G,\theta)} \prod_{i=1}^N
             n_{s_i} \prod_{\langle i,j\rangle}  \left( c_{s_i s_j}
             \right)^{A_{ij}} \left( 1- \frac{c_{s_i s_j}}{N}
             \right)^{1-A_{ij}} \label{post_clust}
\ee
where the second product is over all distinct pairs, and we included a
constant $N^{-M}$ into the partition function. We need to evaluate the
maximum-mean-overlap (MMO) estimator (\ref{eq:MMO}) in order to find a configuration
that maximizes the number of nodes that are assigned into the same
group as the planted configuration. In a physics language, we would rather think of the corresponding
Potts model Hamiltonian
\be
 {\cal H} (G,\theta,\bs)= \sum_{i=1}^N
            \log{ n_{s_i}}  + \sum_{\langle i,j\rangle} \left[ A_{ij} \log{
              c_{s_i s_j}} + (1-A_{ij}) \log{ \left( 1- \frac{c_{s_i s_j}}{N}
             \right)} \right] \label{post_clust_Ham}
\ee
and the MMO estimator turns out to be computed from the local
magnetizations of this Potts model.

When the parameters $\theta^*$ are not known, we need to learn them
according to the lines of section \ref{sec:learn}. We write the
posterior distribution on the parameters $P(\theta|G)$ and notice that
it is concentrated around the maximum of the partition function
$Z(G,\theta)$. We then iteratively maximize this partition function by
imposing the Nishimori conditions \bea n_a &=& \frac{1}{N}
\sum_{i=1}^N {\mathbb E}(\delta_{s_i,a}) \, ,
     \label{EM_clust1}\\
    c_{ab} &=& \frac{1}{N n_a n_b} \sum_{(ij)\in E} {\mathbb
      E}(\delta_{a,s_i} \delta_{b,s_j}) \quad {\rm when} \quad a\neq b
    \, ,
    \\
    c_{aa} &=& \frac{2}{N n^2_a} \sum_{(ij)\in E} {\mathbb
      E}(\delta_{a,s_i} \delta_{a,s_j})   \, . \label{EM_clust3}
\eea
This learning algorithm was called expectation maximization learning in
\cite{decelle2011asymptotic}. The expectations were computed using
belief propagation or Gibbs sampling as will be discussed
subsequently. But even if the expectations were computed exactly, this
parameter learning algorithm depends strongly on the initial choice
for the parameters $\theta$. It is easily blocked in a local maximum
of the partition function. The size of the space of
initial conditions that need to be explored to find the global maximum
is independent of $N$, but in practice this is a problem, and
completely general models $c_{ab}$ are not so easy to learn. A~considerable improvement can be obtained when the learning is
initialized based on the result of spectral clustering. This procedure
was suggested in \cite{zhang2012comparative} and works nicely, in
particular in conjunction with the best known spectral clustering
techniques \cite{krzakala2013spectral} as will be discussed in sec.~\ref{sec:NB}

Two algorithms that are (in the Bayes-optimal setting) asymptotically
exact for computing the marginals and the expectations in
(\ref{EM_clust1}-\ref{EM_clust3}) were proposed in
\cite{decelle2011inference,decelle2011asymptotic}. One is the Gibbs
sampling (or Markov Chain Monte Carlo), another one is belief
propagation (BP). We present here the belief propagation algorithm, as
we have introduced it in sec.~\ref{sec:cavity}.  It can also be used
very directly to understand the behavior of the system and the phase
transitions in the thermodynamic limit. The BP algorithm for the SBM
was first written by \cite{hastings2006community} but in that work the
parameters were fixed ad-hoc, not in the Bayes optimal way, nor by
learning. Moreover, the algorithm was tested on a network of only $128$
nodes, which is too small to observe any phase transition.

Repeating the steps of sec.~\ref{sec:cavity}, we now derive BP. For
the posterior distribution (\ref{post_clust}) BP is written in terms of
the so-called messages $\psi_{s_i}^{i\to j}$ that are defined as
probabilities that node $i$ is assigned to group $s_i$ conditioned on
the absence of edge $(ij)$. Assuming asymptotic conditional
independence of the probabilities $\psi_{s_k}^{k\to i}$ for
$k\neq j,i$, the iterative equations read \be \psi_{s_i}^{i\to j} =
\frac{1}{Z^{i\to j}} \, n_{s_i} \prod_{k\neq i,j} \left[ \sum_{s_k}
  c^{A_{ik}}_{s_i s_k} \left(1-\frac{c_{s_i s_k}}{N}\right)^{1-A_{ik}}
  \psi_{s_k}^{k\to i} \right] \, , \label{BP_iter_exact} \ee where
$Z^{i\to j}$ is a normalization constant ensuring that
$\sum_{s=1}^q \psi_{s}^{i\to j}=1$.  We further notice that because of
the factor $1/N$, messages on non-edges $(ij) \notin E$ depend only
weakly on the target node and hence at the leading order the BP
equations simplify to \be \psi_{s_i}^{i\to j} = \frac{1}{Z^{i\to j}}
\, n_{s_i} e^{-h_{s_i}} \prod_{k\in \partial i\setminus j} \left[
  \sum_{s_k} c_{s_k s_i} \psi_{s_k}^{k\to i}\right] \,
, \label{BP_iter} \ee where we denote by $\partial i$ the set of
neighbors of $i$, and define an auxiliary external field as \be
h_{s_i} = \frac{1}{N}\sum_{k=1}^N \sum_{s_k}c_{s_k s_i}
\psi_{s_k}^{k}\, , \label{ex_field} \ee where $\psi_{s_i}^{i} $ is the
BP estimate of the marginal probability of node $i$ \be \psi_{s_i}^{i}
= \frac{1}{Z^{i}} \, n_{s_i} e^{-h_{s_i}} \prod_{j\in \partial i}
\left[ \sum_{s_j} c_{s_j s_i} \psi_{s_j}^{j\to i}
\right] \label{BP_marg} \, , \ee with $Z^i$ being again a
normalization. In order to find a fixed point of Eq.~(\ref{BP_iter})
we update the messages $\psi^{i\to j}$, recompute $\psi^j$, update the
field $h_{s_i}$ by adding the new contribution, subtracting the old
one, and repeat.

Once a fixed point of the BP equations is reached, its associated free
energy (log-likelihood) $f_{\rm BP} = -\log{Z(G,\theta)}/N$ reads
\be
    f_{\rm BP}(G,\theta) =  - \frac{1}{N} \sum_i \log{Z^i} + \frac{1}{N}\sum_{(i,j) \in E}  \log{Z^{ij}} - \frac{c}{2} \label{Bethe_fe}\, ,
\ee
where 
\bea
    Z^{ij} &=& \sum_{a < b} c_{ab} ( \psi^{i\to j}_a \psi_b^{j\to i}+ \psi^{i\to j}_b \psi_a^{j\to i}) + \sum_a c_{aa}  \psi^{i\to j}_a \psi_a^{j\to i} \quad {\rm for} \quad (i,j)\in E  \label{Z_ij} \\
    Z^i&=& \sum_{s_i} n_{s_i} e^{-h_{s_i}} \prod_{j\in \partial i}
    \sum_{s_j}  c_{s_j s_i} \psi_{s_j}^{k\to i}  \, .
\eea

In \cite{decelle2011inference,decelle2011asymptotic} the author argued
that the above equations can be used to analyze the asymptotic
performance of Bayes optimal inference in the SBM,
$\theta=\theta^*$. The analysis goes as follow: Let us iterate BP
(\ref{BP_iter}) from two different initializations:
\begin{itemize}
     \item Random initialization, for each $(ij)\in E$ we chose
       $\psi_{s}^{i\to j} = n_a+ \epsilon^{i\to j}_{s}$ where
       $\epsilon_s^{i\to j}$ is a small zero-sum perturbation, $\sum_s
       \epsilon_s^{i\to j} =0$. 
     \item Planted initialization, each message is initialized as
       $\psi_{s}^{i\to j} = \delta_{s,s_i^*}$, where $s_i^*$ is the
       actual assignment of node to groups (that is known only for the
       purpose of analysis). 
\end{itemize}
The free energy (\ref{Bethe_fe}) is computed for the two corresponding
fixed points. The fixed point with smaller free energy (larger
likelihood) corresponds asymptotically to the performance of the
Bayes-optimal inference. Here ``asymptotically'' is meant in the sense that the
difference between the Bayes-optimal and the BP marginals goes to zero
as $N\to \infty$. This is an important and nontrivial property,
because in general the analysis of Bayes optimal inference in such a
high-dimensional setting is very difficult (sharp-P hard). Note that
the asymptotic exactness of BP is so far only a conjecture. Rigorous
results showing the asymptotic exactness of BP are so far only
partial, either at relatively high average degrees
\cite{mossel2014belief}, on in related problems for very low average
degrees \cite{montanari2007counting}, or e.g. in ferromagnets
\cite{dembo2010ising}.

Another property of belief propagation that makes it very convenient
for the analysis of phase transitions and asymptotic properties is the
fact that BP, in a sense, ignores some part of the finite size
effects. Let us compare BP to Markov chain Monte Carlo methods. 
In the large size limit the performance and behaviour of BP and MCMC
are perfectly comparable, as demonstrated for the stochastic block
model in \cite{decelle2011asymptotic}. But for finite sizes there are
several remarkable differences.  
Consider a set of parameters where the two initializations of BP do not give the same
fixed point. Even if we iterate BP for infinite time we will have two
different fixed points. On the other hand, in Monte Carlo Gibbs
sampling for a system of size $N$, we will eventually find the
lower free energy state (just as for every ergodic Markov chain satisfying
detailed balance). The time needed is in general exponentially
large in $N$, but decreases as the free energy barrier gets smaller and therefore
locating the spinodal threshold precisely from finite system size
simulations is computationally more demanding compared to BP. Another manifestation of this property is
that in the paramagnetic phase on a graph with $N$ nodes and $M$ edges
the free energy given by BP is always the same number. Whereas the
exact (finite $N$) free energy typically has some fluctuations.

In the SBM, belief propagation provides an asymptotically exact analysis of
Bayes-optimal inference. In the case where the model or its
parameters $\theta$ are mismatching the ones that generated the graph,
this is not true in general. The most striking example here is the MAP
estimator, in other words the ground state of the corresponding Potts
model. Modularity maximization is a special case of this setting. The
ground state could also be estimated by BP if we introduced a
temperature-like parameter $1/\beta$ in the posterior and tuned this
temperature to zero. A related algorithm was discussed recently in the
context of modularity-based community detection in
\cite{zhang2014scalable}. However, as we decrease this temperature we
will typically encounter a point $\beta_{\rm SG}>1$ above which the
replica symmetric approximation fails and the corresponding belief
propagation marginals are not any longer asymptotically exact. When
this phase transition at $\beta_{\rm SG}$ is continuous this will
manifest itself as non-convergence of the BP algorithm. Note that
depending on the set of parameters, this spin glass phase can be
correlated to the orignal assignment (mixed phase) or not (pure spin
glass phase). This behavior is qualitatively the same as the one seen in the phase digram
presented in Fig.~\ref{fig:phase_planted}. In the above sense it is hence easier to compute the
marginals at the Bayes-optimal value of the parameters (that may need
to be learned as described above) than to compute the ground state (or
maximize the modularity) that typically is glassy.

\subsection{Examples of phase transitions in the SBM}

The asymptotic analysis of the stochastic block model as described
above leads to the discovery of phase transitions. The most striking
picture arises in the case when the parameters of the model are such
that the average degree is the same in every group. In a sense this is
the hardest case for inference, because if groups differ in their
average degree, then a simple histogram of degrees gives some
information about the correct assignment into groups. 

Mathematically,
the same average degree condition is expressed as 
\be
      \sum_{b} c_{ab} n_b = c  \quad \quad \forall b=1,\dots,q \, ,\label{av_degree}
\ee
where $c$ is the average degree.  It is straightforward to see that
when (\ref{av_degree}) holds, then  
\be
    \psi_s^{i\to j} = n_s \,
\ee
is always a (so-called {\it uniform}) fixed point (in general not the only one) of belief
propagation (\ref{BP_iter}). 

\begin{figure}[!ht]
\begin{center}
\includegraphics[scale = 0.56]{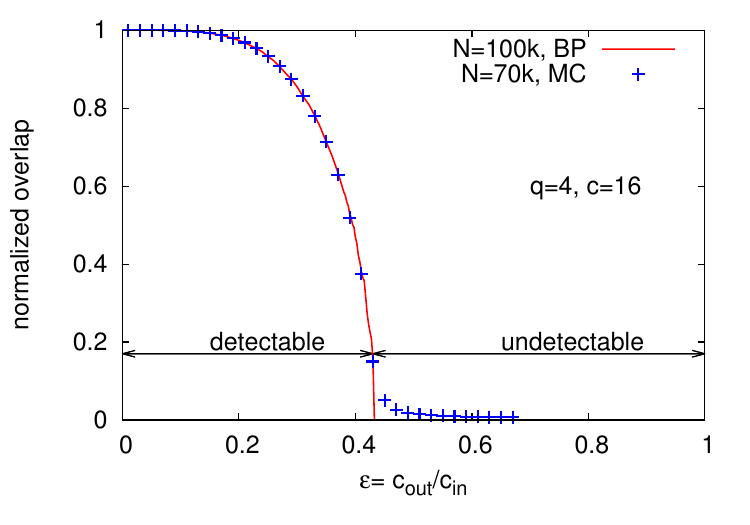}
\includegraphics[scale = 0.56]{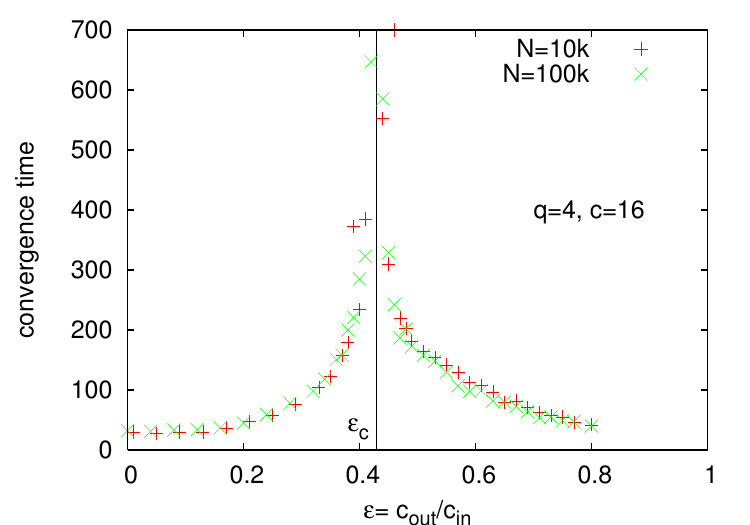}
\end{center}
  \caption{ \label{fig:transition_q4} Left: Normalized overlap between
    the MMO estimator and the original assignment to groups. We
    compare the BP result to Monte Carlo Gibbs sampling. Both agree
    well, but we see the absence of finite size corrections in BP in the
    undetectable phase.  Right: The BP convergence time, illustrating
    the critical slowing down around the phase transition. Figures are
  both taken from \cite{decelle2011asymptotic}.}
\end{figure}

Let us now return to the concept of quiet planting explained for the planted spin
glass in sec.~\ref{quiet_planting} and note that all we needed for quiet
planting to work was precisely the existence of such a uniform fixed
point. It is hence straightforward that the phase transitions described
in sections \ref{sec:second} and \ref{sec:Potts} also exist in Bayes-optimal
inference in the stochastic block model. Let us give the example of
assortative community structure with $q=4$ equally sized groups,
$n_a=1/q$ for all $a=1,\dots,q$, average degree $c=16$, and affinity parameters such that
$c_{aa}=c_{\rm in}$ and $c_{ab}=c_{\rm out}$ for all $a\neq b$. We
define a noise-like parameter $\epsilon=c_{\rm out}/c_{\rm in}$. When
$\epsilon=0$ the model generates four entirely disconnected groups, and when $\epsilon=1$
the generation process corresponds to an Erd\H{o}s-R\'enyi graph with no
information on the community structure. Without having any experience
with phase transition and quiet planting, one might anticipate that
whenever $\epsilon<1$ the Bayes optimal inference will give a better
overlap than a random assignment of nodes into groups. This
expectation is wrong and the phase transition that happens instead is illustrated in
Fig.~\ref{fig:transition_q4}.

For other values of parameters we can observe the triplet of
discontinuous phase transition discussed in sec.~\ref{sec:Potts}. Actually, the planted coloring example, that was presented in Fig.~\ref{fig:planted_col}, is a case of the stochastic block model with
$q=5$ equally sized groups and extremely dis-assortative structure
$c_{\rm in}=0$. 

For the general case of the sparse SBM, the order of the phase transition,
and the locations of the dynamical $c_d$ and detectability $c_c$ phase transitions, have to
be analyzed case by case mostly numerically. However, the spinodal
transition has a neat analytical expression in the case of the same
average degree in every group (\ref{av_degree}). This expression is
obtained as in sec.~\ref{sec:random} by analyzing the
local stability of the uniform fixed point. Belief propagation (\ref{BP_iter})
linearized around this fixed point reads 
\be
     \epsilon_a^{i\to j} = \sum_{b=1}^q \sum_{k\in \partial i
       \setminus j} T_{ab} \,  \epsilon_b^{k\to i}\, ,
\ee 
where the $q\times q$ matrix $T_{ab}$ is defined as 
\be
       T^{ab}  =  n_a\left(\frac{c_{a b}}{c} - 1\right)\, . \label{matrix_T}
\ee

The phase transition where the paramagnetic fixed point ceases to be
locally stable happens when 
\be
            c \lambda_T^2= 1 \, , \label{spinodal}
\ee
where $\lambda_T$ is the largest eigenvalue of the matrix $T$. 

In case of equally sized groups in the {\it assortative-diagonal} case where
$c_{aa}=c_{\rm in}$ and $c_{ab}=c_{\rm out}$ for $a\neq b$,
the largest eigenvalue of the $T^{ab}$ matrix (\ref{matrix_T}) is
$\lambda_T=(c_{\rm in} - c_{\rm out})/(cq)$ from which
(\ref{spinodal}) implies for the spinodal phase transition \cite{decelle2011inference}
\be
           |c_{\rm in} - c_{\rm out}| = q \sqrt{c} \, . 
\ee

When the difference between $c_{\rm in}$ and $c_{\rm out}$ is larger,
BP detects communities and gives asymptotically the Bayes-optimal
overlap. When the difference is smaller BP fails. Depending on the
parameters and the order of the phase transition the detectability
phase transition itself can either be given by eq.~(\ref{spinodal}) or
happen at strictly smaller $|c_{\rm in}-c_{\rm out}|$, giving rise to the hard
phase discussed in sec.~\ref{sec:hard}.  

For the planted graph coloring (i.e. the extremely
dis-assortative SBM, $c_{\rm in}=0$), the detectability phase
transition is of second order for $q\le 3$ and of first order for
$q\ge 4$. The spinodal transition happens at $c_s =  (q-1)^2$.  In
the large $q$ limit the dynamical and detectability phase transitions
can be deduced from 
\cite{ZdeborovaKrzakala07} while keeping in mind the concept of quiet
planting 
\bea
       c_d   &=&   q ( \log{q} + \log{\log{q}}) + O(1)\, ,\\
       c_c   &=&   2 q \log{q}   - \log{q} - 2\log{2} + o(1)    \, .
\eea

For the assortative case $c_{\rm in}>c_{\rm out}$ the
phase transition is of second order for $q\le 4$ and of first order
(although a very weak one for small $q$) for $q\ge 5$. The limit of dense
graphs (average probability for an edge to exist is $p$, $p_{\rm
  in/out}$ are probabilities of connection within/between groups), and
large number of groups $q\to \infty$ was analyzed in \cite{lesieur2015mmse} and is
related to critical temperatures in the dense Potts glass \cite{gross1985mean,caltagirone2012dynamical}. The
regime in which both the dynamical and detectability phase transitions appear is 
\be
      N (p_{\rm in} - p_{\rm out})^2 = \gamma(q)  p(1-p)     \, ,
\ee
with the dynamical transitions at $\gamma_d(q)= 2q \log{q}[1+o(1)]$,
the detectability transition at $\gamma_c(q)= 4q
\log{q}[1+o(1)]$, and the spinodal threshold at $\gamma_s(q)= q^2$. In
this limit we see that the hard phase between $\gamma_s(q)$ and
$\gamma_c(q)$ is very wide (compared to the overall scale). 

The prediction and location of these phase transitions in the
stochastic block model from \cite{decelle2011inference,decelle2011asymptotic} was followed by a remarkable mathematical
development where the location of the threshold was made rigorous for
$q=2$ \cite{mossel2012stochastic,massoulie2014community,mossel2013proof}. The detectable side was proven
for generic $q$ in \cite{bordenave2015non}.
 
To briefly comment on the case where the average degree of every group
is not the same: BP is still conjectured to be
asymptotically exact, but the phase transitions get
smeared. As recently studied in \cite{zhang2016community},  the second
order phase transitions disappear, and the first order
phase transitions shrink into a multi-critical point beyond which they
also disappear. 
The situation is similar to the semi-supervised clustering, where the
assignment of a fraction of nodes is known \cite{zhang2014phase}.  
From a physics point of view it is
relevant to note that this problem is an inference interpretation of
the particle pinning that was recently used quite extensively for
studies of structural glasses \cite{cammarota2012ideal}.

\subsection{Spectral redemption for clustering sparse networks}
\label{sec:NB}

\begin{figure}[!ht]
\begin{center}
\includegraphics[scale = 0.38]{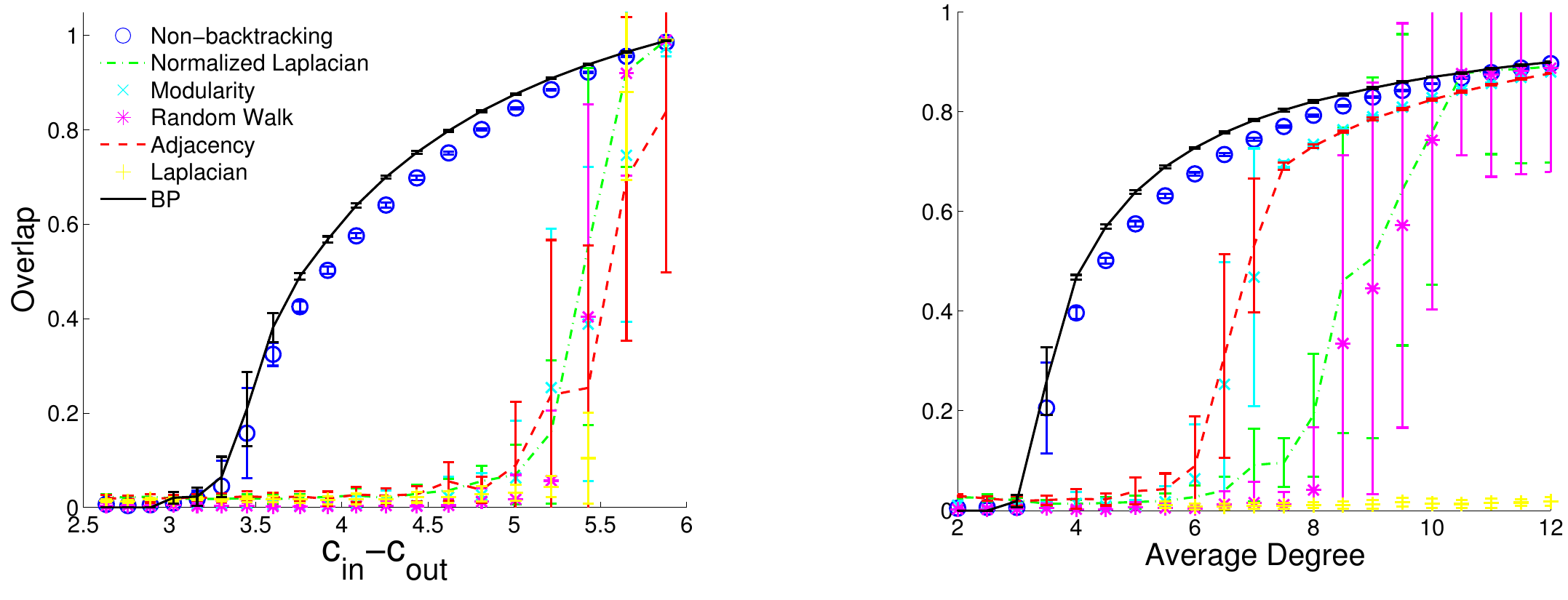}
\end{center}
  \caption{ \label{NB_performance}
The accuracy of spectral algorithms based on different linear
  operators, and of belief propagation, for two groups of equal size.
  On the left, we vary $c_{\rm in}-c_{\rm out}$ while fixing the average degree
  $c=3$; the detectability transition given
  by occurs at $c_{\rm in}-c_{\rm out} = 2 \sqrt{3} \approx
  3.46$.  On the right, we set $c_{\rm out}/c_{\rm in}=0.3$ and vary $c$; the
  detectability transition is at $c \approx 3.45$.  Each point is
  averaged over $20$ instances with $N=10^5$.  The spectral algorithm
  based on the non-backtracking matrix $B$ achieves an accuracy close
  to that of BP, and both remain large all the way down to the
  transition.  Standard spectral algorithms 
  based on the adjacency matrix, modularity matrix, Laplacian, normalized Laplacian, and the
  random walk matrix all fail well above the transition, giving a
  regime where they do no better than chance. Figure is
  taken from \cite{krzakala2013spectral}.}
\end{figure}

Any reader that did no skip section \ref{sec:NB_planted} is now clearly anticipating
the use of the non-backtracking matrix for spectral clustering of
networks. Indeed, in
\cite{krzakala2013spectral} the following spectral
clustering method was proposed
\begin{itemize}
     \item Construct the non-backtracking matrix 
      \be
             B_{i\to j, k\to l} = \delta_{il} (1-\delta_{jk}) \, .\label{NB_matrix_clust}          
      \ee 
     and compute its largest (in module) eigenvalues until they start
     to have a non-zero imaginary part. Denote $k$ the number of those
     large real eigenvalues. 
     \item Consider the $k$ corresponding $2M$-dimensional eigenvectors $u_{i\to j}$
       and create a set of $N$-dimensional vectors by resuming
       $u_i=\sum_{k\in \partial i} u_{k\to i}$. Then view the $k$
       vectors $u_i$ as $N$ points in $k$ dimensional space and cluster those points into $k$ groups
       using some off-the-shell clustering algorithm such as
       $k$-means. 
\end{itemize}
Note in particular that $k$ here is the estimator for the number of
groups. Up to this point we were assuming that the number of groups
was known. The performance of this non-backtracking based spectral
algorithms compared to BP, and five other spectral algorithms on clustering
of sparse graphs, is illustrated in Fig.~\ref{NB_performance}. All these
spectral algorithms look for the largest (or smallest) eigenvalue of
the associated matrix and then assign nodes into groups based on the
sign of the corresponding eigenvector. The five matrices that are most
commonly used for clustering are the adjacency matrix $A$, the
Laplacian $L=D-A$ (where $D$ is a diagonal matrix with the node
degrees $d_i$ on the diagonal), normalized Laplacian $\tilde
L=D^{-1/2} L D^{-1/2}$, random
walk matrix $W_{ij}=A_{ij}/d_i$, and the modularity
matrix $Q_{ij}=A_{ij}-d_i d_j /(2M)$. The spectra of all these
matrices on sparse graphs is affected by the existence of large
eigenvalues associated to localized eigenvactors. This fact
deteriorated their performance with respect to both BP and the
non-backtracking based algorithm. 

Ref. \cite{krzakala2013spectral} analyzed the performance of the
above algorithm on networks generated from the stochastic block model
with equal average degree in every group. A large part of results that
were not rigorous in \cite{krzakala2013spectral} were very remarkably made rigorous
in \cite{bordenave2015non}.  

On a random graph taken from the configurational model,
i.e. at random with a given degree distribution $P(d)$, in the large
size limit, the spectrum of the
non-backtracking matrix is (except one eigenvalue) included inside a circle of radius
$\sqrt{\tilde c}$. Here $\tilde c$ is the average excess degree of
the graph computed as 
\be
     \tilde c = \frac{ \sum_d d^2 P(d) }{ \sum_d d P(d)   } -1 \, .
\ee
The one eigenvalue that is not inside this circle is real and has
value $\tilde c$.  This random graph spectrum should be contrasted
with the spectrum of the adjacency matrix (and other related ones)
which for graphs with constant average and unbounded maximum degree is not even bounded in the thermodynamic limit. 

\begin{figure}[!ht]
\begin{center}
\includegraphics[width=0.5\columnwidth]{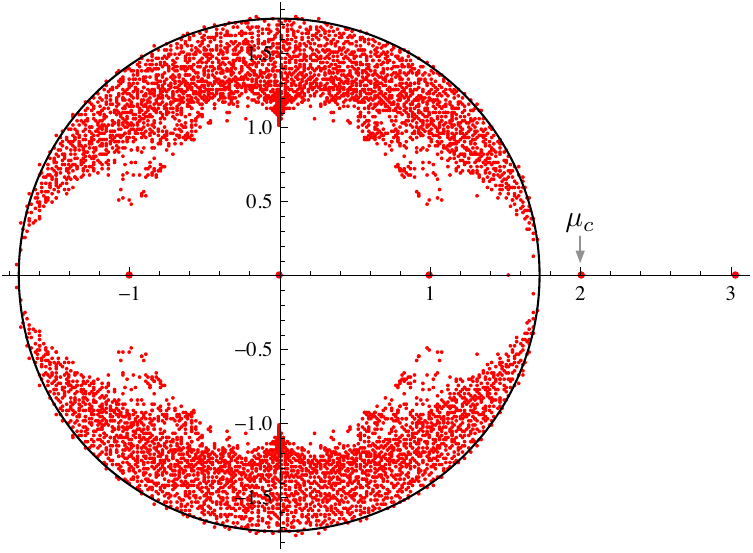} 
\end{center}
\caption{The spectrum of the non-backtracking matrix $B$ for a network generated 
by the block model with $N=4000$,
 $c_{\rm in} = 5$, and $c_{\rm out} = 1$. The leading eigenvalue is at
  $c = 3$, the second eigenvalue is close to $\mu_c=(c_{\rm in}-c_{\rm out})/2 = 2$,
  and the bulk of the spectrum is confined to the disk of radius
  $\sqrt{c} = \sqrt{3}$.  
  The spectral algorithm that labels vertices according to 
  the sign of $B$'s second eigenvector (summed over the incoming edges at each vertex) 
  labels the majority of vertices correctly. Figure is
  taken from \cite{krzakala2013spectral}.
\label{fig:complexplot}}
\end{figure}

Now, consider a graph generated by the SBM with equal average degree for
every group. Denote $\lambda_T$ an eigenvalue of the matrix
$T$, eq. (\ref{matrix_T}), then as long as
$\lambda_T > 1/\sqrt{\tilde c}$ the non-backtracking  matrix will have purely real
eigenvalues at positions $c\lambda_T$. This way we obtain up to $q-1$
additional eigenvalues on the real axes out of the circle of radius
$\sqrt{\tilde c}$. Note that in the  {\it assortative-diagonal} case the matrix
$T$ has $q-1$ degenerate eigenvalues $\lambda_T=(c_{\rm in}-c_{\rm
  out})/(qc)$, and therefore the non-backtracking matrix has a
$(q-1)$-times degenerate eigenvalue at $(c_{\rm in}-c_{\rm
  out})/q$ which merges into the bulk below the spinodal phase
transition. The spectrum of the non-backtracking operator is
illustrated in Fig.~\ref{fig:complexplot}. 

Consequently, when the average degree of every group is the same,
condition (\ref{av_degree}), the
phase transition of the spectral algorithm coincides with the spinodal
phase transition of the belief propagation. In other words as long the
average degree is larger than given by eq.~\ref{spinodal} the eigenvectors
corresponding to the eigenvalues on the real axes out of the circle
contain information about he ground-truth communities. 

We need to stress that the agreement between the spectral and belief propagation thresholds arises only
for the case of equal average degree in every group. 
When the condition  (\ref{av_degree}) is not satisfied there is
information about the groups on the degree and belief propagation
explores this information in an optimal way. What concerns the phase
transitions, they tend to be smeared (second order phase transitions
disappear, and first order ones become weaker) \cite{zhang2016community}. The
spectral methods, on the other hand, always have a threshold
where  the informative eigenvalues merge back into the bulk. At
this spectral phase transition, when (\ref{av_degree}) does not hold, BP does not witness any particular change in performance.

Let us compare properties of belief propagation and the non-backtracking (NB)
based spectral clustering. The positive point for BP is that it gives
a slightly better overlap. The positive points for NB-spectral clustering,
that are shared with other spectral methods, e.g. all those tested in Fig.~\ref{NB_performance},
are that it does not need to get any information about the model
parameters nor the number of groups. It works in time linear in
the number of groups, as opposed to quadratic for BP.
It is not expected to be affected by the presence of small loops which usually cause the
failure of the Bethe approximation. 
An inconvenience that is shared by all the considered spectral methods
is that the performance can deteriorate considerably upon small
adversarial changes to the graph (such as planting a small dense
subgraph). Approaches based on some global optimization, such as BP,
are expected to be robust towards small local changes. 

Another inconvenient property of the non-backtracking matrix is that
it can be rather large. However, this is no obstacle at all because
the spectrum (different from $\pm 1$) of $B$ can be computed from the spectrum of the following
$2N \times 2N$ matrix $B'$ \cite{krzakala2013spectral}
\be
\label{eq:2nby2n}
B' = \begin{pmatrix}
0 & D-\mathds{1} \\
-\mathds{1} & A
\end{pmatrix} \, ,
\ee
where $A$ is the adjacency matrix and $D$ is the diagonal matrix with
$D_{ii}=d_i$ being the degree of node $i$. 

The matrix $B'$ is smaller but
still non-symmetric. The power method for computing the largest
eigenvalues works much better for symmetric matrices. 
In \cite{saade2014spectral}, it was suggested to use the so-called
Bethe Hessian matrix defined as 
\be
   {\rm BH}(r) =(r^2-1) \mathds{1} -rA + D \, ,
\ee
with $r =\sqrt{\tilde c}$ (or more general the squared root of the leading eigenvalue of
(\ref{eq:2nby2n}) which is usually not problematic to compute) as a basis for spectral clustering. In particular, to take
eigenvectors of all of its negative eigenvalues. Ref. \cite{saade2014spectral} showed that this has
all the advantages of the non-backtracking matrix while being $N\times
N$ and symmetric. The name Bethe Hessian is motivated by the relation
between the Hessian of the Bethe free energy and this matrix \cite{mooij2004validity,ricci2012bethe}, which is
particularly apparent when generalized to the weighted case.

\subsection{More on the non-backtracking operator}

The non-backtracking operator as defined above is relatively well
known in mathematics, often called the Hashimoto directed edges
adjacency matrix \cite{hashimoto1989zeta}. It is
used in a number of works as a tool in proofs, see
e.g. \cite{bass1992ihara,alon2007non,sodin2007random,friedman2008proof,karger2011iterative,pfister2013relevance}. However,
using it as a basis for a spectral algorithm was first suggested in
\cite{krzakala2013spectral}. The idea of using the linearization of BP
was inspired by the earlier work of \cite{coja2009spectral}.  
Since then, a considerable number of papers on promising algorithmic applications of the
non-backtracking matrix on sparse networks started appearing, see~\cite{martin2014localization,zhang2014non,morone2015influence}. 
Here, we briefly discuss selected contributions in this direction.

Looking at Fig.~\ref{fig:complexplot}, we observe a very interesting
structure in the bulk spectral density of the non-backtracking
operator. Motivated by this, ref. \cite{saade2014spectraldensity} applied the cavity method designed to
compute spectral densities of non-Hermitian random matrices \cite{rogers2009cavity} to
study the non-backtracking spectral density. A phase
transition in the spectral density was observed to happen on the boundary of the circle of
radius $\sqrt{\tilde c}$. Such a phase transition is absent in the spectral
density of traditional matrices such as the adjacency, Laplacian,
random walks etc. This is no proof since we are only concerned with
spectral density not with sub-dominantly numerous eigenvalues, but from
a physics point of view a phase transition is a strong indication that the
properties of the non-backtracking matrix on a random graph are
fundamentally different from those of traditional matrices. Ref.  \cite{saade2014spectraldensity}  also
observed that the eigenvalues that are close to the center of the
circle are not visible in the asymptotic calculation of the spectral
density. Numerical investigations revealed that eigenvectors
corresponding to these eigenvalues are localized. More detailed understanding
of this localization and its physical consequences is an interesting direction
of future work. 

In \cite{angelini2015spectral}, the non-backtracking spectral clustering method was generalized to clustering of sparse
hypergraphs. These are graphs where instead of connected pairs of nodes,
one connects $k$-uples of nodes. Interestingly, using the
non-backtracking based spectral method detection can be done also
e.g. in planted constraint satisfaction problems without even knowing
the form of the corresponding constraints. Moreover, for the same
reasons as in clustering, the method works
down to previously computed spinodal thresholds \cite{zdeborova2011quiet}. It was also observed that the
overlap (performance) of the spectral method is always a continuous
function of the noise parameter, even if the Bayes optimal belief
propagation has a strongly discontinuous transition. This further
supports our previous suggestion that the spectral
methods are extremely useful to point towards the right direction, but
message-passing based methods should then be used to improve the
precision. Note that the hypergraph clustering can also be formulated
as tensor factorization that became recently popular in connection
with interdependent networks or networks with different kinds of edges
that are referred to as multiplexes in the literature on complex networks.

Another exciting application of the non-backtracking matrix, or the
Bethe Hessian, is the matrix completion problem
\cite{saade2015matrix}. In matrix completion one aims to estimate a
low-rank matrix from a small random sample of its entries. This
problem became popular in connection to collaborative filtering and
related recommendation systems. Of course, the fewer entries are
observed, the harder the problem is.  Ref. \cite{saade2015matrix}
shows that with the use of the non-backtracking matrix the number of
entries from which the low-rank structure is detectable is smaller
than with other known algorithms. The way the matrix completion
problem is approached is by realizing that a Hopfield model with a
Hebb learning rule can be viewed as a rank $r$ structure that is able
to recover the $r$ patterns as distinct thermodynamic phases
\cite{hopfield1982neural}. One views the observed entries of the
low-rank matrix to complete as a sparse Hopfield model and estimate
the patterns from the leading eigenvectors of the corresponding Bethe
Hessian. In a second stage, we use this as a initialization for more
standard matrix completion algorithms in the spirit of
\cite{keshavan2010matrix}, and achieve overall better performance.

Another problem where the leading eigenvalue of the non-backtracking
matrix is relevant is percolation on networks
\cite{karrer2014percolation}. In percolation, we start with a graph $G$
and remove each of its edges with probability $1-p$. Then we look at
the largest component of the remaining graph and ask whether it covers an extensive fraction of all nodes.  The value of $p_c$
below which the giant component does not exist anymore with high
probability is called the percolation threshold. Percolation
thresholds were computed for many geometries including random
graphs. A remarkable result concerns any sequence of dense graphs for
which it was proven that the threshold is given by the inverse of the
leading eigenvalue of the adjacency matrix
\cite{bollobas2010percolation}. This result does not extend to sparse
graphs, where a counter-example is given by random regular graphs
in which the leading eigenvalue is the degree whereas the percolation
threshold is the inverse of the degree minus one. 

Based on a message-passing algorithm for percolation,
\cite{karrer2014percolation} argues that on a large class of sparse
locally tree-like graphs the percolation threshold is given by the
inverse of the leading eigenvalue of the non-backtracking
matrix. Specifying for what exact class of graphs this result holds is
still an open problem. Ref.  \cite{karrer2014percolation} also proved
that the non-backtracking matrix provides a lower bound on the
percolation threshold for a generic graph (with the exception of
trees). An analogous result was obtained independently by
\cite{hamilton2014tight}. Due to these results, the non-backtracking
matrix was used to study percolation on real networks
\cite{rogers2015assessing,morone2015influence,radicchi2015predicting}.


\subsection{The deceptiveness of the variational Bayes method}
\label{sec:MF}

Both in physics and in computer science, a very common and popular method for
approximate Bayesian inference is the so-called naive mean field
method, often called variational Bayesian inference \cite{beal2003variational,opper2001advanced}. For clustering of
sparse networks generated by the stochastic block model this method
was studied e.g. in \cite{hofman2008bayesian,celisse2012consistency}.   
Away from the detectability threshold the variational Bayes method
gives very sensible results. Deep in the undetectable
phase (very small $|c_{\rm in}-c_{\rm out}|$) the predicted
magnetizations are close to zero, reflecting the undetectability. Deep in the detectable phase the variational Bayesian
method is able to estimate the correct group assignment, and is
statistically consistent as proven in \cite{celisse2012consistency}. 
The performance of variational Bayes in the neighborhood of the
detectability transition was studied in \cite{zhang2012comparative}. 
The conclusions of the comparison to belief propagation are quite interesting. 
It is not surprising that BP is more precise and converges faster on sparse graphs.  
What is surprising is that the
naive mean-field inference fails in a very deceptive way: the
marginals it predicts are very biased towards a configuration that is
not correlated with the planted (ground truth) configuration. Ref.~\cite{zhang2012comparative} quantifies this by
introducing the so-called illusive overlap and comparing it to the actual
one. This means that it can be rather dangerous to use naive mean-field
for testing statistical significance of the results it provides and
this could have profound consequences beyond the problem of clustering
networks. Especially when trying to use the variational Bayes method in
the vicinity of the corresponding detectable threshold, i.e. in
regimes where the number of samples is barely sufficient for estimation.
Monte Carlo methods do not have this problem, but are in
general slower.

\subsection{Real networks are not generated by the SBM}
\label{real_nets}

Our study of clustering of networks is asymptotically exact for the
case when the networks were generated by the stochastic block
model. In view of actual data sets and realistic applications this is
never the case.

\begin{figure}[!ht]
\begin{center}
\includegraphics[width=1.0\columnwidth]{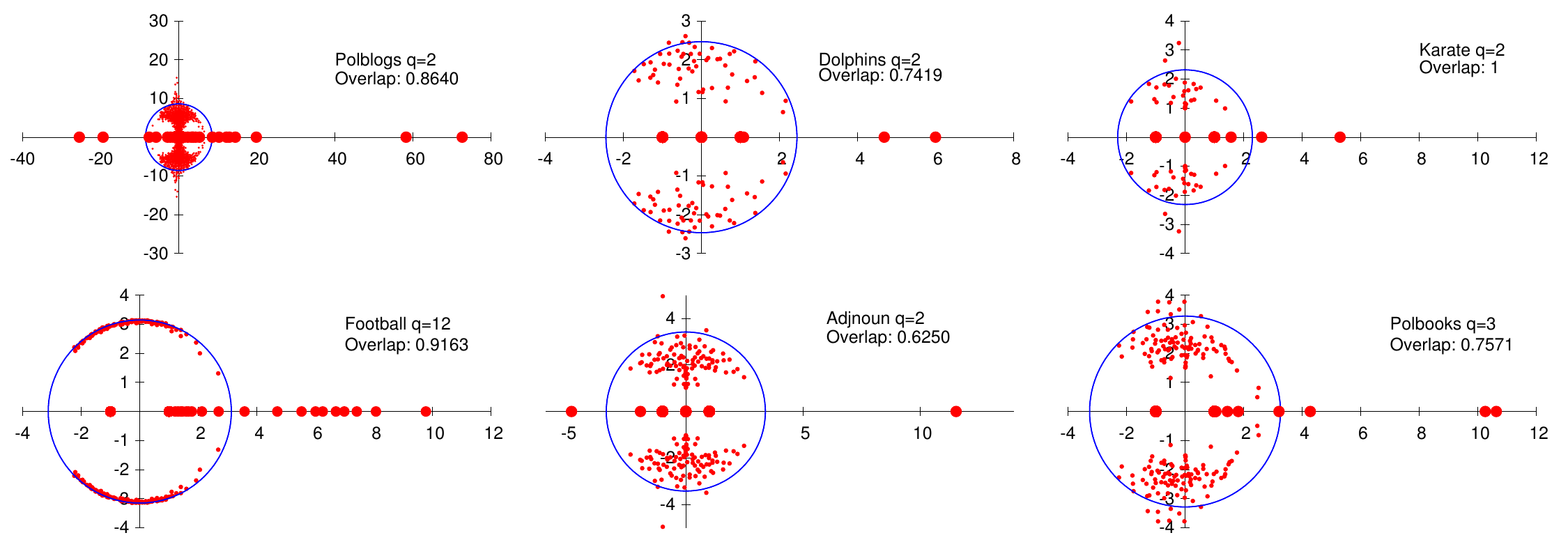}
\end{center}
\caption{Spectrum of the non-backtracking matrix in the complex plane
  for some real networks commonly used as benchmarks for community
  detection, taken
  from~\cite{adamic2005political,zachary1977information,newman2006finding,girvan2002community,lusseau2003bottlenose,krebs2004books}.
  The radius of the circle is the square root of the largest
  eigenvalue, which is a heuristic estimate of the bulk of the
  spectrum. The overlap is computed using the signs of the second
  eigenvector for the networks with two communities, and using
  $k$-means for those with three and more communities. The
  non-backtracking operator detects communities in all these networks,
  with an overlap comparable to the performance of other spectral
  methods.  As in the case of synthetic networks generated by the
  stochastic block model, the number of real eigenvalues outside the
  bulk appears to be a good indicator of the number $q$ of
  communities. Figure taken from
  \cite{krzakala2013spectral}. \label{fig:real_spectrum}}
\end{figure}

One important discrepancy between the stochastic block model and real
networks is that the latter often have communities with a broad degree
distribution, whereas the degree distribution of every group generated by
the SBM is Poissonian. A variant of the stochastic block model that
takes the degree distribution into account is the so-called degree
corrected stochastic block model \cite{karrer2011stochastic}.  An
asymptotically exact belief propagation can also easily be written for
this model, see \cite{yan2014model}. In Ref.~\cite{yan2014model} it
was also shown that the corresponding Bethe free energy is a great tool
for model selection between the canonical and degree corrected
stochastic block models, and that classical model selection criteria
do not apply to this problem because of the limited information per
degree of freedom.  Interestingly, BP for the degree-corrected SBM also
leads to the same non-backtracking matrix upon linearization.

There are of course many more discrepancies between real networks and
those generated by the (degree corrected) SBM. However, when observing
the spectrum of the non-backtracking operator on some of the commonly
used benchmarks for community detection, we conclude that the
properties derived for the SBM basically stay conserved also in these
examples, see Fig.~\ref{fig:real_spectrum}. To conclude this section,
let us quote the British mathematician and Bayesian statistician
George E. P. Box (1919-2013): ``Essentially, all models are wrong, but
some are useful".

\newpage

\section{Linear estimation and compressed sensing}
\label{chap:CS}

The inference problem addressed in this section is the following. We
observe a $M$ dimensional vector $y_\mu$, $\mu=1,\dots,M$ that was
created component-wise depending on a linear projection of an unknown $N$
dimensional vector $x_i$, $i=1,\dots,N$ by a known matrix~$F_{\mu i}$
\be
                y_\mu  = f_{\rm out}\left( \sum_{i=1}^N F_{\mu i }
                  x_i\right) \, , \label{y_mu}
\ee
where $f_{\rm out}$ is an output function that includes
noise. Examples are the additive white Gaussian noise (AWGN) there the
output is given by $f_{\rm
  out}(x)=x+w$ with $w$ being random Gaussian variables, or the linear
threshold output where $f_{\rm out}(x)=\sign(x-\kappa)$, with $\kappa$
being a threshold value.  The goal in linear estimation is to infer $\bx$ from the knowledge of $\by$, $F$ and
$f_{\rm out}$. An alternative representation of the output function
$f_{\rm out}$ is to denote $z_\mu =\sum_{i=1}^N F_{\mu i }  x_i $ and
talk about the likelihood of observing a given vector $\by$ given~$\bz$
\be
           P(\by| \bz) = \prod_{\mu=1}^M P_{\rm out}(y_\mu|z_\mu)\, .
\ee

\subsection{Some examples of linear estimation}
\label{CS_ex}
To put the immense number of important applications of the above setting into perspective, let us give an illustrative example from medical 
diagnostics. Consider we have data about $M$ cancer-patients that took
a given treatment. Then $y_\mu$ is a variable rating the effectiveness
of the treatment for a patient $\mu$. Our goal is to predict
the outcome of the treatment for a new patient that did not undergo treatment
yet. To do that, we collect medical and personal data about each of the $M$
patients, their medical history, family situation, but also for
instance Facebook feeds, where they worked, what sport they did in
their life, what education they have, how they rate their relationship
with their wife, and all kinds of other information that is on first
sight seems totally unrelated to the eventual outcome of the
treatment. This information is collected in the vector $\vec
F_\mu$. We aim to find a set of coefficients $x_i$ such that the
observations $y_\mu$ can be well explained by $z_\mu = \sum_{i=1}^N F_{\mu
  i }  x_i $ for some $P_{\rm out}(y_\mu|z_\mu)$. 

 In the way we set the problem, it is reasonable to think
that most patient-attributes will have no effect on the outcome of the
treatment, i.e. for many $i$'s the $x_i$ will be close to zero. In a
method, called LASSO (which stands for {\it least absolute shrinkage and
  selection operator})~\cite{tibshirani1996regression,hastie2015statistical}, that is used everyday for similar types of problems
(in banking or insurance industry for instance) the coefficients $x_i$
are such that 
\be
        {\rm LASSO:} \quad \quad       \min_{\bx} \left[ \sum_{\mu=1}^M \left(y_\mu -  \sum_{i=1}^N F_{\mu
  i }  x_i \right)^2 + \lambda \sum_{i=1}^N |x_i| \right]  \, , 
\ee   
where $\lambda$ is a regularization parameter that controls the
trade-off between a good fit and the sparsity of the vector $\bx$. The advantage of LASSO is that it is a convex
problem that can be solved efficiently and with strong guarantees
using linear programming. LASSO and related linear regression problems are
well studied in statistics and machine learning, instead of
reviewing the literature we refer the reader to the excellent textbook
\cite{hastie2005elements}.

LASSO can be framed in the Bayesian probabilistic setting by
considering \be P_{\rm out}(y_\mu|z_\mu) = \frac{1}{\sqrt{2 \pi
    \Delta}} \e^{-\frac{ (y_\mu - \sum_{i=1}^N F_{\mu i } x_i
    )^2}{2\Delta}} \label{gauss_out} \ee and the prior of $x_i$ being the
Laplace distribution \be P^{\ell_1}_X(x_i)= e^{-\lambda |x_i|}/2\, .
\ee Another sparsity inducing prior is a combination of strict zeros
and fraction $\rho$ of non-zeros \be P_{X}(x_i) = \rho\phi(x_i) +
(1-\rho) \delta(x_i) \, , \label{G-B} \ee where $\phi(x_i)$ is the
distribution of non-zero elements. Linear programming does not work
for priors such as (\ref{G-B}), but the statistical-physics framework is
more general in this aspect.

In many applications, the output variables $y_\mu$ are binary
(interpreted in the above example as the
treatment worked or did not), then the most commonly considered output
distribution is the
logistic function. The conditional dependence then is
\be
      P^{\rm logistic}_{\rm out}(y_\mu=1|z_\mu)  = \frac{1}{1+e^{-z_\mu}} \, .
\ee

The example of this linear estimation setting that is the best known
in statistical physics is the perceptron
\cite{rosenblatt1958perceptron}, as introduced briefly in sec.~\ref{teacher_student}. In the perceptron setting, components
of the unknown vector $\bx$ are interpreted as synaptic weights (that
are in general not sparse). The rows of the matrix $F$ are patterns
presented to the perceptron.  The measurements $y_\mu$ are binary and
correspond to classifications of the patterns (think of some patterns
being pictures of cats $y_\mu=+1$ and the others being pictures of
dogs $y_\mu=-1$). The goal is to learn the synaptic weights $x_i$ in
such a way that $\sum_i F_{\mu i} x_i >0$ when $y_\mu=+1$, and
$\sum_i F_{\mu i} x_i <0$ when $y_\mu=-1$. The random version of the
problem has been considerably studied by physicists, see for instance
\cite{gardner1988optimal,gardner1989three,KrauthMezard89,watkin1993statistical,shinzato2008perceptron}. 

Another example well known and studied in the statistical physics
literature is the code-division multiple access (CDMA) problem in
which $N$ users simultaneously aim to transmit their binary signal
$x_i \in \{\pm 1\}$ to one base station.  In order to be able to deal
with interference one designs a so-called spreading code associated to
each user $F_{\mu i}\in \{\pm 1\}/\sqrt{N}$, $\mu =1,\dots,M$, that
spreads the signal of each user among $M$ samples. The base station
then receives a noisy linear combination of all the signals $y_\mu$,
eq.~(\ref{y_mu}). The goal is to reconstruct the signal of every user
$x_i$. 

Interestingly, the two examples above are the first inference problems
on which, to our knowledge, the strategy discussed in the present
chapter has been introduced in a statistical physics. The seminal work
of Marc M\'ezard derived the TAP equations for the perceptron problem as early as in 1989
\cite{mezard1989space}. It has then been generalized to many other
cases, mainly by Kabashima and co-workers (see
\cite{kabashima2004bp,kabashima2008inference} again for the perceptron
and \cite{kabashima2003cdma} for randomly spreading CDMA).

In this chapter, we shall follow these pioneering works and adopt the
Bayesian probabilistic framework for the above generalized linear
estimation problem. In general this is algorithmically much more
involved than the convex relaxations of the problem, and consequently
less widely studied. The cases where interesting theoretical analysis
can be done are limited to special classes of matrices $F_{\mu i}$
(mostly with random iid elements, but see the discussion in section
\ref{Sec:structured}). Nevertheless the resulting algorithms can (as
always) be used as heuristic tools for more general data $F_{\mu i}$.
One problem where the above linear estimation problem truly involves a
matrix $F_{\mu i}$ with random iid elements is {\it compressed
  sensing} \cite{donoho2006compressed} to which we dedicate the next
sections.

\subsection{Compressed sensing}

Compressed sensing, a special case of the above linear estimation setting, triggered a major evolution in signal acquisition.
Most signals of interest are compressible and
this fact is widely used to save storage place. But so far compression is made only once the full signal is
measured. In many applications (e.g. medical imaging using MRI or tomography) it is desirable to reduce
the number of measurements as much as possible (to reduce costs, time,
radiation exposition dose etc.). It is hence desirable to measure the
signal directly in the compressed format. Compressed sensing is a
concept implementing exactly this idea by designing new measurement protocols. In compressed sensing one aims to infer a sparse
$N$-dimensional vector $\bx$ from the smallest possible number $M$ of
linear projections $\by=F\bx$. The $M \times N$
measurement matrix $F$ is to be designed. 

Most commonly used techniques in compressed sensing are based on a
development that took place in 2006 thanks to the works of Candes,
Tao, and Donoho
\cite{candes2006near,donoho2006compressed,cande2008introduction}. They
proposed to find the vector $\bx$ satisfying the constraints $\by =
F\bx$ which has the smallest $\ell_1$ norm (i.e., the sum of the
absolute values of its components). This $\ell_1$ minimization problem
can be solved efficiently using linear programming
techniques. They also suggested using a random matrix for $F$. This is
a crucial point, as it makes the $M$ measurements random and
incoherent. Incoherence expresses the idea that objects having a
sparse representation must be spread out in the domain in which they
are acquired, just as a Dirac or a spike in the time domain is spread
out in the frequency domain after a Fourier transform. Rather strong
and generic theoretical guarantees can be derived for the $\ell_1$-based sparse
reconstruction algorithms \cite{hastie2015statistical}. These ideas
have led to a fast, efficient and robust algorithm, and this
$\ell_1$-reconstruction is now widely used, and has been at the origin
of the burst of interest in compressed sensing over the last
decade. 

It is possible to compute exactly the performance of
$\ell_1$-reconstruction in the limit $N\to \infty$, and the analytic
study shows the appearance of a sharp phase transition
\cite{donoho2005sparse}.  For any signal with density $\rho=K/N$,
where $K$ is the number of non-zero elements, the
$\ell_1$-reconstruction gives the exact signal with probability one
if the measurement rate, $\alpha=M/N$, is above the so-called
Donoho-Tanner line, $\alpha>\alpha_{\ell_1}(\rho)>\rho$, see
Fig.~\ref{fig:phase_diag_Bayes} (black dashed line). Interestingly,
the very same transition can be found within the replica method
\cite{kabashima2009typical,ganguli2010statistical} but can also be
computed, rigorously, with the approximate message passing technics
that we discuss in this chapter \cite{donoho2009message}.

If one is willing to forget the (exponentially large) computational
cost, sparse signal reconstruction from random linear projections is
possible whenever $\alpha>\rho$ (this can be done for instance
by exhaustively checking all possible positions of the non-zero elements
and trying to invert the reduced system of linear equations). A
challenging question is whether there is a compressed sensing design
and a computationally tractable algorithm that can achieve
reconstruction down to such low measurement rates. In a series of
works \cite{krzakala2012statistical,krzakala2012probabilistic},
complemented by an impressive rigorous proof of
\cite{donoho2013information}, this question was answered positively
under the assumption that the empirical distribution of the non-zero
elements of the signal is known. This contribution to the theory
of compressed sensing is explained below and uses many of the
statistical physics elements from section~\ref{planted_Ising}.

\subsection{Approximate message passing}
\label{sec:AMP}

Approximate message passing (AMP) is the algorithmic procedure behind
the aforementioned work of \cite{donoho2009message} on linear
estimation and compressed sensing. Origins of AMP trace back to
statistical physics works on the perceptron problem
\cite{mezard1989space} and on the code division multiple
access (CDMA) \cite{kabashima2003cdma}.

What nowadays is known as the generalized approximate message passing
(G-AMP) algorithm, i.e. the generalization to arbitrary element-wise
output functions, was derived for the first time, to our knowledge, in
connection to perceptron neural network in \cite{mezard1989space,kabashima2004bp}. The
algorithm presented in \cite{kabashima2004bp} had the correct
time-indices and good performance was illustrated on the CDMA
problem. However, for an unknown reasons, that paper did not have much
impact and many of its results waited to be rediscovered in the
connection to compressed sensing much later.  The opinion that this
type of algorithm has invincible convergence problems prevailed in the
literature until the authors of \cite{BraunsteinZecchina06}
demonstrated impressive performance of a related message passing
algorithm for learning in binary perceptron. AMP is closely related to
relaxed-belief propagation (r-BP) \cite{rangan2010estimation}. The
algorithm of \cite{BraunsteinZecchina06} can actually be seen as a
special case of r-BP.

Donoho, Maleki and Montanari introduced AMP for compressed sensing
\cite{donoho2009message}, they also established the name. Very
remarkably, the rigorous proof of its performance for random matrices
$F$ was done in \cite{bayati2011dynamics,bayati2012universality}. A
comprehensive generalization to arbitrary priors and element-wise
output function is due to Rangan \cite{rangan2011generalized}, who
established the name {\it generalized approximate message passing} for
the version with a general output channel. Many variants of this
approach have then been considered (see for instance, without being
exhaustive,
\cite{kamilov2012approximate,krzakala2012probabilistic,fletcher2011neural,som2012compressive}).

As is clear from the earlier papers \cite{kabashima2003cdma}, AMP is
actually very closely related to the TAP equations
\cite{ThoulessAnderson77} presented in section \ref{sec:TAP}. The main
difference between TAP and AMP is that TAP was originally derived for
a model with Ising spins and pairwise interactions, whereas AMP is for
continuous variables and $N$-body interactions.

\begin{figure}[!ht]
  \begin{center}
		\includegraphics[width=0.5\linewidth]{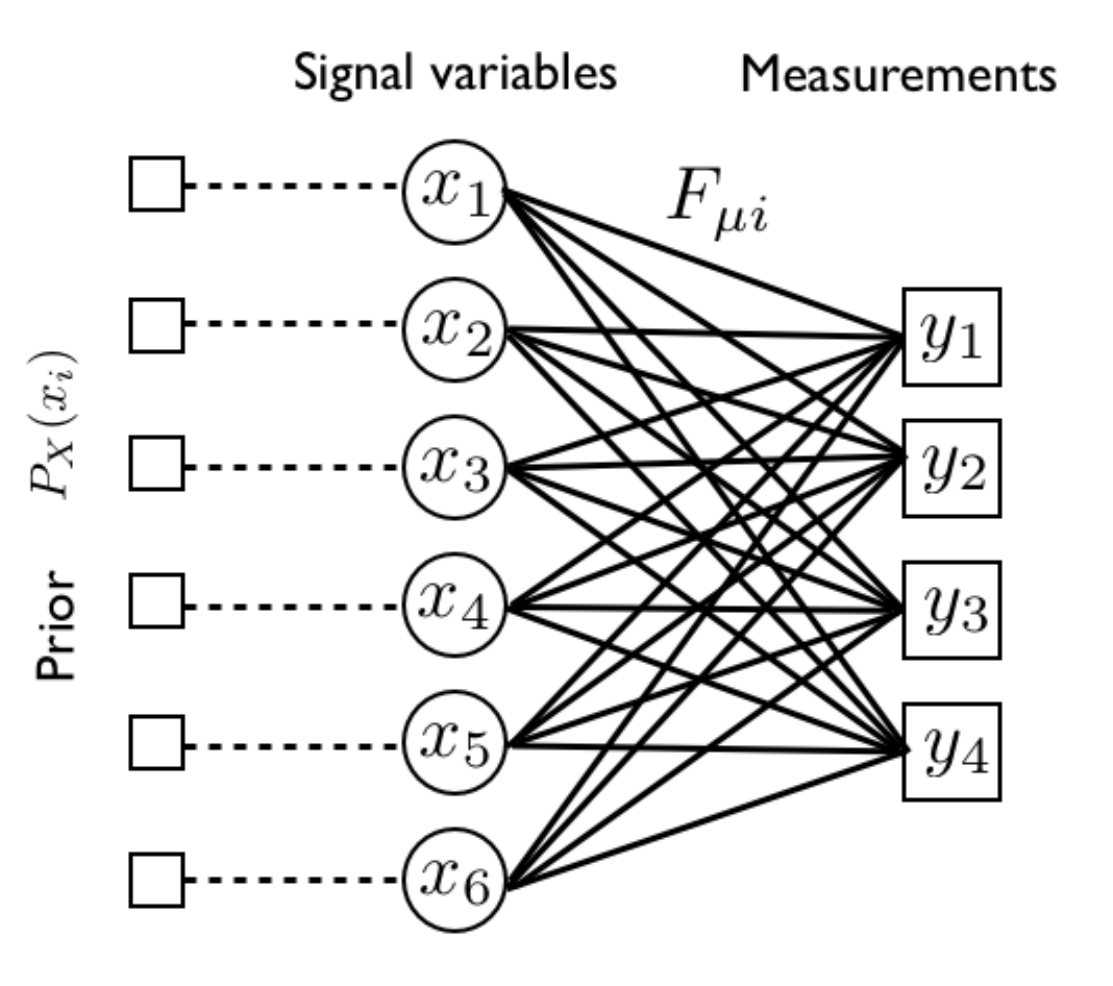}
       \caption{ Factor graph of the linear estimation problem
         corresponding to the posterior probability
         (\ref{CS_post}). Circles represent the unknown variables,
         whereas squares represent the interactions between
         variables. 
  \label{fig:factor}}
\end{center}
\end{figure}

\subsubsection{Relaxed Belief propagation}
\label{sec:rBP}
A detailed statistical physics derivation of the AMP algorithm was
presented in \cite{krzakala2012probabilistic} for the Gaussian output
channel, and for general output as a part of the yet more general
problem of matrix factorization in \cite{kabashima2014phase}. Here we
shall describe the main elements necessary for the reader to
understand where the resulting algorithm comes from.

We are given the posterior distribution 
\be
         P(\bx|\by,F) =  \frac{1}{Z} \prod_{\mu=1}^M P_{\rm out}(y_\mu|
         z_\mu) \prod_{i=1}^N P_X(x_i) \, ,   \quad {\rm where}
         \quad z_\mu = \sum_{i=1}^N F_{\mu i} x_i\, ,   \label{CS_post}
\ee
where the matrix $F_{\mu i}$ has independent random entries (not
necessarily identically distributed) with mean and variance
$O(1/N)$. This posterior probability distribution corresponds to a
graph of interactions $y_\mu$ between variables (spins) $x_i$ called the
{\it graphical model} as depicted in Fig.~\ref{fig:factor}. 

A starting point in the derivation of AMP is to write the belief
propagation algorithm corresponding to this graphical model, as done
originally in \cite{baron2010bayesian}. The
matrix $F_{\mu i}$ plays the role of randomly-quenched
disorder, the measurements $y_\mu$ are planted disorder. As long as
the elements of $F_{\mu i}$ are independent and their mean and variance of
of order $O(1/N)$ the corresponding system is a mean field spin
glass. In the Bayes-optimal case (i.e. when the prior is matching the
true empirical distribution of the signal) the fixed point of belief
propagation with lowest free energy then provides the asymptotically
exact marginals of the above posterior probability distribution.

For model such as (\ref{CS_post}), where the interaction are not pairwise,
the BP equation we presented in sec.~\ref{sec:BPexplained} need to be
slightly modified, for examples see \cite{MezardMontanari07}, but the logic is very
similar. BP still implements a message-passing scheme between nodes in
the graphical model of Fig.~\ref{fig:factor}, ultimately allowing one to
compute approximations of the posterior marginals.
Messages $m_{i \to \mu}$ are sent from the variable-nodes (circles) to the
factor-nodes (squared) and subsequent messages $m_{\mu \to i}$ are sent from
factor-nodes back to variable-nodes that corresponds to algorithm's
current ``beliefs'' about the probabilistic distribution of the
variables $x_i$. It reads
\bea
m_{i \to \mu}(x_i) &=& \frac{1}{z_{i \to \mu}} P_X(x_i) \prod_{\gamma \neq \mu}
m_{\gamma \to i}(x_i) \label{bp-1}\\
m_{\mu \to i}(x_i) &=&\frac{1}{z_{\mu \to i}} \int \prod_{j \neq
  i} \[[ {\rm d}x_j m_{j
  \to \mu}(x_j) \]] P_{\rm out}(y_{\mu}|\sum_ l F_{\mu l} x_l) \label{bp-2}
\eea
While being easily written, this BP is not computationally tractable,
because every interaction involves $N$ variables, and the resulting
belief propagation equations involve $(N-1)$-uple integrals.

%
Two facts enable the derivation of a tractable BP algorithm: the
central limit theorem, on the one hand, and a projection of the
messages to only two of their moments (also used in algorithms such as
Gaussian BP~\cite{shental2008gaussian} or non-parametric BP
\cite{sudderth2010nonparametric}). This results in the so-called
relaxed-belief-propagation (r-BP) \cite{rangan2010estimation}: a form
of equations that is tractable and involves a pair of means and
variances for every pair variable-interaction.

Let us go through the steps that allow to go from BP to r-BP and first
concentrate our attention to eq.~(\ref{bp-2}). Consider the variable
$z_{\mu}= F_{\mu i} x_i + \sum_{j\neq i} F_{\mu j} x_j$. In the
summation in eq.~(\ref{bp-2}), all $x_j$s are independent, and thus
we expect, according to the central limit theorem,
$\sum_{j\neq i} F_{\mu j}^2 x_j$ to be a Gaussian random variable with
mean $\omega_{\mu \to i}=\sum_{j\neq i} F_{\mu j} a_{j \to \mu}$ and
variance $V_{\mu \to i}= \sum_{j\neq i} F^2_{\mu j} v_{j \to \mu}$,
where we have denoted
\bea
  a_{i\to \mu } &\equiv&  \int {\rm d}x_i \, x_i \,  m_{i\to \mu}(x_i) ~~~~~~~~
  v_{i\to \mu } \equiv \int {\rm d}x_i \, x^2_i \, m_{i\to \mu}(x_i) -
  a^2_{i\to \mu } \, . \label{c_imu} 
\eea
We thus can replace the multi-dimensional integral in
eq.~(\ref{bp-2}) by a scalar Gaussian one over the random variable
$z$:
\bea m_{\mu \to i}(x_i) &\propto& \int dz_{\mu} P_{\rm
  out}(y_{\mu}|z_{\mu}) e^{-\frac{(z-\omega_{\mu \to i}-F_{\mu i}x_i)^2}{2
    V_{\mu \to i }}} \, .\eea
We have replaced a complicated multi-dimensional integral
involving full probability distributions by a single scalar one that involves only
first two moments.

One can further simplify the equations: The next step is to rewrite
$(z-\omega_{\mu \to i}-F_{\mu i}x_i)^2= (z-\omega_{\mu \to i})^2+F_{\mu i}^2
x_i^2 -2 (z-\omega_{\mu \to i}) F_{\mu i}x_i F_{\mu i}x_i$
and to use the fact that $F_{\mu i}$ is $O(1/\sqrt{N})$ to expand the
exponential
 \bea
  m_{\mu \to i}(x_i) 
&\propto& 
\int dz_{\mu}  P_{\rm out}(y_{\mu}|z_{\mu})   
e^{-\frac{(z-\omega_{\mu      \to i})^2}{2 V_{\mu \to i }}} \\
&& \times \((1 + F_{\mu
    i}^2 x_i^2 -2 (z-\omega_{\mu      \to i}) F_{\mu i}x_i +\frac 12
  (z-\omega_{\mu      \to i})^2 F^2_{\mu i}x_i^2  + o(1/N) \)) \, .\nonumber
\eea
At this point, it is convenient to introduce the output function
$g_{\rm out}$, defined via the output probability $P_{\rm out}$ as 
\be g_{\rm out} (\omega, y , V) \equiv \frac{ \int {\rm d}z P_{\rm
    out}(y|z)\, (z-\omega) \, e^{-\frac{(z-\omega)^2}{2V}} }{ V \int
  {\rm d}z P_{\rm out}(y|z) e^{-\frac{(z-\omega)^2}{2V}} } \,
. \label{eq:def_gout}\ee
The following useful identity 
holds for the average of
$(z-\omega)^2/V^2$ in the above measure: 
\be
  \frac{ \int {\rm d}z P_{\rm
        out}(y|z) \, (z-\omega)^2\,   e^{-\frac{(z-\omega)^2}{2V}}  }{ V^2  \int {\rm d}z P_{\rm
        out}(y|z) e^{-\frac{(z-\omega)^2}{2V}}  }=
        \frac{1}{V} + \partial_\omega  g_{\rm out} (\omega, y , V)  +
        g^2_{\rm out} (\omega, y , V) \, . \label{eq:id_gout}
\ee 
Using definition (\ref{eq:def_gout}), and re-exponentiating the
$x_i$-dependent terms while keeping all the leading order terms, we 
obtain (after normalization) that the iterative form of
equation eq.~(\ref{bp-2}) reads:
\be {m}_{\mu \to i} (t,x_i) = \sqrt{ \frac{ A^t_{\mu \to i } }{2\pi N}
} e^{ -\frac {x_i^2}{2N}
  A^t_{\mu \to i} +  B^t_{\mu \to i} \frac{x_i}{\sqrt{N}} - \frac{(B^t_{\mu \to      i } )^2}{2  A^t_{\mu \to i} } } \label{eq_def_AB}\\
\ee with 
\bea
B^t_{\mu  \to i } &=&  g_{\rm out} (\omega^t_{\mu \to i}, y_{\mu} ,
V^t_{\mu \to i}) \, F_{\mu i} \text{~~~and~~~}
A^t_{\mu \to i } = - \partial_{\omega} g_{\rm out} (\omega^t_{\mu \to i}, y_{\mu} ,
V^t_{\mu \to i}) \, F_{\mu i}^2\, . \nonumber \eea
Given that $ {m}_{\mu \to i} (t,x_i)$ can be written with a quadratic
form in the exponential, we just write it as a Gaussian distribution.
We can now finally close the equations by writing (\ref{bp-1}) as a
product of these Gaussians with the prior
\bea m_{i \to \mu}(x_i) &\propto& P_X(x_i) e^{-\frac{(x_i-R_{i \to \mu})^2}{2 \Sigma_{i \to \mu}}}
\eea
so that we can finally  give the iterative form of the r-BP algorithm
Algorithm \ref{alg:rBP} (below)
where we denote by $\partial_{\omega}$ (resp. $\partial_R$) the partial derivative with respect to
variable $\omega$ (resp. $R$) and we define the ``input" functions as
\be
    f_a(\Sigma,R) = \frac{    \int {\rm d}x \, x\,  P_X(x)\, 
      e^{-\frac{(x-R)^2}{2\Sigma}}      }{ \int {\rm d}x \,  P_X(x) \, 
      e^{-\frac{(x-R)^2}{2\Sigma}}  }\, , \quad \quad  f_v(\Sigma,R)
    =\Sigma \partial_R f_a(\Sigma,R) \, . \label{fa}
\ee

\begin{algorithm}[H]
    \caption{relaxed Belief-Propagation (r-BP) \label{alg:rBP}}  
  \begin{algorithmic}
    \STATE {\bfseries Input:} $\mathbf{y}$
    \STATE \emph{Initialize}: $\mathbf{a}_{i \to \mu}(t=0)$,$\mathbf{v}_{i \to \mu}(t=0)$, ${\rm t} =1$   
    \REPEAT   
      \STATE r-BP Update of $\{\omega_{\mu \to i},V_{\mu \to i}\}$
      \bea
      V_{\mu \to i}(t) &\gets& \sum_{j\neq i} F^2_{\mu j}    v_{j \to \mu}(t-1) \\
      \omega_{\mu \to i}(t) &\gets& \sum_{j\neq i} F_{\mu j}    a_{j \to \mu}(t-1) \label{eqq_mp}
      \eea
      \STATE r-BP Update of $\{A_{\mu  \to i }, B_{\mu  \to i }\}$
      \bea
B_{\mu  \to i }(t) &\gets&  g_{\rm out} (\omega_{\mu \to i}(t), y_{\mu} ,
V_{\mu \to i}(t)) \,  F_{\mu i} \, ,\label{eq:B} \\
A_{\mu \to i }(t) &\gets& - \partial_{\omega} g_{\rm out} (\omega_{\mu \to i}(t), y_{\mu} ,
V_{\mu \to i}(t)) F_{\mu i}^2\label{eq:A} \eea
      \STATE r-BP Update of $\{R_{\mu \to i},\Sigma_{\mu \to
        i}\}$
       \bea
       \Sigma_{i \to \mu}(t) &\gets& \frac{1}{\sum_{\nu\neq \mu}A_{\nu \to           i}(t)} \\
       R_{i \to \mu}(t) &\gets& \Sigma_{i \to \mu}(t) \sum_{\nu\neq \mu}B_{\nu \to i}(t)
      \eea
      \STATE AMP Update of the estimated partial marginals
      $\mathbf{a}_{i \to \mu}(t)$,$\mathbf{v}_{i \to \mu}(t)$
      \bea
      a_{i\to \mu} (t)&\gets& f_a\left(\Sigma_{i \to \mu}(t),      R_{i \to \mu}(t)   \right)\, , \\
      v_{i\to \mu} (t)&\gets& f_v\left(\Sigma_{i \to \mu}(t),      R_{i \to \mu}(t)   \right)\, .\label{eqq2}
      \eea
      \STATE ${\rm t} \gets {\rm t} + 1$
      \UNTIL{Convergence on $\mathbf{a}_{i \to \mu}(t)$,$\mathbf{v}_{i
          \to \mu}(t)$}
      \STATE {\bfseries output:} Estimated marginals mean and variances:
    \bea
a_{i}&\gets f_a\left(\frac{1}{\sum_{\nu}A_{\nu \to
      i}},\frac{\sum_{\nu}B_{\nu \to i}}{ \sum_{\nu}A_{\nu
      \to
      i}   }\right) \, ,  \label{eqq3} \\
v_{i}&\gets f_v\left(\frac{1}{\sum_{\nu}A_{\nu \to
      i}},\frac{\sum_{\nu}B_{\nu \to i}}{ \sum_{\nu}A_{\nu
      \to i} }\right)\, .  \label{eqq4}
\eea
  \end{algorithmic}
\end{algorithm}

\subsubsection{From r-BP to G-AMP}
While one can implement and run the r-BP, it can be simplified further
without changing the leading order behavior of the marginals by
realizing that the dependence of the messages on the target node is
weak. This is exactly what we have done to go from the standard belief
propagation to the TAP equations in sec.~\ref{sec:TAP}.

After the corresponding Taylor expansion the corrections add up
into the so-called {\it Onsager reaction terms}
\cite{ThoulessAnderson77}. The final G-AMP iterative equations are
written in terms of means $a_i$ and their variances $c_i$ for each of
the variables $x_i$. The whole derivation is done in a way that the
leading order of the marginal $a_i$ are conserved. Given the BP was
asymptotically exact for the computations of the marginals so is
the G-AMP in the computations of the means and variances. Let us define
\begin{eqnarray}
\omega_{\mu}^{t+1} = \sum_i F_{\mu i} a_{i \to \mu}^t\, ,\\
V_{\mu}^{t+1} = \sum_i F_{\mu i}^2 v^t_{i \to \mu}\, ,\\
\Sigma_i^{t+1} = \frac 1{\sum_{\mu} A^{t+1}_{\mu \to i}}\,,\\
R_i^{t+1} = \frac {\sum_{\mu} B^{t+1}_{\mu \to i}}{\sum_{\mu} A^t_{\mu
    \to i}}\, .
\end{eqnarray}
First we notice that the correction to $V$ are small so that
\begin{eqnarray}
V^{t+1}_{\mu} = \sum_i F_{\mu i}^2 v^t_{i \to \mu} \approx \sum_i
F_{\mu i}^2 v^t_i \, .
\end{eqnarray}
From now on we will use the notation $\approx$ for equalities that are
true only in the leading order. 

From the definition of $A$ and $B$ we get
\begin{eqnarray}
\Sigma_i &=& \left[- \sum_{\mu} F_{\mu i}^2 \partial_{\omega} g_{\rm
               out}(\omega_{\mu \to i}^{t+1},y_{\mu},V^{t+1}_{\mu\to i}) \right]^{-1} \approx  \left[- \sum_{\mu} F_{\mu
               i}^2 \partial_{\omega} g_{\rm
               out}(\omega^{t+1}_{\mu},y_{\mu},V^{t+1}_{\mu})
               \right]^{-1} \nonumber \\
R_i &=& \left[- \sum_{\mu} F_{\mu i}^2 \partial_{\omega} g_{\rm
    out}(\omega^{t+1}_{\mu},y_{\mu},V^{t+1}_{\mu}) \right]^{-1} \times
\left[  \sum_{\mu} F_{\mu i} g_{\rm
    out}(\omega^{t+1}_{\mu\to i},y_{\mu},V^{t+1}_{\mu\to i})  \right] \nonumber
\end{eqnarray}
Then we write that
\begin{eqnarray}
g_{\rm out}(\omega^{t+1}_{\mu \to i},y_{\mu},V^{t+1}_{\mu \to i})  &\approx& 
g_{\rm out}(\omega^{t+1}_{\mu},y_{\mu},V^{t+1}_{\mu}) - F_{\mu i} a^t_{i \to \mu} \partial_{\omega} g_{\rm out}(\omega^{t+1}_{\mu},y_{\mu},V^{t+1}_{\mu}) \\
&\approx& 
g_{\rm out}(\omega^{t+1}_{\mu},y_{\mu},V^{t+1}_{\mu}) - F_{\mu i} a^t_{i} \partial_{\omega} g_{\rm out}(\omega^{t+1}_{\mu},y_{\mu},V^{t+1}_{\mu}) 
\end{eqnarray}
So that finally
\begin{eqnarray}
(\Sigma_i)^{t+1} &=& \left[- \sum_{\mu} F_{\mu i}^2 \partial_{\omega} g_{\rm out}(\omega^{t+1}_{\mu},y_{\mu},V^{t+1}_{\mu}) \right]^{-1}\\
R^{t+1}_i &=& (\Sigma_i)^{t+1} \times \left[  \sum_{\mu} F_{\mu i} g_{\rm out}(\omega_{\mu},y_{\mu},V_{\mu}) - F_{\mu i}^2 a_{i} \partial_{\omega} g_{\rm out}(\omega_{\mu},y_{\mu},V_{\mu})  \right] \\
&=& a^t_i + (\Sigma_i)^{t+1} \sum_{\mu} F_{\mu i} g_{\rm out}(\omega^{t+1}_{\mu},y_{\mu},V^{t+1}_{\mu}) 
\end{eqnarray} 
Now let us consider $a_{i \to \mu}$:

\begin{eqnarray}
a^t_{i \to \mu}&=&f_{a}(R^t_{i \to \mu},\Sigma_{i \to \mu}) \approx f_{a}(R^t_{i \to \mu},\Sigma_i) \\ &\approx& f_{a}(R^{t}_i,\Sigma_i)-B^{t}_{\mu \to i}  f_{v}(R^{t}_i,\Sigma_i) \\
&\approx& a^t_i- g_{\rm out}(\omega^{t}_{\mu},y_{\mu},V^{t}_{\mu}) F_{\mu i} v^t_i 
\end{eqnarray} 

So that 
\begin{eqnarray}
\omega^{t+1}_{\mu} = \sum_i F_{\mu i} a^t_i 
- \sum_i g_{\rm out}(\omega^{t}_{\mu},y_{\mu},V^{t}_{\mu}) F_{\mu i}^2 v^t_i = \sum_i F_{\mu i} a_i 
- V^{t}_{\mu} g_{\rm out} (\omega^{t}_{\mu},y_{\mu},V^{t}_{\mu})
\end{eqnarray}

\begin{algorithm}[H]
    \caption{Generalized Approximate Message Passing (G-AMP) \label{alg:GAMP}}  
  \begin{algorithmic}
    \STATE {\bfseries Input:} $\mathbf{y}$
    \STATE \emph{Initialize}: $\mathbf{a}^0$,$\mathbf{v}^0$, $g^{0}_{\rm out, \mu}$, ${\rm t} =1$   
    \REPEAT   
    \STATE AMP Update of $\omega_{\mu},V_{\mu}$
    \bea
    V^{t}_{\mu} &\gets& \sum_i F^2_{\mu i} v_i^{t-1} \\
    \omega^{t}_{\mu} &\gets& \sum_i F_{\mu i} a_i^{t-1} - V^{t}_{\mu}
    g_{\rm out,\mu}^{t-1}\label{omega}
    \eea
    \STATE AMP Update of $\Sigma_i,R_i, g_{\rm out, \mu}$ 
    \bea
    g_{\rm out,\mu}^{t} &\gets&   g_{\rm out}(\omega^{t}_{\mu},y_{\mu},V^{t}_{\mu})\\
    \Sigma_i^t &\gets& \left[- \sum_{\mu} F_{\mu i}^2 \partial_{\omega} g_{\rm out}(\omega^{t}_{\mu},y_{\mu},V^{t}_{\mu}) \right]^{-1}\\
    R_i^{t} &\gets& a^{t-1}_i + \Sigma_i^{t} \sum_{\mu}
    F_{\mu i}     g_{\rm out,\mu}^{t} \eea \STATE AMP Update of the
    estimated marginals $a_i,v_i$ \bea
    a_i^{t} &\gets& f_a(\Sigma^t_i ,R^{t}_i) \\
      v_i^{t} &\gets& f_v(\Sigma^t_i ,R^{t}_i) \eea \STATE
      ${\rm t} \gets {\rm t} + 1$ \UNTIL{Convergence on
        $\mathbf{a}$,$\mathbf{v}$} \STATE {\bfseries output:}
      $\mathbf{a}$,$\mathbf{v}$.
\end{algorithmic}
\end{algorithm}

An important aspect of these equations to note is the index $t-1$ in
the Onsager reaction term in eq.~(\ref{omega}) that is crucial for
convergence and appears for the same reason as in the TAP equations in
sec. \ref{sec:TAP}. Note that the whole algorithm is comfortably
written in terms of matrix multiplications only, this is very useful
for implementations where fast linear algebra can be used.

\subsubsection{The potential}

Using the usual Bethe approach, one can also compute the corresponding
potential, and show that these message passing equations actually aim to minimize
the following Bethe free energy
\cite{rangan2014convergence,krzakala2014variational,vila2014adaptive}:
\begin{align}
& {\cal F}^{\rm Bethe} \left(\{\Sigma_i\},\{R_i\}\right)
= \sum_{i} D_{\rm KL} (Q_i||P_X)  +\sum_{\mu} D_{\rm KL}( {\cal
    M}_{\mu}|| P_{\rm out} )  \nonumber \\  &+ \frac 12 \sum_{\mu}\left[ \log{2\pi V_\mu^*}
+1+V_\mu^*\partial_{\omega} g_{\rm out}(\omega^*_\mu,y_\mu,V^*_\mu) \right]  , \label{eq:Bethe}
\end{align}
with $V_\mu^*$ and $\omega_\mu^*$ satisfying their respective fixed-point
conditions, and $D_{\rm KL}(Q||P)=\int  [\log{Q(x)} - \log{P(x)}]  {\rm d}Q(x)$ being the Kullback-Leibler
divergences between two probability distributions. 
The two Kullback-Leibler divergences above are taken with
respect to probability distributions
\bea {\cal M}_{\mu}(z;\omega_{\mu}^*,y_\mu,V_{\mu}^*) &=&
\frac{1}{{\cal    Z}_{\cal M}} P_{\rm
  out}(y_{\mu}|z) \frac{1}{\sqrt{2\pi V_{\mu}^*}}e^{-\frac{(z-\omega_{\mu}^*)^2}{2V_{\mu}^*}} \\
Q_i(x;T_i,\Sigma_i) &=& \frac 1{{\cal Z}_{\cal Q}} P_X(x)
e^{-\frac{(x-T_i)^2}{2\Sigma_i}} \, ,
\eea
where ${\cal    Z}_{\cal M}$ and ${\cal Z}_{\cal Q}$ are the
corresponding normalizations. Clearly these two probability
distributions are related to the denominators in (\ref{fa}) and (\ref{eq:def_gout}).

\subsubsection{Note on the convergence and parameter learning}
When G-AMP converges, its performance in terms of mean-squared error is usually
considerably better than the performance of other existing algorithms. The AMP equations are proven to converge in the idealized setting of large systems, random
iid matrices~$F$ with zero mean entries, and when the prior $P_X$ either
corresponds to the empirical distribution of the actual signal $\bx^*$
or to the convex relaxation of the problem
\cite{bayati2011dynamics,bayati2012universality}. However, for
mismatched signal distributions, for non-random matrices, positive mean
random matrices or small system-sizes they might have unpleasant convergence
issues. 

Justified by the promise of a strikingly better
performance, it is an important subject of current research how to make the G-AMP
algorithm the most robust as possible, while keeping its performance and
speed. Several recent works followed this line of research. In
\cite{caltagirone2014convergence} it was identified why AMP has
convergence problems for the simple case of iid matrices with elements
of non-zero mean. For means of the matrix elements larger than some
critical value, there is an anti-ferromagnetic-like
(i.e. corresponding to a negative eigenvalue of the Hessian matrix) instability of the Nishimori
line. Finite size effects always cause a slight mismatch between the true and
the empirical distribution of the signal, and hence a slight deviation
from the Nishimori line. The instability amplifies the deviation and
causes the iteration of the AMP equation to fail and go far away from
the Nishimori line. Even though we know that in the Bayes-optimal case
the correct fixed point must satisfy the Nishimori conditions.  

The simplest way to fix this problem for matrices of non-zero mean is
the so-called mean removal as used in many of the early
implementations and discussed in \cite{vila2014adaptive}. However, the
instability of the Nishimori line is a problem that can cause failure
of the algorithm in a wider range of settings than the non-zero mean matrices, e.g. for some
of the structured matrices. It is therefore useful to study other
strategies to stabilize the Nishimori line in order to improve the
convergence.  We observed that randomization of the update is doing
just that. Combining the sequential update while keeping track of the correct
time-indices, Ref. \cite{manoel2015swept} designed the so-called Swept-AMP (SwAMP). 
Ref. \cite{krzakala2012probabilistic} studied the expectation-maximization
learning of parameters for the compressed sensing problem when the number of parameters is small,
typically finite while the system size grows. Since the parameter-learning is precisely imposing the validity of the Nishimori
conditions, this is another way that greatly stabilizes the Nishimori
line and improves the convergence. 

An additional approach to stabilize the Nishimori line is to slow down the
iterations via damping, i.e. taking a linear combination of the new and old values when updating the messages.
Such damping can, however, slow
down the iteration-process considerably. In order to avoid a drastic slow-down
we take advantage of the following related development. 

The authors of \cite{krzakala2014variational} studied the Bethe free energy associated to
G-AMP and wrote a variational form of the Bethe free energy (\ref{eq:Bethe}) for which
the G-AMP fixed points are local minima (not only generic stationary points
such as saddles). This free
energy can be optimized directly leading to a provably converging (but considerably
slower) method to find the fixed
points. This variational Bethe free energy can be used to slow down the iterations in a way to
keep decreasing the free energy - this we call adaptative damping in
\cite{vila2014adaptive}. Other groups suggested several closely related
convergence improvements \cite{rangan2014convergence,rangan2015inference}.

Combining the above implementation features - adaptive damping, sequential update, mean
removal and parameter learning - leads to the current state-of-the art
implementation of the G-AMP algorithm for linear estimation problems \cite{GAMP_implementation}.  

\subsubsection{Examples of priors and outputs}
The G-AMP algorithm is written for a generic prior on the signal $P_X$
(as long as it is factorized over the elements) and a generic
element-wise output channel $P_{\rm out}$.  The algorithm depends on
their specific form only trough the function $f_a$ and $g_{\rm out}$
defined by (\ref{fa}) and (\ref{eq:def_gout}). It is useful to give a couple
of explicit examples.

The sparse prior that is most commonly considered in probabilistic
compressed sensing is the Gauss-Bernoulli prior, that is when in
(\ref{G-B}) we have $\phi(x)={\cal N}(\overline x,\sigma)$ Gaussian
with mean $\overline x$ and variance $\sigma$. For this
prior the input function $f_a$ reads \be f^{\rm
  Gauss-Bernoulli}_a(\Sigma,T) =\frac{ \rho\, e^{-\frac{(T-\overline
      x)^2}{2(\Sigma+\sigma)}}
  \frac{\sqrt{\Sigma}}{(\Sigma+\sigma)^{\frac{3}{2}}} (\overline x
  \Sigma + T \sigma) }{ (1-\rho) e^{-\frac{T^2}{2\Sigma}} + \rho
  \frac{\sqrt{\Sigma}}{\sqrt{\Sigma+\sigma}} e^{-\frac{(T-\overline
      x)^2}{2(\Sigma+\sigma)}} } \, , \ee The most commonly considered
output channel is simply additive white Gaussian noise (AWGN)
(\ref{gauss_out}). The output function then reads \be g^{AWGN}_{\rm
  out}(\omega,y,V) = \frac{y - \omega}{ \Delta + V }\,
. \label{eq:out_GN} \ee

As we anticipated above, the example of linear estimation that was most
broadly studied in statistical physics is the case of the perceptron
problem discussed in detail e.g. in \cite{watkin1993statistical}. In
the perceptron problem each of the $M$
$N$-dimensional patterns
$F_{\mu}$ is multiplied by a vector of synaptic weights $x_i$ in
order to produce an output $y_{\mu}$ according to 
\bea
            y_\mu &=& 1    \quad \quad {\rm if}  \quad \quad
            \sum_{i=1}^N F_{\mu i} x_i > \kappa \, ,  \\ 
           y_\mu &=& -1    \quad \quad {\rm otherwise} \, ,
\eea
where $\kappa$ is a threshold value independent of the pattern. 
The perceptron is designed to classify patterns, i.e. one
starts with a training set of patterns and their corresponding outputs
$y_\mu$ and aims to learn the weights $x_i$ in such a way that the
above relation between patterns and outputs is satisfied. 
To relate this to the linear estimation problem above, let us consider the
perceptron problem in the teacher-student scenario where the teacher
perceptron generated the output $y_\mu$ using some ground-truth set of
synaptic weights $x_i^*$. The student perceptron knows only
the patterns and the outputs and aims to learn the weights. How many
patterns are needed for the student to be able to learn the synaptic
weights reliably? What are efficient learning algorithms?

In the simplest case where the threshold is zero, $\kappa=0$ one can
redefine the patterns $F_{\mu i} \leftarrow F_{\mu
i} y_\mu$ in which case the corresponding redefined output is
$y_\mu=1$. The output function in that case reads 
\be
       g^{\rm perceptron}_{\rm out}(\omega,V) =
  \frac{1}{\sqrt{2\pi V}}   \frac{ e^{-\frac{\omega^2}{2V}}}{ H\left(-\frac{\omega}{\sqrt{V}}\right)}
     \, ,
\label{gout:perceptron}
\ee
where 
\be
   H(x) =\int_x^{\infty}  \frac{dt}{\sqrt{2\pi}}e^{-\frac {t^2}2}\, .
\ee
Note also that the perceptron problem formulated this way is closely
related to what is called 1-bit compressed sensing in the signal
processing literature \cite{boufounos20081,xu2014bayesian}.

In physics a case of a perceptron that was studied in detail is that of binary
synaptic weights $x_i\in \{\pm 1\}$
\cite{gardner1988optimal,gardner1989three,gyorgyi1990first}. To take
that into account in the G-AMP we consider the binary prior $P_X(x) = [\delta(x-1)
+ \delta(x+1)]/2$ which leads to the input function 
\be
   f^{\rm binary}_a(\Sigma,T) = \tanh{\left(\frac{T}{\Sigma} \right)} \, . 
\ee
\subsection{Replica free energy, state evolution and phase transitions} 

\subsubsection{State Evolution} 
The performance of the G-AMP algorithm above can be analyzed in the
large-size limit, for matrices with independent entries. The
corresponding analysis was baptized {\it state evolution} in
\cite{DonohoMaleki09}.  Rigorous results on the state evolution are
provided in
\cite{bayati2011dynamics,bayati2012universality,javanmard2013state}
for both AMP and G-AMP.

A statistical physics derivation of the state evolution in the spirit
of the cavity method is given in \cite{krzakala2012probabilistic} and
is formally very related to the state evolution of the TAP equation in
sec.~\ref{sec:TAP}. The main point in the derivation of the state
evolution is not to start from the G-AMP algorithm but from its
message passing form, eqs.~(\ref{eqq_mp}-\ref{eqq2}). One then recalls
that in belief propagation we assumed conditional independence of
incoming messages which translates into independence of terms present
in sums of an extensive number of terms in the message passing.  One
then utilizes the central limit theorem, enabling a reduction to the
mean and variance of these terms matters. One simply needs to keep
track of the evolution of this mean and variance.

In statistical physics terms, the state evolution exactly corresponds
to the derivation of the replica symmetric equations via the cavity
method. Since the cavity and replica methods always yield the same
result, the state evolution equations are exactly equivalent to the
equations stemming from the replica method for the corresponding
problem. For compressed sensing this is illustrated in
\cite{krzakala2012probabilistic}. For the perceptron problem all the
replica symmetric equations and related results, e.g. in
\cite{watkin1993statistical}, can be obtained as a special case of the
G-AMP state evolution using the perceptron output.

Let us sketch briefly the derivation of state evolution for G-AMP. We
follow the derivation of \cite{kabashima2014phase} for the specific
case of linear estimation. 
In order to write the
equations for the MMSE for a general output functions, we
need to rewrite the output function using $y=h(z,w)$ where $w$ is a
random variable distributed according to $P_W(w)$. Then 
\be 
P_{\rm
  out}(y|z) = \int {\rm d}w P_W(w) \delta[y-h(z,w)] \, . 
\ee 

Now we consider the random
variable $V_{\mu}=\sum_i F_{\mu i}^2 v_i$. On average its value is
$\sum_i v_i/N$ whereas its is variance $o(1)$. It thus concentrates
around its mean as $N \to
\infty$ and therefore
\begin{eqnarray}
V^{t+1}=\frac 1N \sum_i v_i \, . 
\end{eqnarray}
Let us now turn our attention to
$\omega_{\mu \to i}=\sum_{j \neq i} F_{\mu i} a_{i \to j}$ and
$z_{\mu \to i}=\sum_{j \neq i} F_{\mu i} s_i$. These quantities are, according
the the assumptions of the cavity method, sums of uncorrelated variables, so they
must converge to (correlated) Gaussian variables with co-variances reading
\begin{eqnarray}
\E[\omega^2] &=& \E[a^2] = q\, ,  \label{def_q} \\
\E[z \omega] &=& \E[s a] = m \, , \label{def_m}\\
\E[z^2] &=& \E[s^2] = q_0 \, ,
\end{eqnarray}
where we defined the order parameters $m$ and $q$ as the average
correlation between the estimator $a$ and the ground-truth signal $s$,
and as a self-correlation for the estimator $a$, respectively. 

Let us now see how $R_i$ behaves (note the
difference between the letters $\omega$ and $w$):
\begin{eqnarray}
\frac{R_i}{\Sigma_i} &=&   \sum_{\mu} B_{\mu \to i} = \sum_{\mu} F_{\mu i} g_{\rm out}(\omega_{\mu \to i},y_{\mu},V_{\mu \to i}) \\
&=& \sum_{\mu} F_{\mu i} g_{\rm out}(\omega_{\mu \to i},h(\sum_{j\neq i} F_{\mu j} s_j +F_{\mu i} s_i,w_{\mu}),V)\\
&=&\sum_{\mu} F_{\mu i} g_{\rm out}(\omega_{\mu \to i},h(\sum_{j\neq i} F_{\mu j} s_j,w_{\mu}),V)
+ s_i {\alpha} \hat m\, ,
\end{eqnarray}
where we define
\begin{eqnarray}
\hat m &=& E_{\omega,z,w} \left [ \partial_{z} g_{\rm
    out}(\omega,h(z,w),V)\right]\, .
\end{eqnarray}

%
%
We further write 
\begin{eqnarray}
\frac{R_i}{\Sigma_i} 
&=& {\cal{N}}(0,1)  \sqrt{\alpha \hat q}  + s_i {\alpha} \hat m
\end{eqnarray}
with ${\cal{N}}(0,1) $ being a random Gaussian variables of zero mean
and unit variance, and where
\begin{eqnarray}
\hat q &=& \E_{\omega,z,w} \left [ g^2_{\rm out}(\omega,h(z,w),V) \right]
\end{eqnarray}
while the computation of $\Sigma$ simply gives, thanks to concentration,
\begin{eqnarray}
\frac{1}{\Sigma}= \alpha \hat q \, .
\end{eqnarray}

Now we can close all equations by computing $q$ and $m$ from their
definitions (\ref{def_q}-\ref{def_m}) as
\begin{eqnarray}
q &= \E_{s} \left[ \E_{R,\Sigma} f^2_a(\Sigma,R) \right] \\
m &= \E_{s} \left[ \E_{R,\Sigma} s f_a(\Sigma,R) \right] 
\end{eqnarray}
In the Bayes-optimal case, the Nishimori consition imposes $q=m$ and
thus provides
further simplifications. In fact, one can show that $\hat q$ is also
equal to $\hat m$ \cite{kabashima2014phase}. In this case, the state evolution leads to the
asymptotic Bayes-optimal minimum mean squared error (MMSE) for the
corresponding inference problem. Just as it did for the planted spin
glass or the clustering of networks above. Since in linear estimation
we are working with a fully connected graphical model the asymptotic
analysis can be written in terms of only 2 scalar parameters
\cite{krzakala2012probabilistic}. 

Finally, writing explicitly the integrals, the state evolution 
reads 
\bea m^{t+1} &=& \int {\rm d}x \, P_X(x) \int {\rm d}\xi
\frac{e^{-\frac{\xi^2}{2}}}{\sqrt{2\pi}} f_a^2\left( \frac{1}{\alpha
    \hat m^t}, x + \frac{\xi}{\sqrt{\alpha \hat m^t}}
\right)  \, ,  \\
\hat m^t &=& - \int {\rm d}w \, P_W(w) \int {\rm d}p \, {\rm d}z \,
\frac{e^{-\frac{p^2}{2m^t}} e^{-\frac{(z-p)^2}{2(\rho_0 - m^t)}}
}{2\pi \sqrt{m^t (\rho_0 - m^t)}} \partial_p g_{\rm out}(p,h(z,w),
\rho_0 - m^t ) \, , \eea where we denote
$\rho_0 = N E_F(F^2) E_{P_X}(x^2)$. For the sparse prior (\ref{G-B})
where the non-zero elements have the second moment equal to
$\overline{x^2}$ and the matrix $F$ having iid elements of zero mean
and variance $1/N$ we get $\rho_0=\rho \overline{x^2}$. For the usual
AWGN output (\ref{gauss_out}) the second equation simplifies
considerably into \be \hat m^t = \frac{1}{\Delta + \rho \overline{x^2}
  - m^t} \, .  \ee The mean-squared error of the AMP estimator is then
${\rm MSE} = \rho \overline{x^2} - m^t$. To evaluate the performance
of AMP initialized in a way unrelated to the unknown signal, the state
evolution is iterated until convergence starting from
$m^{t=0} = \rho^2 {\overline x}^2$, where $\overline x$ is the mean of
the distribution $\phi(x)$ in (\ref{G-B}).

\subsubsection{Free energy} 

The same procedure we use to derive the state evolution can be used to
write a scalar expression for the Bethe free energy
(\ref{eq:Bethe}). Alternatively, the same expression can be obtained using the
replica method, as e.g. in \cite{kabashima2014phase}. The average Bethe
free energy  for the generalized linear estimation problem can
be written as $f^{\rm Bethe}  = \min_{m,\hat m} \Phi(m,\hat m)$ where 
\bea
              \Phi(m,\hat m) = &\alpha \int {\rm d} y\, {\cal D}\xi
           \,    {\cal D} s P_{\rm out}(y|s
              \sqrt{\rho_0 -m}+ \xi \sqrt{m}) \log{\left[   \int {\cal D}x   P_{\rm out}(y|x
              \sqrt{\rho_0 -m}+ \xi \sqrt{m})  \right]} \nonumber \\ \hspace{-1cm}
          - &\frac{\alpha m\hat m}{2}
            + \int {\cal D}\xi {\rm d}s P_X(s) e^{-\frac{\alpha \hat m
                s^2}{2} + \xi s \sqrt{\alpha \hat m}} \log{\left[  \int  {\rm d} x
              P_X(x)  e^{-\frac{\alpha \hat m
                x^2}{2} + \xi x \sqrt{\hat m}} \right]}
\eea

The free energy can be also written as a function of the MSE $E$. In
the case of the AWGN channel with variance $\Delta$, for instance, it
reads $f^{\rm Bethe} = \min_E \Phi(E)$ where \cite{krzakala2012probabilistic}:
\bea
\Phi(E) &=& -\frac{\alpha}{2} - \frac{\alpha}{2}
\log{(\Delta+E)} - \frac{\alpha(\rho \overline{x^2} -E)}{2(\Delta+E)}
\nonumber \\ &+& \int {\rm d}s  P_X(s) \int
{\cal D}\xi  \log{\left\{  \int {\rm d}x \,
    e^{\frac{\alpha}{\Delta+E} x
      (s-\frac{x}{2}) + \xi x \frac{\sqrt{\alpha}}{\sqrt{\Delta+E}}} 
P_X(x)
 \right\}}\, .\label{free_rep_NL}
\eea
Actually, this formula is, again, very close to the one discussed in
the context of CDMA by Tanaka \cite{tanaka2002statistical} and Guo and
Verd\'u~\cite{guo2005randomly}. An equivalent form has been derived for
compressed sensing by Verd\'u~\cite{wu2012optimal}. Much more can be done
with the replica method in this context (see,
e.g. \cite{NIPS2009_3635,takeda2010statistical}).

\subsubsection{Phase transitions}

What we want to discuss here is again the appearance of the very
generic phenomena we have discussed already in the first chapter, and
illustrated on Fig.~\ref{fig:schema}: the presence of two different
transitions when trying to solve the Bayesian optimal problem.

\begin{figure}[!ht]
  \begin{center}
		\includegraphics[width=0.6\linewidth]{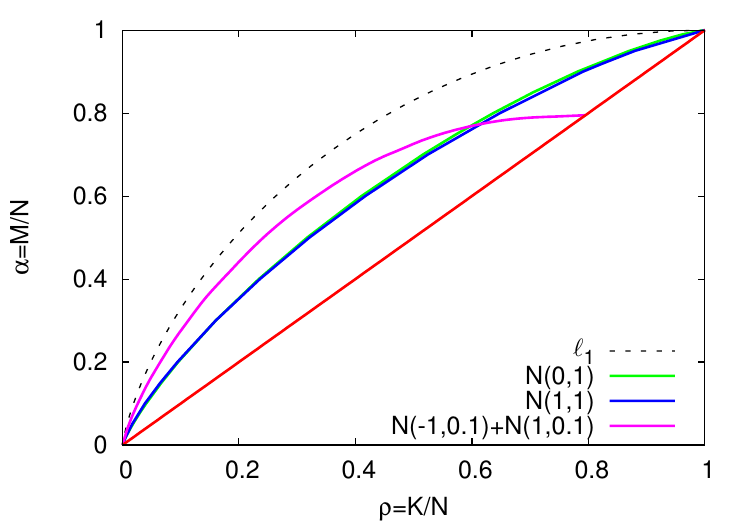}
       \caption{Phase diagram for the AMP reconstruction in the optimal Bayesian
                 case when the signal model is
                  matching the empirical distribution of signal
                  elements. The elements of the $M\times N$
                  measurement matrix $F$ are iid variables with zero mean and variance~$1/N$.
                The spinodal transition $\alpha_{s}(\rho)$ is computed
                with the state evolution and plotted for the following signal
                  distributions: $\phi(x)={\cal N}(0,1)$ (green),
                  $\phi(x)={\cal N}(1,1)$ (blue)
                  $\phi(x)=[{\cal N}(-1,0.1)+{\cal N}(1,0.1)]/2$
                  (magenta).
              The data are compared to the Donoho-Tanner phase
              transition $\alpha_{\ell_1}(\rho)$ (dashed) for $\ell_1$ reconstruction that does not
              depend on the signal distribution, and to the
              theoretical limit for exact reconstruction $\alpha=\rho$
              (red). Figure taken from \cite{krzakala2012probabilistic}.
  \label{fig:phase_diag_Bayes}}
\end{center}
\end{figure}

In Fig.~\ref{fig:phase_diag_Bayes} we plot down to what measurement
rate $\alpha=M/N$ AMP recovers a signal of density $\rho$ for several
distributions of non-zero signal elements $\phi(x)$. We observe a
first order phase transition that depends on the distribution of
non-zeros.

Interestingly, the corresponding line is always better than the
$\ell_1$ transition, as shown in \cite{donoho2013accurate}, where the
distribution maximizing the spinodal line for Bayes-optimal inference
was computed. This is to be expected, since we might hope that the
spinodal for the Bayesian optimal problem is the ``best" spinodal,
though there is no generic proof of this.

One might wonder how a measurement noise is affecting this diagram?
Ref. \cite{krzakala2012probabilistic} further studied how the optimal
inference performance is influenced by measurement noise, mismatch
between the prior and the actual signal distribution, how the
performance changes under expectation maximization learning of the
parameters, such as $\rho$ or the mean $\overline x$ and the variance
$\sigma$ of the signal distribution. Ref.
\cite{barbier2012compressed} explored the role of approximately sparse
signals when the $\delta(x_i)$ in (\ref{G-B}) is replaced by a narrow
Gaussian. The influence of an uncertainty in the elements of the
matrix $F$ was studied in \cite{krzakala2013compressed}. As expected
from statistical physics, where in the presence of disorder first
order phase transitions become weaker, second order or disappear
entirely, the same happens to the first order phase transition in
compressed sensing under presence of various kinds of
noise. Qualitative results are illustrated in
\cite{krzakala2012probabilistic,barbier2012compressed,krzakala2013compressed}. 



Existence of phase transition is not restricted to compressed sensing. In fact,
the very same phenomenology has been discussed and identified as early
as in the 90's in the context of learning a rule via a perceptron with
binary weights
\cite{gardner1989three,mezard1989space,seung1992statistical,watkin1993statistical}. In
the present setting, this corresponds to a binary prior for the
signal, and a perceptron channel given by eq.~(\ref{gout:perceptron}). In this
case, it becomes possible to identify perfectly the signal when the
ratio between the number of examples $M$ and the number of variables $N$ is
larger than $\alpha\approx 1.245$, a result first derived by Derrida
and Gardner \cite{gardner1989three}. It is only beyond $\alpha\approx 1.49$, however,
that the G-AMP algorithm actually succeed in finding the hidden
rule (i.e. the planted weights). All these classic results can thus be recovered by the present,
unifying, approach. Note that, for this problem, the G-AMP (or TAP)
equations has been written and proposed as early as in 1989 in 
\cite{mezard1989space}.  The more recent application of either relaxed
belief propagation \cite{BraunsteinZecchina06} or G-AMP to actual data
\cite{ziniel2014binary} has also been considered.

Another classical example of this approach with a similar
phenomenology is given by the classical work of Tanaka and Kabashima
in the context of CDMA \cite{tanaka2002statistical,kabashima2003cdma}.


\subsection{Spatial coupling}
\label{sec:spatial}

As discussed in section \ref{sec:hard}, if nothing is changed about the
graphical model corresponding to compressed sensing, the 1st order
phase transition presented in Fig.~\ref{fig:phase_diag_Bayes} causes
a barrier that is conjectured to be unbeatable by any polynomial
algorithm. In an idealized setting of compressed sensing, however, the
design of the measurement matrix $F_{\mu i}$ is entirely up to
us. Thanks to this, compressed sensing belongs to the class of problems
where spatial coupling can be successfully applied. This was the main
point of Ref.
\cite{krzakala2012statistical}, where 
spatially coupled measurement matrices were designed that allowed AMP
to reconstruct the signal in compressed sensing anytime the
measurement rate $\alpha$ is larger than the density of non-zero
elements in the signal $\rho$, i.e. all the way above the red line in
Fig.~\ref{fig:phase_diag_Bayes}. This {\it threshold saturation} was
soon after proven rigorously in \cite{donoho2013information}. 
We should note that spatial coupling for compressed sensing was
considered even before in \cite{KudekarPfister10}, but in
that work the observed improvement was limited and it did not seem
possible to achieve the information theoretical limit $\alpha=\rho$.

\begin{figure}[ht]
  \begin{center}
    \begin{tabular}{cc}
    \includegraphics[scale=0.42]{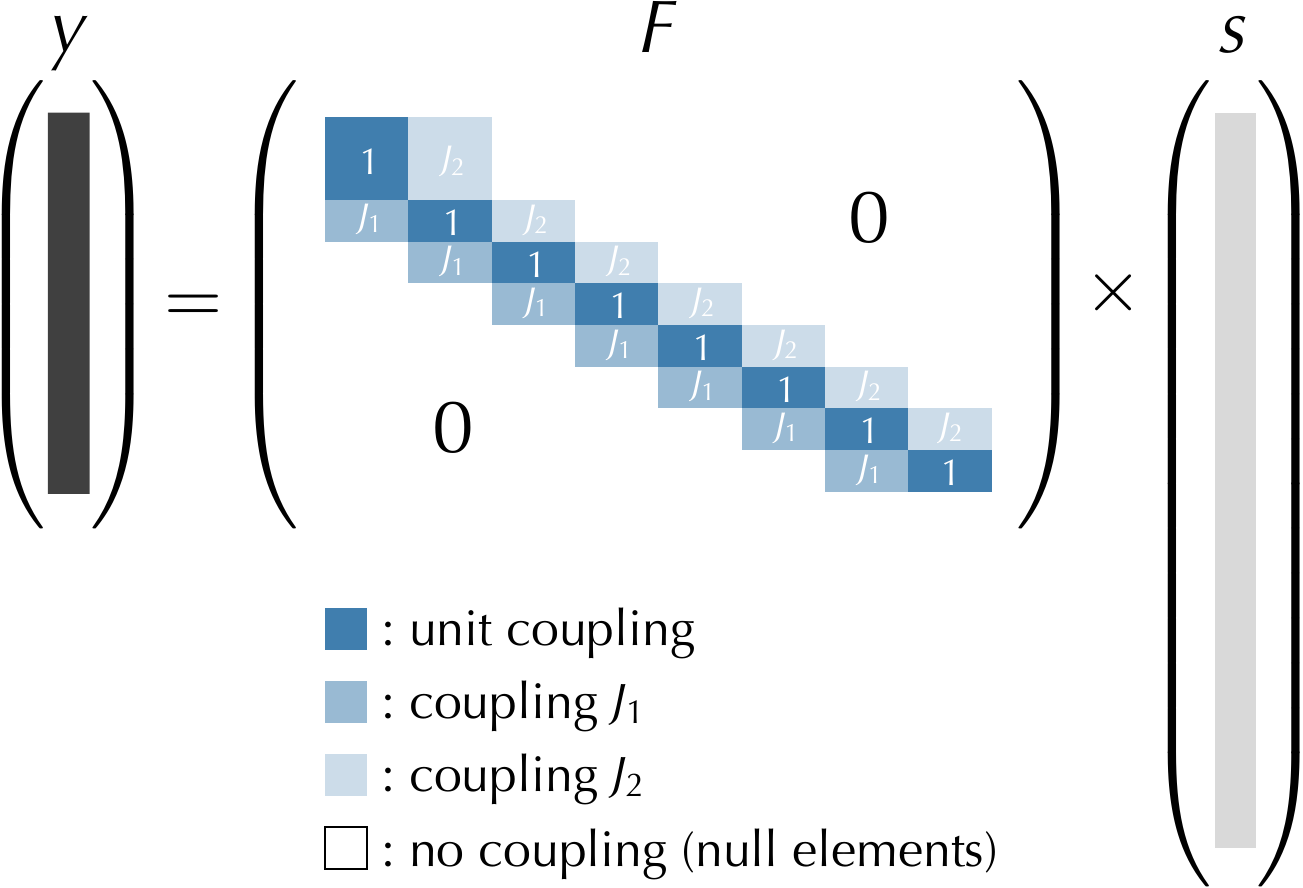}
    \end{tabular}
    \caption{Example of a spatially coupled measurement matrix $F$ for
      compressed sensing. The elements of the matrix are taken
      independently at random, their variance has the block diagonal
      structure as illustrated in the figure. The figure it taken from
   \cite{krzakala2012probabilistic}.}
    \label{fig:1d_matrix}
  \end{center}
\end{figure}

A spatially coupled matrix suggested in
\cite{krzakala2012statistical} is illustrated in
Fig.~\ref{fig:1d_matrix}. The physical mechanism of why the spinodal transition disappears
with such  spatially-coupled measurement matrices is exactly the same as for
the spatially coupled Currie-Weiss model discussed in section
\ref{sec:coupling}. The first block of the matrix has a larger effective
measurement rate and plays the role of a nucleation seed. The
interactions in the blocks next to the diagonal then ensure that the
seed grows into the whole
system. Ref. \cite{krzakala2012probabilistic} further discusses the choice of
    parameters of the spatially coupled measurement matrices, other
    related designs and the effect of noise. 
With the goal of optimizing the spatially coupled matrices the properties of the nucleation and necessary properties of
the seed in compressed sensing were studied in \cite{caltagirone2014properties}. 

%
There are several problems in the literature closely formally related
to compressed sensing to which spatial coupling was applied. One of
them is, again, the CDMA defined in sec.~\ref{CS_ex}. When the spreading
code is not random iid but correspond to a spatially coupled random
matrix then the performance of the CDMA system can be further improved
\cite{takeuchi2011improvement,schlegel2011multiple}.

Another problem formally very closely related to compressed sensing is
the real valued error correction where the codeword is corrupted by
gross errors on a fraction of entries and a small noise on all the
entries. In \cite{barbier2013robust} it was shown that the error
correction and its robustness towards noise can be enhanced
considerably using AMP and spatial coupling. 

A similar application of these ideas has allowed for the development
of the spatially coupled sparse superposition
codes~\cite{joseph2012least,barbier2014replica,barbier2015approximate,rush2015capacity,barbier2015approximate2}. These
are efficient coding schemes using exactly the compressed sensing
setting with a twist: the variables are $L$-dimensional and are forced
to point into one corner of the $L$-hypercube. These codes, with the
associated AMP decoder, have been shown to be fast and reliable, and
actually reach the Shannon limit for the AWGN channel in the large $L$
limit. It is actually expected that they are Shannon achieving
for any memoryless channel.

\subsection{Non-random matrices and non-separable priors}
\label{Sec:structured}

\subsubsection{Structured priors}
AMP as described above uses the empirical distribution of signal
elements. A natural question is whether its performance can be
improved further if more information is known about the signal. 
There are works that have
investigated this question and combined AMP with more complex, structured
priors. Utilizing such structured priors is the key to leveraging many of the
advancements recently seen in statistical signal representation. For
instance techniques such as hybrid AMP
\cite{schniter2010turbo,rangan2011hybrid} have shown promising results. In those works, an additional graphical model is
introduced in order to take into account the correlation between
variables. The improvement, for instance for image denoising, can be
spectacular.

Another possible approach to structured priors is not to model the correlations
directly, but instead to utilize a bipartite construction via hidden
variables, as in the restricted Boltzmann machine (RBM)
\cite{Hin2002,Hin2010}. If a binary RBM can be trained to model the
support pattern of a given signal class, then the statistical
description of the RBM admits its use within AMP, as was shown
recently in \cite{tramel2015approximate}. This is particularly
interesting since RBMs are the building blocks of ``deep belief
networks'' \cite{bengio2009learning} and have recently sparked a surge
of interest, partly because of the efficient algorithms developed to
train them (e.g. contrastive divergence (CD) \cite{Hin2010}). The
possibility, shown in \cite{tramel2015approximate}, to incorporate
deep learned priors into generalized linear problems such as
compressed sensing appears very promising. 

\subsubsection{Orthogonal matrices}
Using non-random matrices for the linear projection is crucial to
explore the range of applicability. The $\ell_1$-minimization based
approaches provably work well for a much larger class of matrices than
random ones \cite{hastie2015statistical}. It is therefore important to
explore performance of the approach stemming from statistical physics
for more general matrices. There has been a
collection of works in this direction, for instance in the analysis
of the so-called perceptron learning with correlated pattern (which
form the sensing
matrix) \cite{kabashima2008inference,shinzato2008perceptron}. 

As we have mentioned in Sec.~\ref{sec:TAP}, there are many works in the
context of the Ising model that have extended the TAP approach for more
generic random matrices. Similar contributions have recently emerged
in the context of AMP as well: the recent work on the so-called S-AMP
algorithm that extends the approximate message-passing algorithm to
general matrix ensembles when there is a certain well-defined large
system size limit, based on the so-called $S$-transform of
the spectrum of the measurement matrix~\cite{cakmak2014s}. Older
works, in fact, already used a similar approach for the CDMA
problem~\cite{takeda2006analysis}. Analysis of performances in this
direction is also doable
\cite{kabashima2014signal,liu2015generalized}. This shows that there
are still interesting open problems in this direction and we feel that
there is still much more to be done in the context of AMP beyond random
matrices.

\subsubsection{Reconstruction in discrete tomography}
\label{sec:tomo}
Signal reconstruction in x-ray computed tomography (CT) is another problem
that belongs to the class of linear estimation. The matrix $F_{\mu i}$
corresponding to computed tomography has a very particular structure. 
Classical reconstruction algorithms such as filtered back-projection 
require the corresponding linear system to be sufficiently determined. Therefore, the number of
measurements needed is generally close to the number of pixels to be reconstructed. Nevertheless,
many applications would benefit from being able to reconstruct from a smaller number
of angles. Ref. \cite{gouillart2013belief} designed and tested a message passing
algorithm that is adapted to the measures performed in CT and is able
to provide a reliable reconstruction with a number of measurements that
is smaller than what was required by previous techniques. This
algorithm is another example of how inspiration from physics can
provide a promising algorithmic approach. 

Each measurement in x-ray CT is well approximated by a sum of
absorption coefficient of the material under investigation along a
straight line of the corresponding x-ray. Let us discretize the object
under study into small elements, and assign a discrete value $x_i\in \{1,\dots,q\}$ of the absorption coefficient
to every small element $i$. One component of the measurement in CT is then 
\be
 y_{\mu} = \sum_{i\in \mu } x_i  \, ,
\ee 
where the sum is over all elements that lie on the line $\mu$. The
structural information that in general allows us to decrease the
number of measurement is that two pixels that lie next to each
other mostly have the same value. This expectation is represented in the
following posterior distribution 
\be
     P(\bx| \by) = \frac{1}{Z} \prod_{\mu =1 }^M\left[ \delta\left( y_\mu -
       \sum_{i \in \mu} x_i \right) e^{J_\mu \sum_{(ij)\in
           \mu}\delta_{x_i,x_j}} \right]\, ,
\ee
where the notation $(ij)\in \mu$ means the elements $i$ and $j$ are
next to each other on the line corresponding to measurement $\mu$, $J_\mu$
is the interaction constant. 

Ref. \cite{gouillart2013belief} writes the belief propagation algorithm
for estimating marginals of this posterior distribution. However, one
update of this belief propagation involves intractable sums over all
configurations of variables on one line. We realize that these sums
correspond to the partition functions of one-dimensional Potts model
with fixed magnetization. A one-dimensional line is a tree and for trees
partition functions can be computed exactly with belief propagation. We
use this to design an algorithm that uses BP within BP. The resulting
implementation is comparably fast to the competing convex relaxation based
algorithms and provides better reconstruction and robustness to
noise. We believe that overall the algorithm suggested in
\cite{gouillart2013belief} provides a promising perspective for
reconstruction in x-ray computed tomography. 

\subsection{An application in physics: Optics in complex media}
We would like to close the loop with physics and mention one
explicit physics experiment in the field of compressed optics where
AMP and related tools have been especially useful.

Wave propagation in complex media is a fundamental problem in physics,
be it in acoustics, optics, or electromagnetism. In optics, it is
particularly relevant for imaging applications.  Indeed, when light
passes through a multiple-scattering medium, such as a biological
tissue or a layer of paint, ballistic light is rapidly attenuated,
preventing conventional imaging techniques, and random scattering
events generate a so-called speckle pattern that is usually considered
useless for imaging.

Recently, however, wavefront shaping using spatial light modulators
has emerged as a unique tool to manipulate multiply scattered
coherent light, for focusing or imaging in scattering
media~\cite{mosk2012controlling}.  This makes it possible to measure
the so-called transmission matrix of the
medium~\cite{popoff2010measuring}, which fully describes light
propagation through the linear medium, from the modulator de
vice to
the detector. Interestingly, these transmission matrices, due to the
fact that they account for the multiple scattering and interference, are essentially iid random
matrices~\cite{popoff2010measuring}, which is exactly what AMP techniques
have been designed for.

An important issue in most such recent experiments lies in
accessing the amplitude and phase of the output field, that in optics
usually requires a holographic measurement, i.e., the use of a second reference
beam. The phase and amplitude of the measured field can then be
extracted by simple linear combinations of interference patterns with
a phase shifted or off-axis reference. This however poses the
unavoidable experimental problem of the interferometric stability of
the reference arm. In order to be able to avoid entirely the use of
the reference
beam, we need to solve
a problem that looks like compressed sensing: finding $\bf x$ (sparse
or binary) such that ${\bf y}=|F {\bf x}|$ where $F$ is complex and
iid.  Combining the G-AMP and the SwAMP approach for phase retrieval
\cite{schniter2015compressive,manoel2015swept}, it has been possible
to precisely do this, to calibrate efficiently, and to perform a
reference-less measurement of the transmission matrix of a highly
scattering material.  Within this framework, the work of
\cite{dremeau2015reference} showed that the transfer matrix can still be
retrieved for an opaque material (a thin layer of white paint), and
can then be used for imaging or focusing, using the opaque complex
material as a lens. This was experimentally validated.

As a result the full complex-valued transmission matrix of a
strongly scattering material can be estimated, up to a global phase,
with a simple experimental setup involving only real-valued inputs and
outputs in combination with the techniques presented in this review.
We believe this illustrates the power of the techniques
presented in this review, and shows that statistical physics does not only provide
interesting perspective to inference problems, but also
that inference techniques can be useful for physics experiments in return.

\subsection{Beyond linear estimation: matrix factorization}
The technics that have been used in the context of (generalized)
linear estimation are not limited to this setting. In fact, AMP can also be applied to the
more challenging setting of matrix factorization.

Decomposing a matrix into a product of two matrices of given
properties is another problem with an huge range of
applications. Mathematically stated, we are given a noisy measurement
of a matrix~$Y$ of dimension $M \times N$ and we aim to decompose it
into a product $Y=DX$, where $D$ is a $M \times R$ matrix, and $X$ is
a $R \times N$ matrix. Depending on the application the problem is
subject to various constraints such as sparsity of $D$ or $X$, low
rank $R$, or specific
models for the noise. Such constraints turn the matrix factorization
into an algorithmically extremely challenging problem. Let us list
several examples of potential applications

\begin{itemize}
\item{Dictionary learning: Many modern signal processing devices
take a great advantage of the fact that almost every class of signals
of interest has a sparse representation, i.e. there exists a basis
such that the observed signal can be written as a sparse linear
combination of atoms (i.e. columns of the corresponding matrix) from
this basis. Widely known, studied and used examples include the
wavelet basis for images, of Fourier basis for acoustic signals. For
other examples of interest, especially those which utilize uncommon or
novel data types, a suitable basis for sparse representation may not
yet be known. The goal of dictionary learning
\cite{mallat1993matching,elad2006image,mairal2009online} is to learn
such a basis purely from samples of the data. Naturally the more
samples are available the easier is this task. The challenge is hence
to design an algorithm that would be able to learn the dictionary
efficiently from the smallest possible number of samples.

In the mathematical statement of the problem the $N$ columns of
matrix~$Y$ represent the $N$ samples of the data for which we aim to
find a sparse representation. $M$ is then the dimension of each of the
data points. The dictionary learning problem is a decomposition of the
matrix~$Y$ into a product $Y=DX + W$, where $D$ is the dictionary of
$R$ so-called atoms, and $X$ is the sparse representation of the data,
having only a fraction $\rho$ of non-zero elements in each column. $W$
is a possible noise that may be taking into account the approximative
nature of the sparse representation.}
\item{Feature learning in neural network: Arguably one of the
most impressive engineering success of the last decade is the design
of the so called ``deep-learned" neural networks
\cite{hinton2006fast,hinton2006reducing,bengio2007greedy}. Using
these, computers are now able to recognize people on video, to tell a
dog from a cat on a picture, to process speech efficiently and even to
answer complicated questions. At the roots of this efficiency is the
ability to learn features. The simplest building block of
deep neural networks
can be seen as a matrix factorization problem with an
output $P_{\rm out}(y|z)$ corresponding to the activation
functions.
}
\item{Blind source separation:  A typical example of blind
source separation \cite{belouchrani1997blind,jung2000removing} are $M$
sensors recording noise at a party during time~$N$. We then aim to
separate the speech of the $R$ participants of the party based on
these recordings. Recording at one of the sensors is an unknown linear
combination of the speech of all the present people. A particularly
challenging, but also very interesting application-wise, is when the
number of sensors is smaller than the number of people, $M<R$. The
individual speeches may then be in principle recovered only if they
can be represented as sparse in some known basis.
}
\item{Sparse PCA:  In PCA the aim is to approximate the
matrix~$Y$ by a low-rank matrix, i.e. columns of $Y$ are written as a
linear combination of only few elements. This reduces the information
needed to describe the data contained in $Y$. This is usually done via
the singular value decomposition (SVD). In some applications it is
desirable that the linear combination to be sparse
\cite{johnstone2004sparse,zou2006sparse}, thus further reducing the
necessary information. The sparsity constraint cannot be
straightforwardly included in the SVD and hence alternative
algorithmic approaches are needed.
}
\end{itemize}

\subsubsection{Learning random sparse dictionaries}
The matrix factorization problem can again be studied within the
teacher-student scenario: The teacher generates a the matrices $F$ and
$X$ with random iid entries, from some probability distribution
(possibly sparse) and provides to the student element-wise measurement
of the product $FX$ via an output function $P_{\rm out}(Y|FX)$. Let us
consider the dimensions $M,N,R$ to be all of the same order.  Both the
AMP
\cite{krzakala2013phase,kabashima2014phase,vila2013hyperspectral,parker2014bilinear}
and the replica approach
\cite{sakata2013statistical,sakata2013sample,krzakala2013phase,kabashima2014phase} can
be applied to such problems. In this the problem of identifiability of
$F$ and $X$ from $Y$, the corresponding sample complexity and the
associated Bayes-optimal mean squared error can be derived. We refer
to these works for the phase diagrams for special cases of the problem
such as the sparse dictionary learning, blind source separation,
sparse PCA, robust PCA or matrix completion.

These results are very striking because in a number of relevant cases
(including the dictionary learning and source separation) it is
observed that the ideal performance predicted by this approach is far
better than the one achieved by current algorithms. This suggests that
there is a lot to gain in algorithmic efficiency in such
problems. Making the algorithm of
\cite{parker2014bilinear,kabashima2014phase} more robust along the
lines described above for AMP for linear estimation and applying it to
more realistic data is a very promising subject of future work.

\subsubsection{Low rank decomposition}
In some part of the applications mentioned above (e.g. the PCA) the
dimension $R$, corresponding to the rank of matrix~$Y$, is considered
as very small. In those cases it is reasonable to consider the limit
where $M$ and $N$ are comparable and both $N,M\to \infty$ whereas
$R=O(1)$. The graphical model corresponding to this case has
$R$-dimensional variables and pairwise interactions. Problems that can
be represented via this graphical model include variations of
PCA, submatrix localization problem, biclustering, clustering.

Again AMP type of algorithm can be written for a general element-wise
output function $P_{\rm out}(Y|FX)$, and the associated state
evolution leads to an analysis of the phase diagram, phase transitions
and optimal mean squared errors. Interestingly in this approach the
prior distribution can be any $R$-variate distribution, thus allowing
for instance for priors representing membership to one or more
clusters.

A problem corresponding to the
rank one matrix factorization, clustering in the Gaussian mixture
model, was studied long time ago using the replica method \cite{watkin1994optimal,biehl1994statistical,barkai1994statistical}. It was
presented as a prototypical example of unsupervised learning, see
e.g. chapter 8 in the textbook \cite{Engel:2001:SML:558792}. Note,
however,  that many more settings closer to current unsupervised
learning architectures should be analyzed. The revival of interest in
low-rank matrix factorization was linked with the development of the
corresponding approximate message passing algorithm 
by Rangan and Fletcher for rank $R=1$ \cite{rangan2012iterative}, and Matsushita and
Tanaka for general rank \cite{matsushita2013low} followed by
rigorous work of \cite{deshpande2014information}. Recently, a version for generic rank and
generic output (as in G-AMP) has been presented and analysed in
\cite{lesieur2015phase,lesieur2015mmse}. The resulting algorithm when
tested numerically has good convergence properties and these works
hence open the stage to further studies and applications along the
line discussed in this review.

As a final comment, it is to be noted that the phase transitions
observed in the matrix factorization setting are
again of the same kind that we have described all along this review,
schematically in Fig.~\ref{fig:schema}. In many problems, the presence
of distinct algorithmic and information theoretic thresholds is
observed. In this setting, it is particularly interesting to
characterize when spectral methods~\cite{baik2005phase}, the most
widely used approach for low rank factorization, are optimal, and when
they are not. As shown in,
e.g. \cite{deshpande2015finding,lesieur2015phase,lesieur2015mmse,montanari2015limitation},
there are a number of situations where they are sub-optimal. 
This is a valuable information one can get from the physics
approach, that should motivate the creation of new algorithms
approaching the best possible performance.

\section{Perspectives}
This review discusses recent progress in the statistical physics
approach to understanding of different problems of statistical
inference. Just as theoretical physics often focuses on understanding
of idealized models that represent in a simplified way the salient
features of realistic settings, we focus on the teacher-student
setting under which the Bayes-optimal inference can be analyzed
without ambiguity, and the phase transition can be identified and
discussed.

The main concept that we presented in detail is the mathematical
relation between the Bayes-optimal inference and properties of
disordered systems on the co-called Nishimori line. We discussed in
detail various phase transitions that arrise in the Bayes-optimal
inference and related them to statistical and computational thresholds
in the corresponding inference problems.  The resulting picture, that
we explain in detail on the simple example of the {\it planted} spin
glass, is relevant for a large class of inference problems. In later
sections we illustrate this picture on the problem of clustering of
sparse networks and on the problem of generalized linear estimation.
In today's scientific landscape the number of very interesting
data-driven problems is considerable and, in the authors' opinion, in
a considerable number of them the strategy presented in this review
will bring exciting new results.

Let us recall here the main theme of the present review. For all the presented
problems, the central scientific questions that one can attempt to
answer are: (1) Under what conditions is the information contained in
the measurements sufficient for a satisfactory inference to be
possible?  (2) What are the most efficient algorithms for this task?
In the probabilistic setting discussed here, what is striking is that all
of these problems have a very similar phenomenology, that also
appeared earlier when physicists studied error correcting codes and
perceptron learning, and that often these two questions can be precisely
answered. When the amount of information is too low a successful
inference of the signal is not possible {\it for any algorithm}: the
corresponding information is simply insufficient.  On the other hand,
for large enough amount of data, inference is possible, and the two
regime are separated by a sharp phase transition. Finally, and perhaps
more importantly, there is often (as in first order transition in
physics), an intermediate regime where a successful inference is in
principal possible but algorithmically hard. We have shown many
example where these transitions can be computed. A first order phase
transition is always linked to appearance of computational hardness,
whereas a second order phase transition is not. 

There is a deep value in this knowledge: If these transitions were not
known, one could not know how good are the state-or-the-art
algorithms. To give an example, it is the very knowledge of the
precise Shannon bound that is driving research in information theory: if we know that we could be
better in principle, when it is worth working on better algorithms. As
we have discussed in the very last chapter, for instance, present day
matrix factorization algorithms are often very far from optimality.

It is thus natural that, next to the theoretical understanding, an
important outcome of the research reviewed in this manuscript are new
algorithmic ideas that are applicable beyond the idealized setting for
which the presented theoretical results hold. This translation of
theoretical physics calculations into algorithmic strategies is very
fruitful, and promising direction of research. What is, perhaps, more
challenging is to push such algorithmic ideas into actual
applications. From a point of view of a physicist, some of the
problems presented here may seem very applied, but in reality the
research presented here is positioned on the theoretical side of the
fields of statistics, machine learning or signal processing.  There is
still a large step towards actual applications. Surely this is a very
exciting direction to explore and many of the researchers active in
this field do explore it.

Let us now discuss some of the limitations and challenges related to
the results and concepts presented in this review. 

On the algorithmic side a large part of the algorithms that stem from
the research presented in this review involve some kind of message
passing derived under various assumptions of correlation decay. These
assumptions rarely hold of real datasets and hence sometimes
this causes a failure of the corresponding algorithm. In our opinion
the theoretical potential of these message passing algorithms is very
large, but much work needs to be done in order to render these
algorithms more robust and suitable for a generic problem.

Computer science and related fields mostly aim at understanding the
worst possible setting or to prove something under the weakest
possible conditions. Analysis in a setting such as the teacher-student
scenario usually does not seem generic enough. In machine learning and
more applied parts of computer science the focus is on a couple of
benchmark-datasets (such as e.g. the MNIST database). Simple models
that aim to represent reasonably well the generic situation are a
domain of physical sciences. This is how progress in physics is made,
and in our opinion this is transferable to other fields as well.

Physicists can use with large effectiveness tools and methods from
statistical physics of disordered systems such as glasses and spin
glasses. The advantage of these methods is that, unlike any other
existing generic method, they are able to describe the Bayes-optimal
probabilistic inference. The disadvantage is that they are not fully
rigorous and are therefore sometimes viewed with dismissal or
skepticism by mathematically oriented researchers. On the other hand,
rigorous proofs of results obtained with the cavity and replica method
are a great mathematical challenge and many partial results were
obtained that have led to new mathematical tools making the whole
probability theory more powerful. One of the goals of the presented
research is to transfer the physics reasoning and insight to
mathematicians in order to help them to craft plausible proof
strategies.

\begin{table*}[!ht]
\begin{center}
\begin{tabular}[t]{|l||c|c|} \hline
 {\bf abbreviation}    & definition section & meaning     \\ \hline
 \hline
  MMO  & \ref{sec:estimators}  &  maximum mean overlap \\ \hline 
  MMSE & \ref{sec:estimators}  & minimum mean squared error \\ \hline 
  MAP   & \ref{sec:estimators}   & maximum a posterior estimator
  \\ \hline 
  MCMC   & \ref{sec:MC_VMF}   & Monte Carlo Markov chain \\  \hline 
  BP      & \ref{sec:MC_VMF}  & belief propagation \\ \hline 
  RS      & \ref{sec:RSB}       & replica symmetry \\ \hline 
  RSB    & \ref{sec:RSB}       & replica symmetry breaking \\ \hline 
  1RSB   & \ref{sec:RSB}      & one-step replica symmetry breaking \\ \hline 
  d1RSB & \ref{sec:RSB}      & dynamical one-step replica symmetry breaking \\ \hline 
  FRSB   & \ref{sec:RSB}      & full-step replica symmetry breaking \\  \hline 
 XOR-SAT & \ref{sec:foll}   & XOR-satisisfiability problem \\ \hline 
  ER      & \ref{sec:RandomGraphs}   & Erd\H{o}s-R\'enyi  \\ \hline 
  V-B    & \ref{sec:random}  & Viana-Bray model  \\ \hline 
  LDPC  & \ref{phase_various}  & low-density parity-check codes  \\ \hline 
  CSP    & \ref{phase_various}   & constraint satisfaction problem \\ \hline 
  K-SAT & \ref{phase_various}   &  K-satisfiability problem\\ \hline  
  PCA    & \ref{sec:NB_planted}   & principal component analysis \\ \hline 
  C-W   & \ref{sec:coupling}  & Curie-Weiss model \\ \hline 
  TAP    & \ref{sec:TAP}   & Thouless-Anderson-Palmer \\ \hline 
  SBM    & \ref{sec:community}  & stochastic block model \\ \hline 
  AWGN & \ref{chap:CS}  & additive white Gaussian noise \\ \hline 
  CDMA & \ref{CS_ex}   & code-division multiple access \\ \hline 
  LASSO  & \ref{CS_ex}    &   least absolute shrinkage and
  selection operator \\ \hline 
  AMP    & \ref{sec:AMP}  &  approximate message passing  \\  \hline 
  G-AMP   & \ref{sec:AMP}    & generalized approximate message passing  \\ \hline 
  r-BP & \ref{sec:rBP}   & relaxed belief propagation \\ \hline 
  RBM    & \ref{Sec:structured}  & restricted Boltzmann machine  \\ \hline 
 CT    & \ref{sec:tomo}  & computed tomography \\ \hline 
\end{tabular}  
\caption{\label{table_glossary} Glossary of abbreviations with the number of
  the section where the term is defined.}
\end{center}
\end{table*}


\section*{Acknowledgements}

We would like to express our gratitude to all our colleagues with whom
many of the results presented here have been obtained. In particular,
we wish to thank Maria-Chiara Angelini, Jean Barbier, Francesco
Caltagirone, T. Castellani, Michael Chertkov, Andrea Crisanti, Leticia
F. Cugliandolo, Laurent Daudet, Aurelien Decelle, Angelique Dr\'emeau,
Laura Foini, Silvio Franz, Sylvain Gigan, Emmanuelle Gouillart, Jacob
E. Jensen, Yoshyiuki Kabashima, Brian Karrer, Lukas Kroc, Srinivas
Kudekar, Jorge Kurchan, Marc Lelarge, Antoine Liutkus, Thibault
Lesieur, Luca Leuzzi, Andr\'e Manoel, David Martina, Marc M\'ezard,
Andrea Montanari, Cristopher Moore, Richard G. Morris, Elchanan
Mossel, Joe Neeman, Mark Newman, Hidetoshi Nishimori, Boshra Rajaei,
Sundeep Rangan, Joerg Reichardt, Federico Ricci-Tersenghi, Alaa Saade,
David Saad, Ayaka Sakata, Fran\,cois Sausset, Christian Schmidt, Christophe Sch\"ulke,
Guilhem Semerjian, Cosma R. Shalizi, David Sherrington, Alan Sly, Phil
Schniter, Eric W. Tramel, Rudiger Urbanke, Massimo Vergassola, Jeremy
Vila, Xiaoran Yan, Sun Yifan, Francesco Zamponi, Riccardo Zecchina and
Pan Zhang. 

A part of this manuscript has been written during a visit to UC
Berkeley as part of the semester long program on "Counting Complexity
and Phase Transitions", at the Simons Institute for the Theory of
Computing, which we thanks for the kind hospitality.

Finally, we also wish to thank Alena and Julie for their continuous
support and for letting us work, if only from time to time, on this
review.

\section*{Fundings}
Part of the research leading to the presented results has received
funding from the European Research Council under the European Union's
$7^{th}$ Framework Programme (FP/2007-2013/ERC Grant Agreement
307087-SPARCS).

\bibliographystyle{plain}
\bibliography{myentries}

\end{document}